\documentclass[aps,twocolumn,pra,superscriptaddress,tightenlines,pdflatex,longbibliography]{revtex4-1}
\usepackage{amssymb}
\usepackage{amsmath}
\usepackage{dcolumn}
\usepackage{multirow}
\usepackage{graphicx}
\usepackage{mathrsfs}
\usepackage{appendix}
\usepackage{booktabs}
\usepackage{color}

\usepackage[position=top,caption=false]{subfig}

\usepackage{url}
\usepackage[colorlinks]{hyperref}
\hypersetup{%
    plainpages=true,
    breaklinks=true,
    hypertexnames=false,
    pageanchor=true,
    colorlinks=true,
    linkcolor={blue},
    citecolor={red},
    urlcolor={blue},
    anchorcolor={black}
    }

\def \AND{{~~~{\rm and}~~~}}
\def \<{{\langle}}
\def \>{{\rangle}}
\def \tr{{\rm Tr}}

\newcommand{\mean}[1]{\mbox{$\langle #1\rangle$}}
\newcommand{\ket}[1]{\mbox{$|#1\rangle$}}
\newcommand{\bra}[1]{\mbox{$\langle#1|$}}

\newcommand{\cor}[1]{{g^{(#1)}(0)}}

\newcommand{\etal}{{\em et al.}}
\newcommand{\Arg}{{\rm arg\,}}
\newcommand{\Nthermal}{{n_{\rm th}}}
\newcommand{\EP}{{\rm EP}}

\begin{document}


\title{Two-photon blockade and photon-induced tunneling generated by squeezing}




\author{Anna Kowalewska-Kud\l{}aszyk}
\affiliation{ Faculty of Physics, Adam Mickiewicz University,
61-614 Pozna\'{n}, Poland}

\author{Shilan Ismael Abo}
\affiliation{ Faculty of Physics, Adam Mickiewicz University,
61-614 Pozna\'{n}, Poland}

\affiliation{Department of Physics, University of Duhok, Duhok,
Kurdistan Region, Iraq}

\author{Grzegorz Chimczak}
\affiliation{ Faculty of Physics, Adam Mickiewicz University,
61-614 Pozna\'{n}, Poland}

\author{Jan Pe\v{r}ina Jr.}
\affiliation{Joint Laboratory of Optics, Faculty of Science,
Palack\'{y} University, 17. listopadu 12, 771~46 Olomouc, Czech
Republic}

\author{Franco Nori}
\affiliation{RIKEN Center for Emergent
Matter Science (CEMS), Wako, Saitama 351-0198, Japan}
\affiliation{Department of Physics, University of Michigan, Ann
Arbor, Michigan 48109-1040, USA}

\author{Adam Miranowicz}
\affiliation{ Faculty of Physics, Adam Mickiewicz University,
61-614 Pozna\'{n}, Poland} \affiliation{RIKEN Center for Emergent
Matter Science (CEMS), Wako, Saitama 351-0198, Japan}

\date{\today}

\begin{abstract}
Inspired by the recent experiment of Hamsen \emph{et al.} [Phys.
Rev. Lett. {\bf 118}, 133604 (2017)], which demonstrated
two-photon blockade in a driven nonlinear system (composed of a
harmonic cavity with a driven atom), we show that two-photon
blockade and other nonstandard types of photon-blockade and
photon-induced tunneling can be generated in a driven harmonic
cavity without an atom or any other kind of nonlinearity, but
instead coupled to a nonlinear (i.e., squeezed) reservoir. We also
simulate these single- and two-photon effects with squeezed
coherent states and displaced squeezed thermal states.
\end{abstract}

\maketitle

\section{Introduction}
\label{sec1}

\subsection{Squeezed states of light}
\label{sec1A}

Squeezed states of light~\cite{Dodonov2002}, which have less
quantum noise in one quadrature than a coherent state, are a
powerful resource for quantum technologies. These include quantum
communication, improving the precision of optical measurements,
and fundamental spectroscopic tests of general relativity and
quantum mechanics~\cite{Walls1983, Loudon1987, DodonovBook,
DrummondBook, Andersen2016}. Although squeezed states were already
studied in 1927 by Kennard~\cite{Kennard1927} and the squeezing
operator was introduced in 1955~\cite{Infeld1955, Plebanski1956},
these states had not been attracting much attention for 50 years.
A real practical interest in squeezed states has been triggered
only 40 years ago by finding their first applications for
detecting gravitational waves via supersensitive
interferometry~\cite{Hollenhorst1979, Caves1980, Dodonov1980,
Caves1981}. Since the pioneering experimental generation of
squeezed states via four-wave mixing in 1985 by Slusher
\etal~\cite{Slusher1985}, shortly followed by two other
experiments~\cite{Wu1986,Shelby1986}, various methods of
squeezed-light generation have been implemented experimentally not
only for optical fields~\cite{Andersen2016}, but also for
microwave fields using superconducting quantum
circuits~\cite{Gu2017}. The first long-term practical applications
of squeezed-vacuum states were demonstrated in 2013 for increasing
the astrophysical limits of gravitational-wave detectors including
those in the Laser Interferometer Gravitational-Wave Observatory
(LIGO)~\cite{LIGO2013} and the Gravitational-Wave Observatory (GEO
600)~\cite{Grote2013}. Among many applications of squeezing, we
mention also recent proposals of an exponential enhancement of
light-matter interactions via squeezing~\cite{Bartkowiak2014,
Lu2015, Lemonde2016, Qin2018, Leroux2018, Qin2019} (for a review
see Ref.~\cite{Kockum2019}). Such increased interactions at the
single-photon level can fundamentally change nonlinear optical
effects, including photon blockade
(PB)~\cite{Ridolfo2012,LeBoite2016}. (This and other abbreviations
used in this paper are also defined in Table~\ref{table1}.) Here
we study multiphoton correlations in squeezed coherent states
(SCS), displaced squeezed thermal states (DSTS), and in light
generated by a driven harmonic cavity coupled to a squeezed
reservoir for generating (or simulating) various kinds of PB.

\begin{table}
\begin{tabular}{l|l}
    \hline\hline
    Full Name & Abbreviation \\
    \hline
    photon blockade     &  PB\\
    nonstandard photon blockade & NPB\\
    single-photon (two-photon) blockade     &  1PB (2PB)\\
    photon-induced tunneling  &  PIT \\
    two-photon (three-photon) tunneling &  2PT (3PT) \\
    squeezed coherent states & SCS \\
    displaced squeezed thermal states & DSTS \\
    photon antibunching & PAB \\
    \hline
    \hline
\end{tabular}
\caption{Abbreviations used in this paper.} \label{table1}
\end{table}

\begin{figure}[t]
\includegraphics[width=\columnwidth]{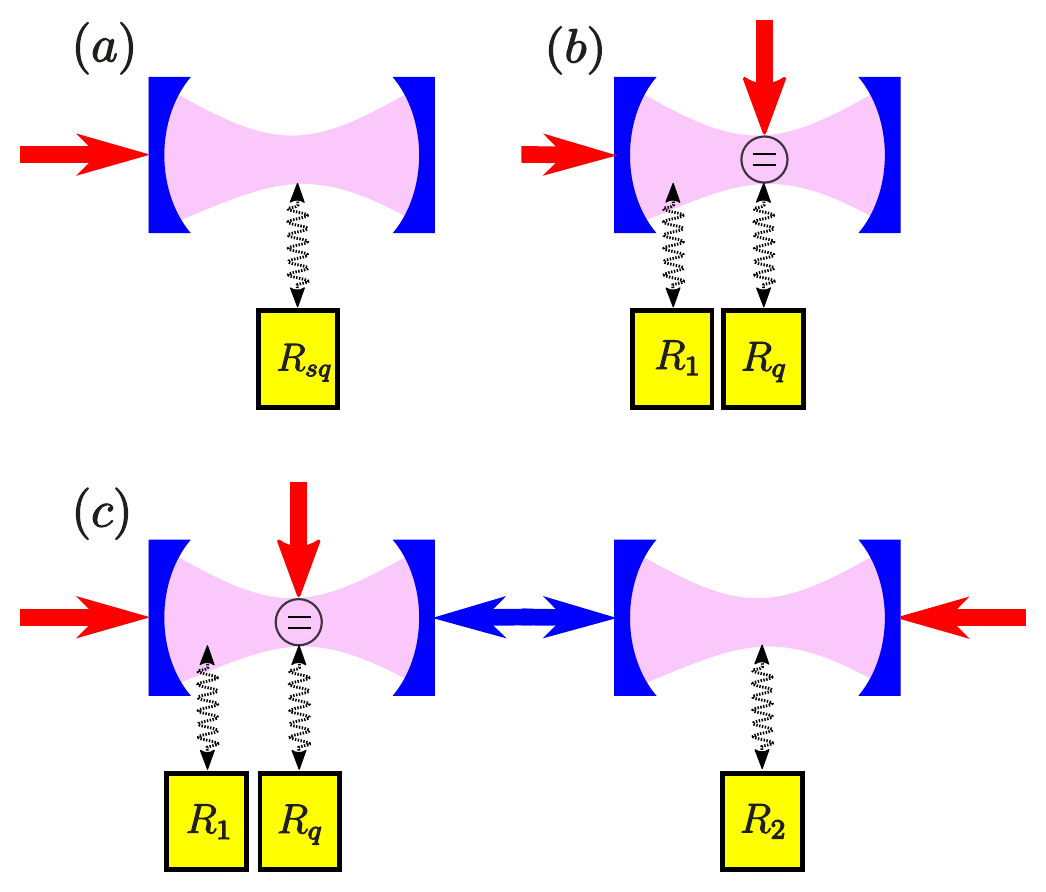}
\caption{Schematics of three prototype systems for observing
photon blockade and photon-induced tunneling: (a) an unusual
photon blockade device, described in Sec.~\ref{sec3A}, which is
composed of a driven harmonic cavity coupled to a quantum
(squeezed) reservoir $R_{\rm sq}$. Panel (a) is shown in contrast
to the common photon blockade devices (see Appendix~\ref{appA} for
more details): (b) a driven anharmonic cavity (due to the atom)
coupled to a harmonic reservoir $R_1$ and (c) a two-cavity system,
which is the anharmonic resonator shown in panel (b) coupled to a
harmonic (or anharmonic) resonator linked to a harmonic reservoir
$R_2$. The anharmonicity can be induced in a harmonic resonator by
its coupling to a two-level atom (qubit) as shown in panels (b)
and (c). This qubit is coupled to a reservoir $R_q$. Red arrows
denote classical coherent drives applied to a cavity or a qubit.
Note that in setup (a), the cavity anharmonicity is replaced by
the reservoir anharmonicity.} \label{fig01}
\end{figure}
\subsection{Single-photon blockade}
\label{sec1B}

The phenomenon of Coulomb's blockade has its optical analog, known
as PB~\cite{Imamoglu1997} (also referred to as nonlinear quantum
scissors~\cite{Leonski2011}). PB (or more precisely single-photon
blockade, 1PB) refers to the effect in which a single photon
generated in a driven nonlinear system (as those schematically
shown in Fig.~\ref{fig01}) can block the generation of more
photons in the system. This effect was first predicted by Tian and
Carmichael~\cite{Tian1992}, Leo\'nski and
Tana\'s~\cite{Leonski1994}, and later by Imamo{\u g}lu \emph{et
al.}~\cite{Imamoglu1997}, who coined the term \emph{photon
blockade} and studied the effect in the steady-state limit.
Indeed, Ref.~\cite{Tian1992} predicted PB by demonstrating a
two-state behavior in an driven optical cavity containing one
atom, as shown in Fig.~\ref{fig01}(b) and discussed in
Appendix~\ref{appA}, applying the quantum trajectory method to the
Jaynes-Cummings model; while Ref.~\cite{Leonski1994} predicted the
PB effect in a driven Kerr nonlinear cavity and showed its
application for the generation of the single-photon Fock state.
Note that the Jaynes-Cummings model in the dispersive limit (i.e.,
far off-resonance) becomes equivalent to the Kerr Hamiltonian,
which shows the correspondence of the PB predictions of
Refs.~\cite{Tian1992,Leonski1994}. We also mention that PB has a
mechanical analog referred to as phonon blockade, i.e., blockade
of quantum excitations of mechanical
oscillators~\cite{Liu2010,Didier2011,Adam2016,XinWang2016}.

PB has been experimentally generated in a number of driven systems
of single~\cite{Birnbaum2005, Faraon2008, Lang2011, Hoffman2011,
Reinhard2011, Peyronel2012, Muller2015, Hamsen2017} and
two~\cite{Snijders2018, Vaneph2018} resonators with a
nonlinearity, as shown schematically in Figs.~\ref{fig01}(b) and
\ref{fig01}(c), respectively. Such a nonlinearity can be induced
by a two-level atom (or atoms) coupled to one or both cavities. In
the dispersive regime, such atom-cavity interaction can
effectively lead to a Kerr-type nonlinearity as mentioned above.
Note that PB can be generated not only in a Kerr-nonlinear driven
cavity, but also other types of nonlinearities enable the
generation of PB. The occurrence of PB is usually experimentally
characterized by the second-order correlation function
$\cor{2}<1$, which means that the PB generated state exhibits the
sub-Poissonian photon-number statistics, also referred to as
(single-time) photon antibunching (PAB). PB and the generation of
Bell states in two-cavity driven nonlinear systems, as shown in
Fig.~\ref{fig01}(c) and discussed in Appendix~\ref{appA}, were
first demonstrated in Refs.~\cite{Leonski2004,Adam2006}. It was
later shown in~\cite{Liew2010,Bamba2011} that the nonlinear system
of Fig.~\ref{fig01}(c) can exhibit surprisingly strong single-time
PAB for weak nonlinearities or, equivalently, weak
atom-cavity-field couplings. This effect is now usually referred
to as unconventional PB~\cite{Flayac2018review}.

\begin{figure}[t]
\includegraphics[width=0.9\columnwidth]{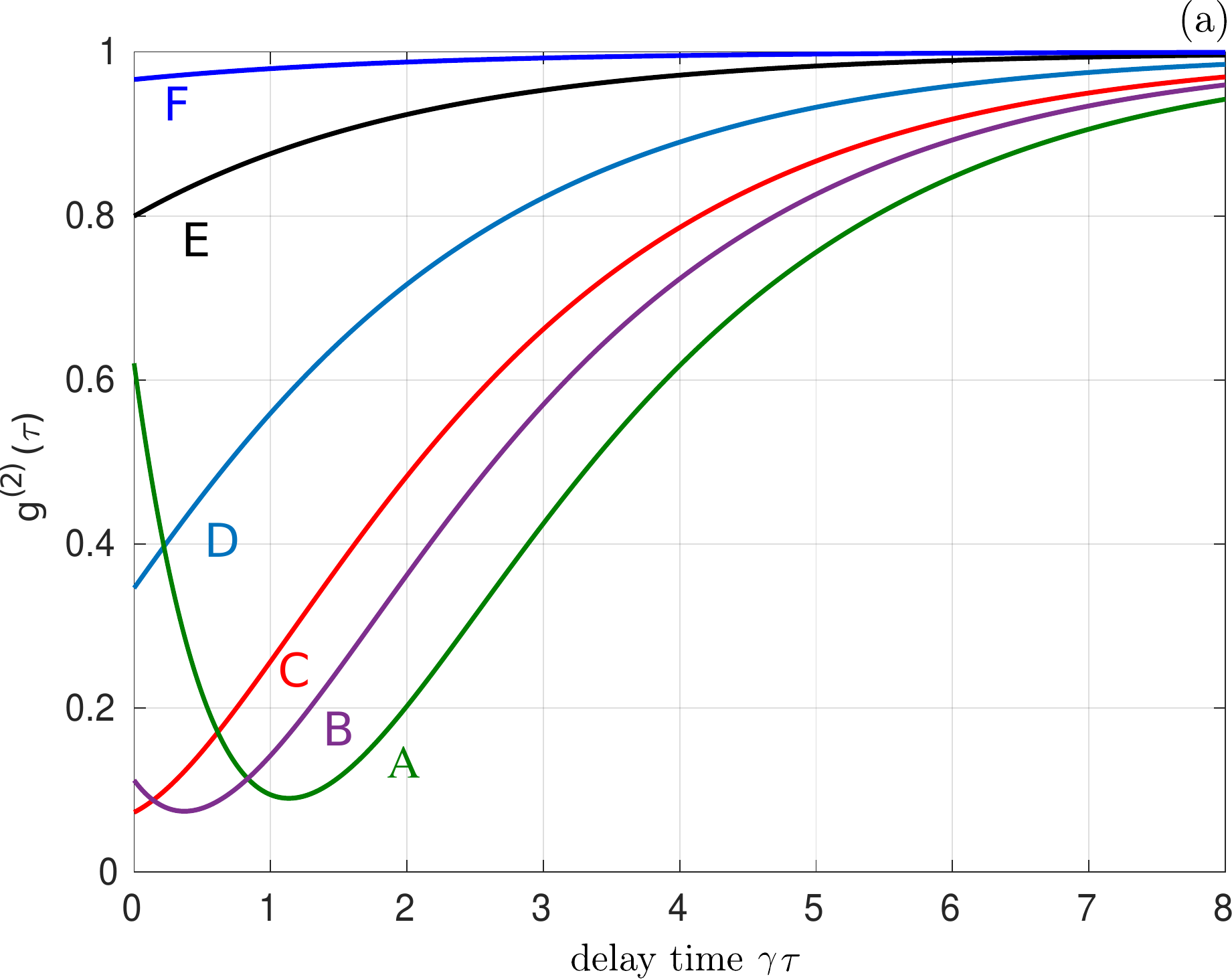}\\
\includegraphics[width=0.9\columnwidth]{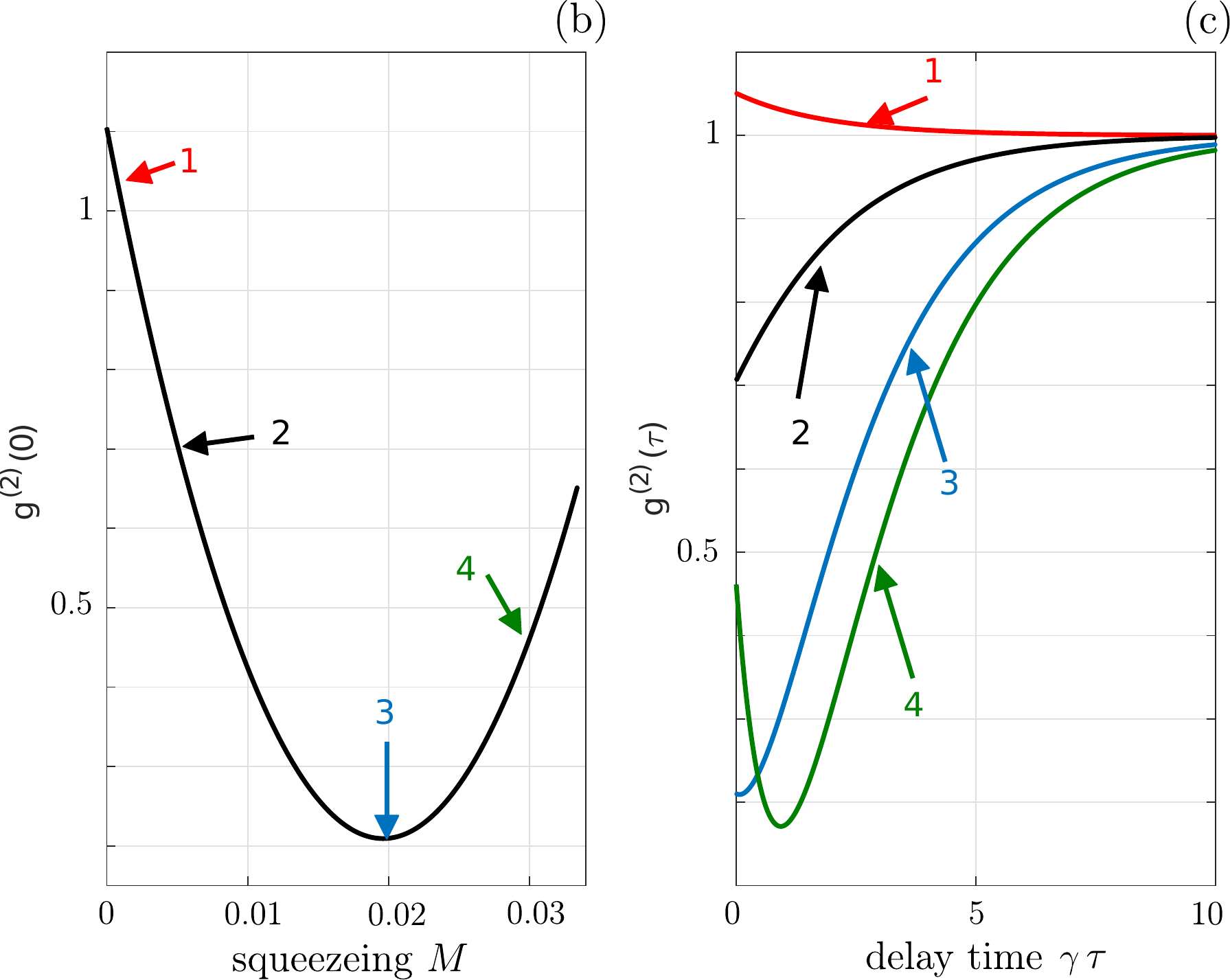}
\caption{Squeezed-reservoir model: Steady-state second-order
correlation function $g^{(2)}(\tau)$ vs.: (a,c) the (rescaled)
delay time $\gamma\tau$ between the measurements of subsequent
photons and (b) the reservoir squeezing parameter $M$ for various
values of the external field strength $\varepsilon$ at the
resonance, $\Delta=0$, between the cavity and external fields,
with the damping rate $\gamma=1$. Moreover, in panel (a) {we set
$\varepsilon/\gamma=$0.05 (curve A), 0.06 (B), 0.07 (C), 0.1 (D),
0.2 (E), and 0.5 (F), and assume} that the reservoir is maximally
squeezed with the reservoir mean photon number $n=0.003$, which
corresponds to $M=0.017$. The $\tau$-dependences for the four
specific points in panel (b) are shown in panel (c). In panels
{(b) and (c) we set $\varepsilon/\gamma=0.07$ and $n=0.001$.} It
is evident that the curves 2 and 3 (1 and 4) show two-time photon
antibunching (bunching) in panel (c). This implies that only the
points 2 and 3 in panel (b) can correspond to ``true''
single-photon blockade states.}
 \label{fig02}
 \end{figure}

Note that this single-time PAB, should not be confused with
standard two-time PAB, defined by $g^{(2)}(\tau)>\cor{2}$ for
small delay times $\tau$, which is another important feature of
PB. Indeed, if one considers single-PB as a true source of single
photons, one would be required to satisfy not only single-time
PAB, but also two-time PAB, characterized by a local minimum of
the second-order correlation function,
\begin{equation}
g^{(2)}(\tau )=\lim_{t\rightarrow \infty} \frac{\langle \hat
a^{\dag }(t)\hat a^{\dag }(t+\tau) \hat a(t+\tau)\hat a(t)\rangle
} {\langle \hat a^{\dag }(t)\hat a(t)\rangle \langle \hat a^{\dag
}(t+\tau)\hat a(t+\tau)\rangle}, \label{g2tau}
\end{equation}%
as a function of the delay time $\tau\approx 0$, where $\hat a$
($\hat a^\dagger$) is the annihilation (creation) operator of an
optical mode. Thus, at least the following conditions should be
satisfied for ``true'' single-PB:
\begin{equation}
  \cor{2}<1 \quad {\rm and} \quad \cor{2}< g^{(2)}(\tau),
  \label{type0}
\end{equation}
for small $\tau$. For brevity, we analyze two-time PAB only in
Sec.~\ref{sec3} and Fig.~\ref{fig02}. Otherwise we limit our
characterization of PB to single-time correlation functions.

\subsection{Multi-photon blockade}
\label{sec1C}

Single-PB has been generalized to include two-PB and multi-PB
effects~\cite{Shamailov2010Carmichael, Adam2013, Hovsepyan2014,
Carmichael2015, Deng2015, Zhu2017, Felicetti2018a, Felicetti2018b,
Huang2018, Li2019}. Two-PB was first experimentally demonstrated
by Hamsen \etal~in 2017~\cite{Hamsen2017}. We also note earlier
theoretical works on multi-PB in dissipation-free driven Kerr
systems~\cite{Adam1996, Leonski1996} (for reviews see
Refs.~\cite{Leonski2001, Leonski2011}). Multi-\emph{phonon}
blockade, which is a mechanical analog of multi-PB, was studied in
Ref.~\cite{Adam2016}. Multi-PB in dissipation-free systems enables
generation of quantum optical states in a finite-dimensional
Hilbert space including finite-dimensional analogs of coherent and
squeezed states of light~\cite{Adam1996, Leonski1997, Adam1994,
Adam2001, Leonski2001}.

Intuitively, two-PB (and analogously multi-PB) occurs if the
single and two-photon Fock states, which are generated in a driven
nonlinear system, block the generation of more photons in the
system. This paper is focused on the \emph{study of two-PB and
other kinds of single- and two-photon correlations.}

For any classical states, the second-order equal-time correlation
function satisfies $\cor{2}\geq 1$, which is a property of
classical intensity fluctuations. The states for which $\cor{2}<1$
have the sub-Poissonian photon-number statistics and, thus, are
\emph{nonclassical} (see Appendix~\ref{appC}). This condition is
also used for identifying the presence of single-photon blockade
(1PB). The analysis of higher-order correlations is necessary to
characterize multi-PB or other types of nonstandard PB (NPB).

Thus, in our study of multi-PB, we apply the $k$th-order
equal-time correlation functions, $\cor{k}=
\langle(\hat{a}^{\dagger})^k\hat{a}^k\rangle/\langle\hat{a}^{\dagger}\hat{a}\rangle^k$,
describing the probability of measuring simultaneously $k$
photons. In PB experiments, the second-order correlation functions
$\cor{2}$ and $g^{(2)}(\tau)$ are usually measured, except the
experiment of Hamsen \etal~\cite{Hamsen2017}, where also the
third-order correlation functions $g^{(3)}(0)$ and $g^{(3)}(\tau)$
were measured to confirm two-PB.

We note that experimental tests of PB are not limited to measuring
$\cor{k}$ and $g^{(k)}(\tau)$. Indeed, the occurrence of PB can
also be revealed by showing, e.g., a staircase dependence of the
total transmitted power through a driven nonlinear system for
different incident photon bandwidths, which was experimentally
demonstrated by Hoffman \etal~\cite{Hoffman2011} or a staircase
dependence of the mean photon number in the ground state of a
given Kerr nonlinear system on a rescaled detuning~\cite{Liu2014}.
Such dependences are photonic analogs of a Coulomb-blockade
staircase. This paper is focused on characterizing multi-PB via
$\cor{k}$ and $g^{(2)}(\tau)$ only.

\subsection{Photon-induced tunneling}
\label{sec1D}

Photon-induced tunneling (PIT) refers to a photon-number
correlation effect, which \emph{enhances} the probability of
subsequent photons (from a coherent drive) to enter the driven
cavity~\cite{Faraon2008, Majumdar2012a, Majumdar2012b, Xu2013,
Rundquist2014, Huang2018, Li2019, Zhai2019}. Evidently, this
process is inverse to PB, in which the probability, that the
subsequent photons of a drive enter the driven cavity, is
\emph{decreased} (or even essentially vanishing). PIT has been
observed experimentally in Refs.~\cite{Faraon2008, Majumdar2012a,
Rundquist2014}.

Standard two-photon tunneling (two-PT), where the simultaneous
arrival of two photons is enhanced compared to single-photon
arrivals, is usually characterized by the super-Poissonian
photon-number statistics (i.e., single-time photon bunching), when
$1<\cor{2}$~\cite{Majumdar2012a, Majumdar2012b, Xu2013}.
Analogously, standard three-photon tunneling (three-PT) is a
photon-number correlation effect, in which the simultaneous
arrival of three photons is enhanced compared to the two-photon
and single-photon arrivals. Thus, three-PT can be characterized by
the conditions~\cite{Xu2013,Huang2018}:
\begin{equation}
  1<\cor{2}<\cor{3}. \label{3PT}
\end{equation}
Note that other definitions of PIT are used in the literature (see
Ref.~\cite{Huang2018} for a comparison), e.g., those based on a
local maximum of $g^{(2)}(\tau)$ at $\tau=0$ (i.e., corresponding
to two-time photon bunching)~\cite{Faraon2008} or the requirement
that $\cor{3}>\cor{2}$, i.e., the simultaneous arrival of three
photons is enhanced compared to the simultaneous two-photon
arrivals~\cite{Rundquist2014} without specifying whether
$g^{(2)}(0)$ exhibits the super- or sub-Poissonian statistics.
Various types of PIT in comparison to PB are listed in
Table~\ref{table2}.

\begin{table}[t]
\begin{tabular}{cccc}
\hline
 Case & Permutation & Inequalities\quad & \hspace{7mm} Effect \hspace{7mm}  \\
 \hline
 (a) & (1 2 3) & $1<\cor{2}<\cor{3}$ & 3PT \\[2pt]
 (b) & (1 3 2) & $1<\cor{3}<\cor{2}$ & 2PT \\[2pt]
 (c) & (2 1 3) & $\cor{2}<1<\cor{3}$ & 1PB (type 2) \\[2pt]
 (d) & (2 3 1) & $\cor{2}<\cor{3}<1$ & 1PB (type 3) \\[2pt]
 (e) & (3 1 2) & $\cor{3}<1<\cor{2}$ & 2PB \& 2PT\\[2pt]
 (f) & (3 2 1) & $\cor{3}<\cor{2}<1$ & 1PB (type 1) \\[2pt]
\hline
\end{tabular}
\caption{Different types of photon blockade and photon tunneling
classified via $\cor{2}$ and $\cor{3}$. Four of these types of
photon-number correlations can be exhibited by the steady-state
light generated by the squeezed-reservoir system, as well as
squeezed coherent states and displaced squeezed thermal states,
which are shown in Figs.~\ref{fig03}, \ref{fig04}, and
\ref{fig05}, respectively.} \label{table2}
\end{table}

\begin{figure}[t]
\begin{center}
\subfloat[$(123)$ three-PT]{\includegraphics[scale=0.40]{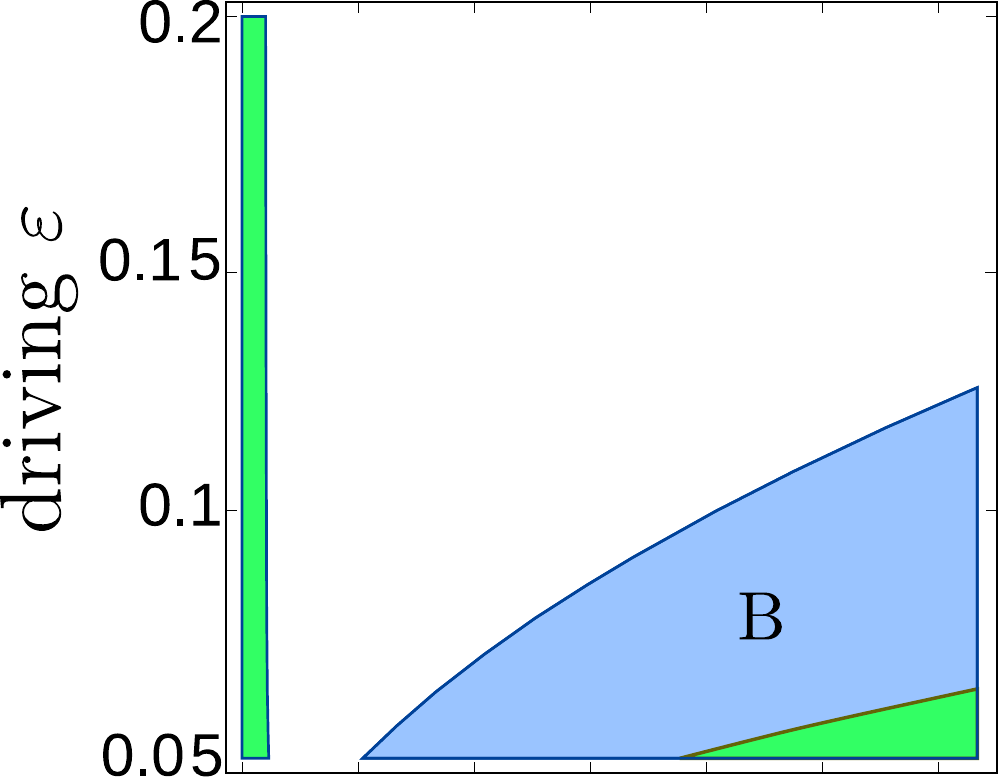}}\hspace*{2pt}                 
\subfloat[$(132)$ no two-PT]{\includegraphics[scale=0.40]{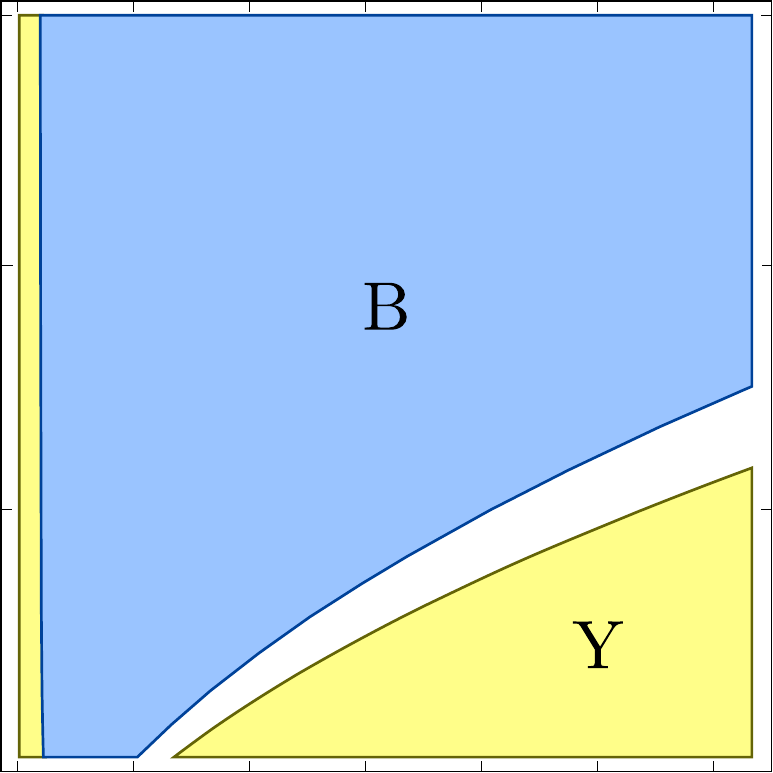}}\\               
\subfloat[$(213)$ single-PB (type 2)]{\includegraphics[scale=0.40]{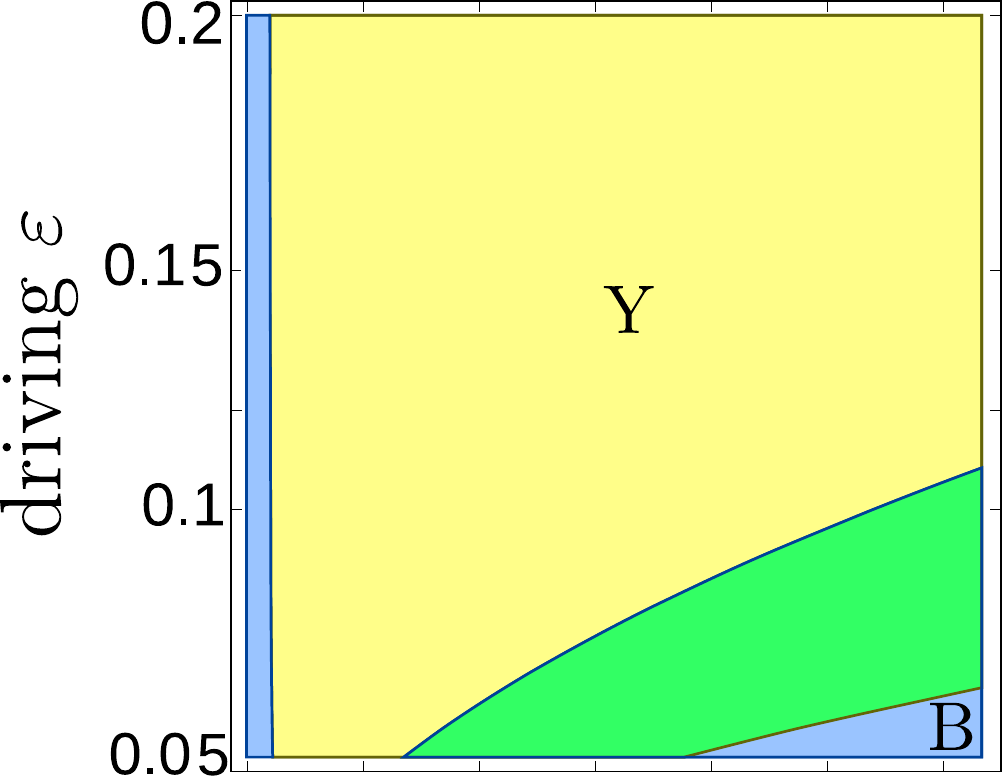}}\hspace*{2pt}     
\subfloat[$(231)$ single-PB (type 3)]{\includegraphics[scale=0.40]{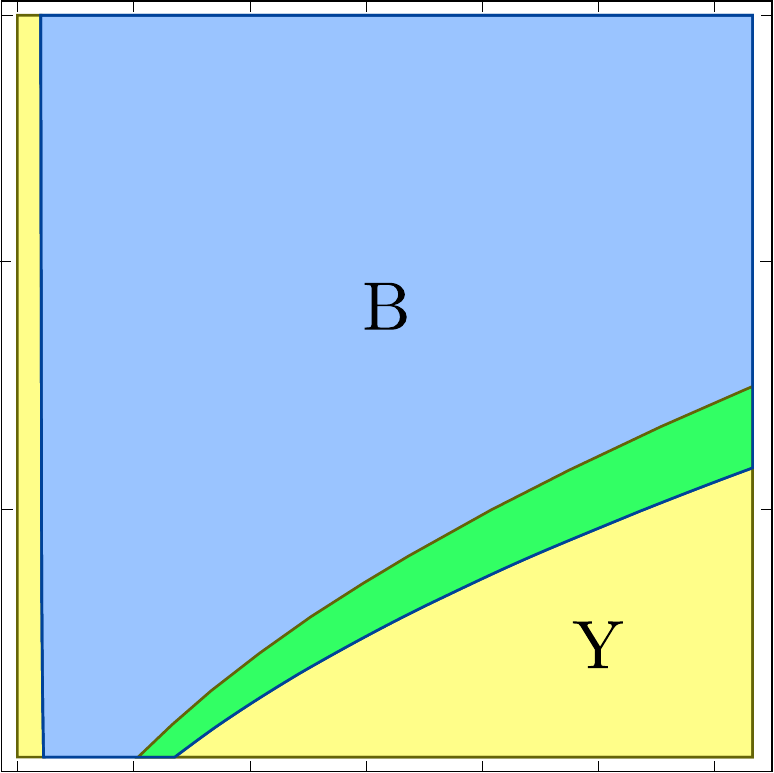}}\\                   
\subfloat[$(312)$ no two-PB]{\includegraphics[scale=0.40]{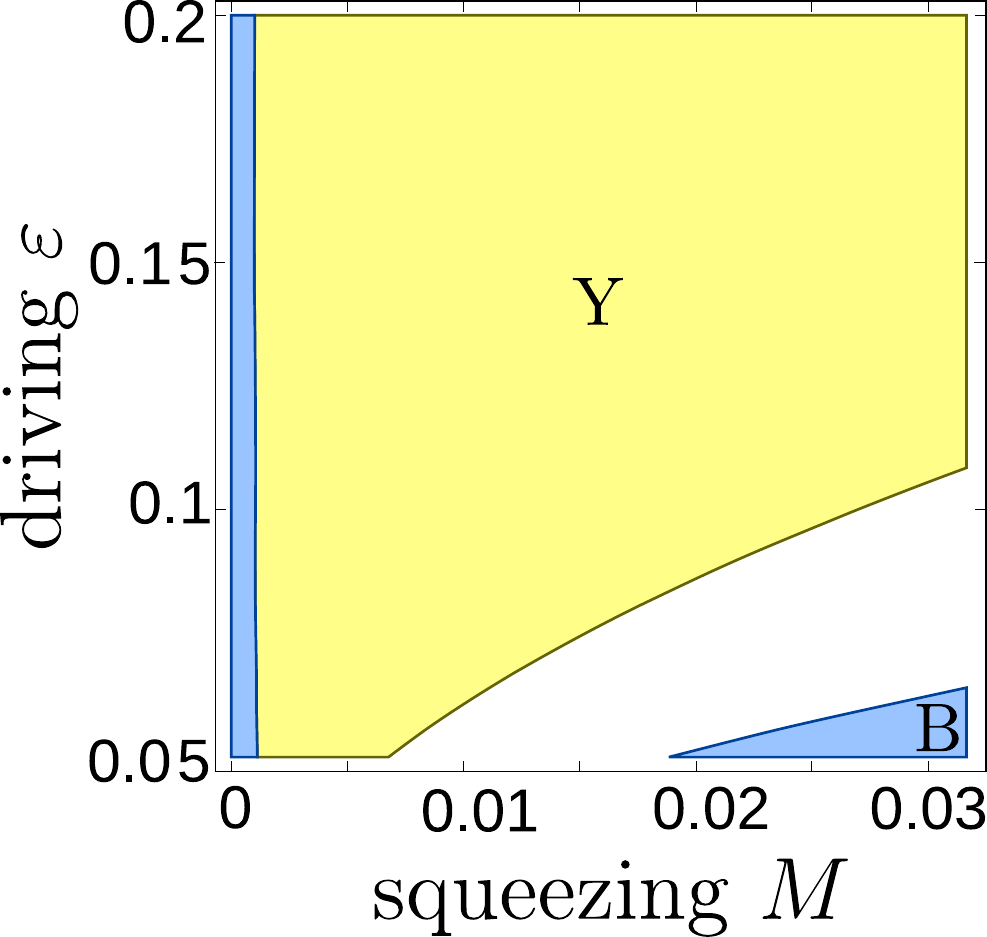}}\hspace*{1pt}        
\subfloat[$(321)$ single-PB (type 1)]{\includegraphics[scale=0.40]{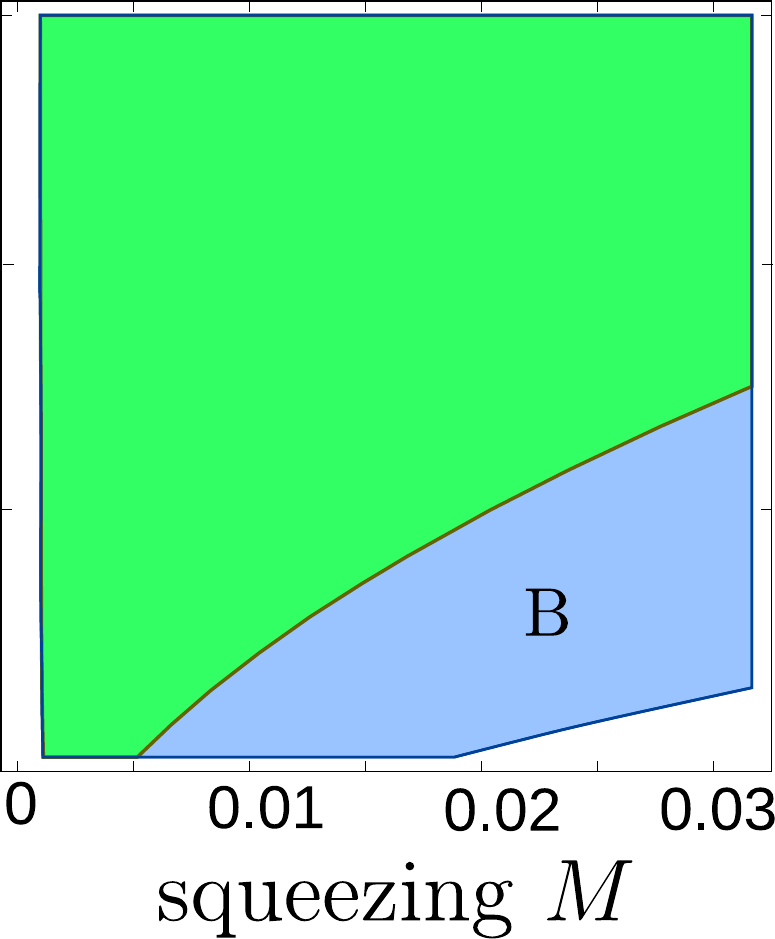}}                            
\end{center}
\caption{Squeezed-reservoir model: Photon-number correlations of
light generated in a driven harmonic cavity coupled to a squeezed
reservoir, assuming $n=0.01$ and $\Delta=0$. The regions of the
driving strength $\varepsilon$ and the reservoir squeezing
parameter $M$ satisfying the six conditions, which are listed in
Table~\ref{table2} for the correlation functions $\cor{2}$ and
$\cor{3}$, are shown here in yellow (Y) and blue (B),
respectively. The regions in green show the ranges of the
parameters $M$ and $\varepsilon$ for which given criteria are
satisfied simultaneously by $\cor{2}$ and $\cor{3}$ indicating a
specific type of photon blockade or photon-induced tunneling. {In
grayscale: yellow is the brightest, and green looks slightly
darker than blue. Yellow is also indicated by ``Y'', and blue by
``B''.}} \label{fig03}
\end{figure}
\begin{figure}[t]
\begin{center}
\subfloat[$(123)$ three-PT]{\includegraphics[scale=0.35]{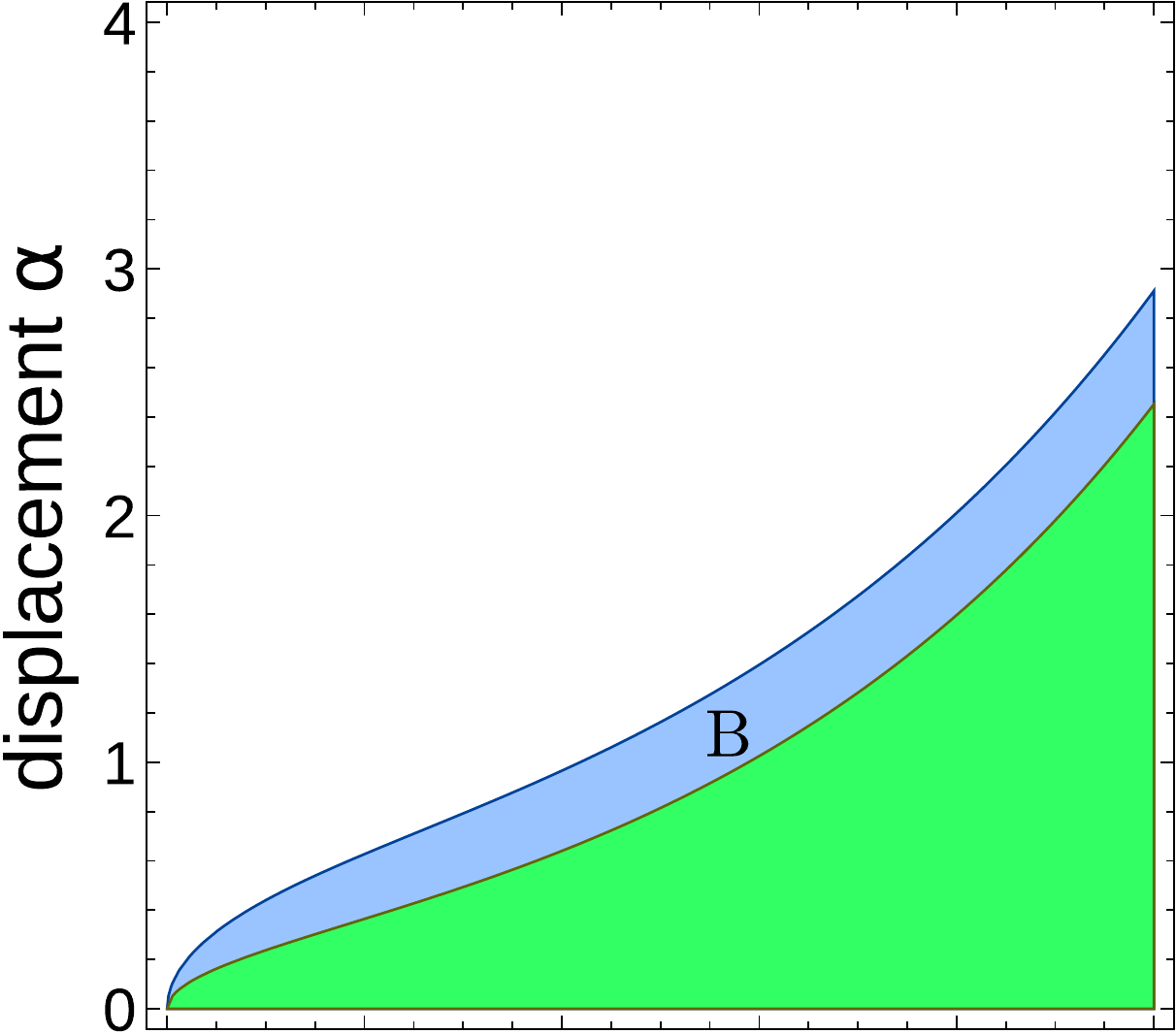}}\hspace*{1pt}                 
\subfloat[$(132)$ no two-PT]{\includegraphics[scale=0.35]{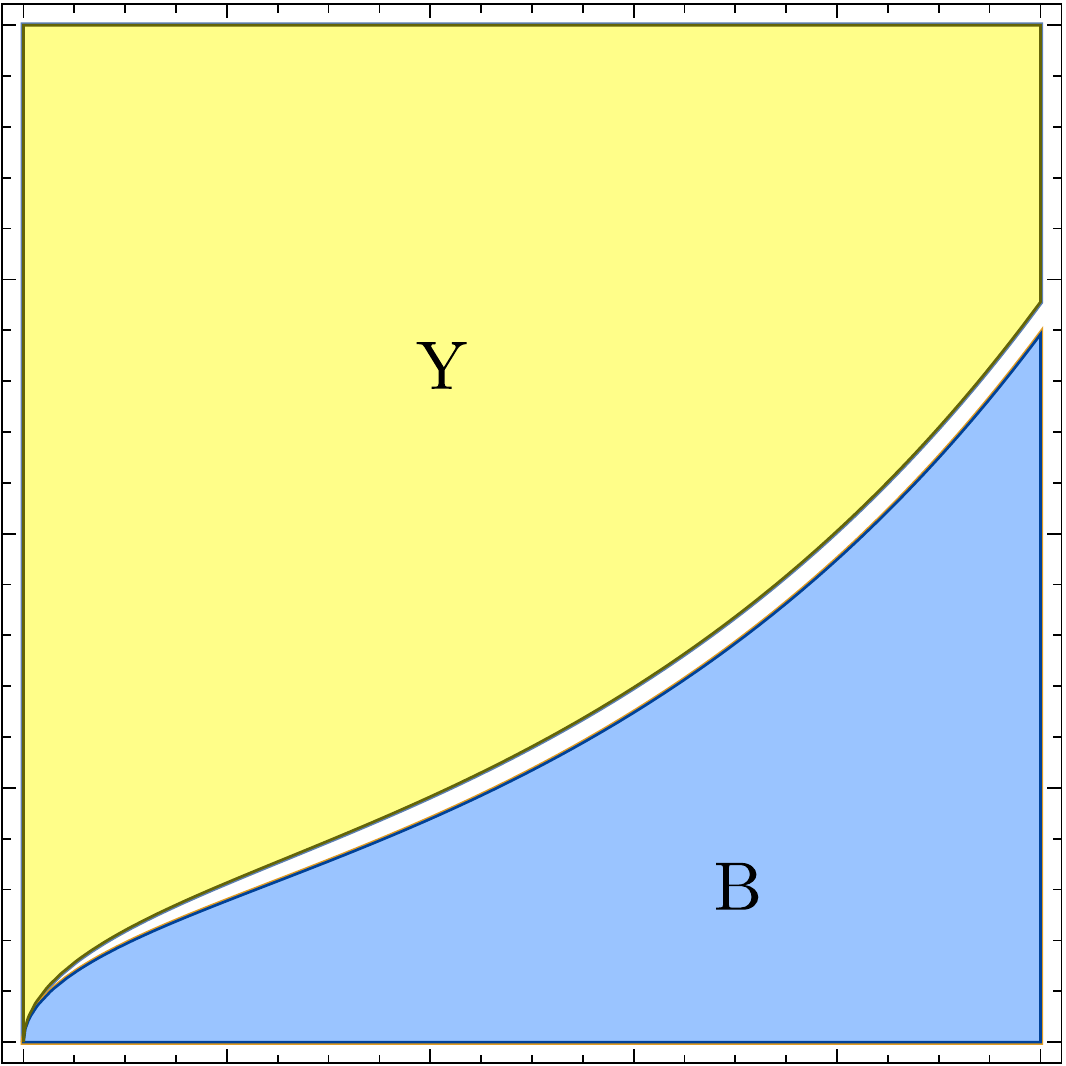}}\\               
\subfloat[$(213)$ single-PB (type 2)]{\includegraphics[scale=0.35]{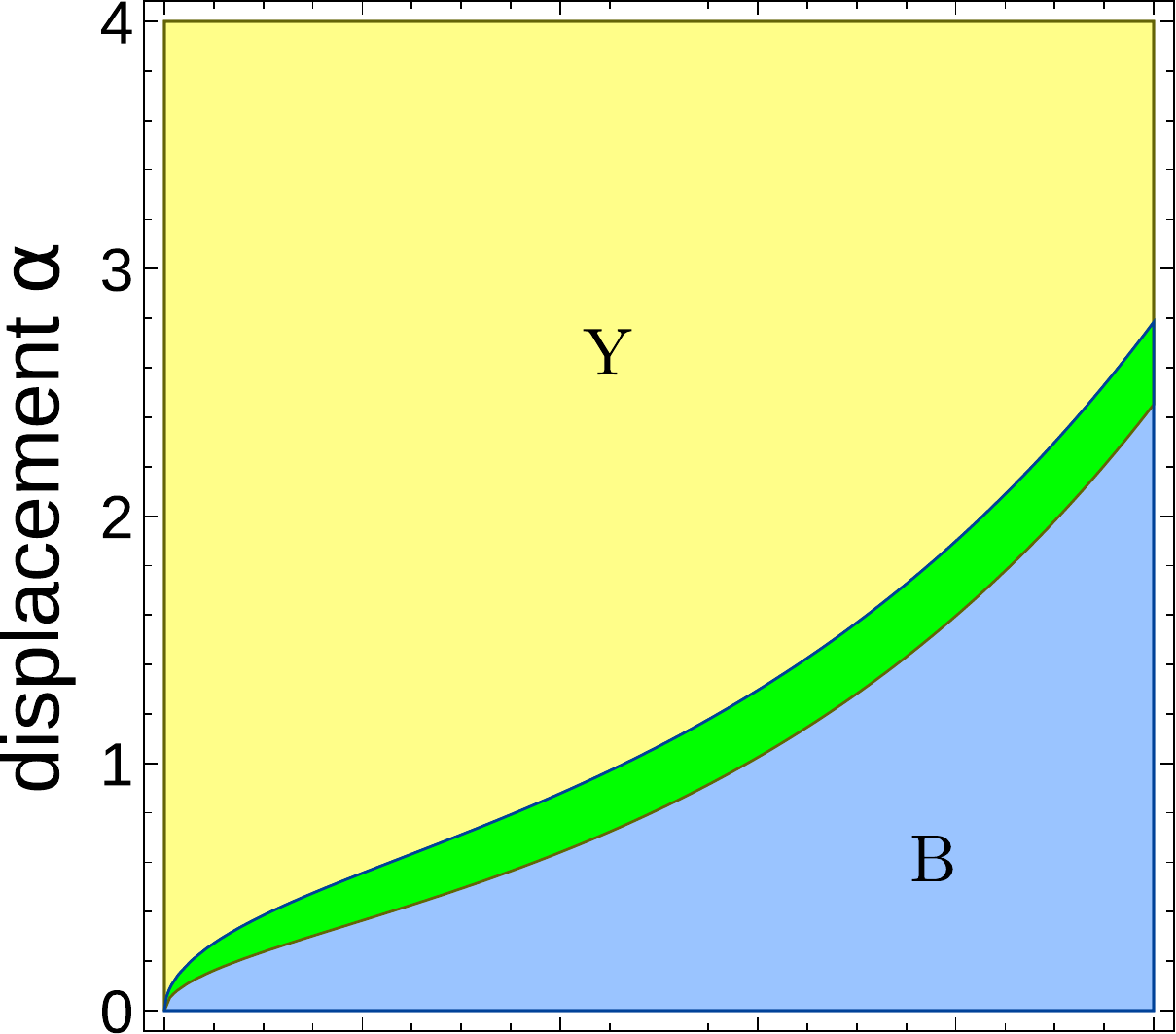}}\hspace*{3pt}     
\subfloat[$(231)$ single-PB (type 3)]{\includegraphics[scale=0.35]{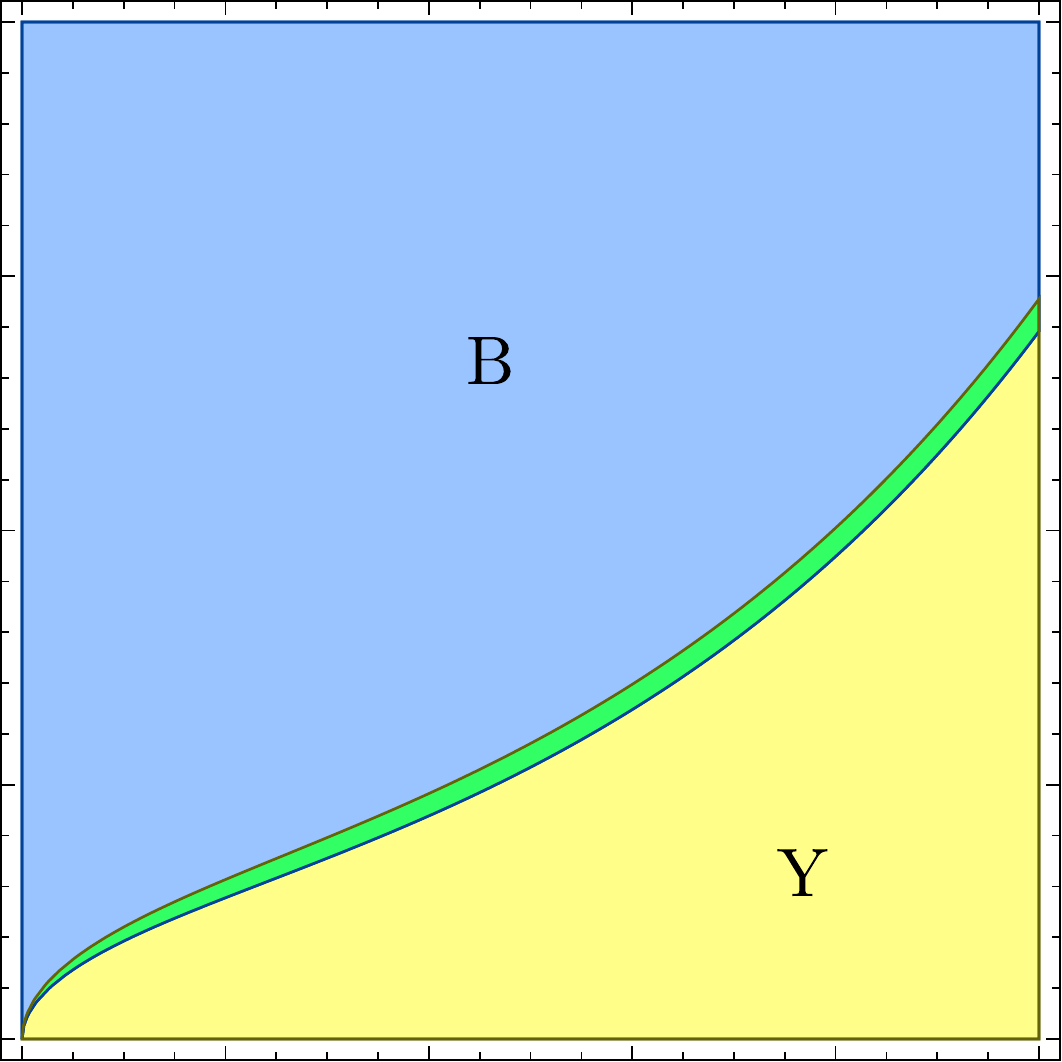}}\\                   
\subfloat[$(312)$ no two-PB]{\includegraphics[scale=0.35]{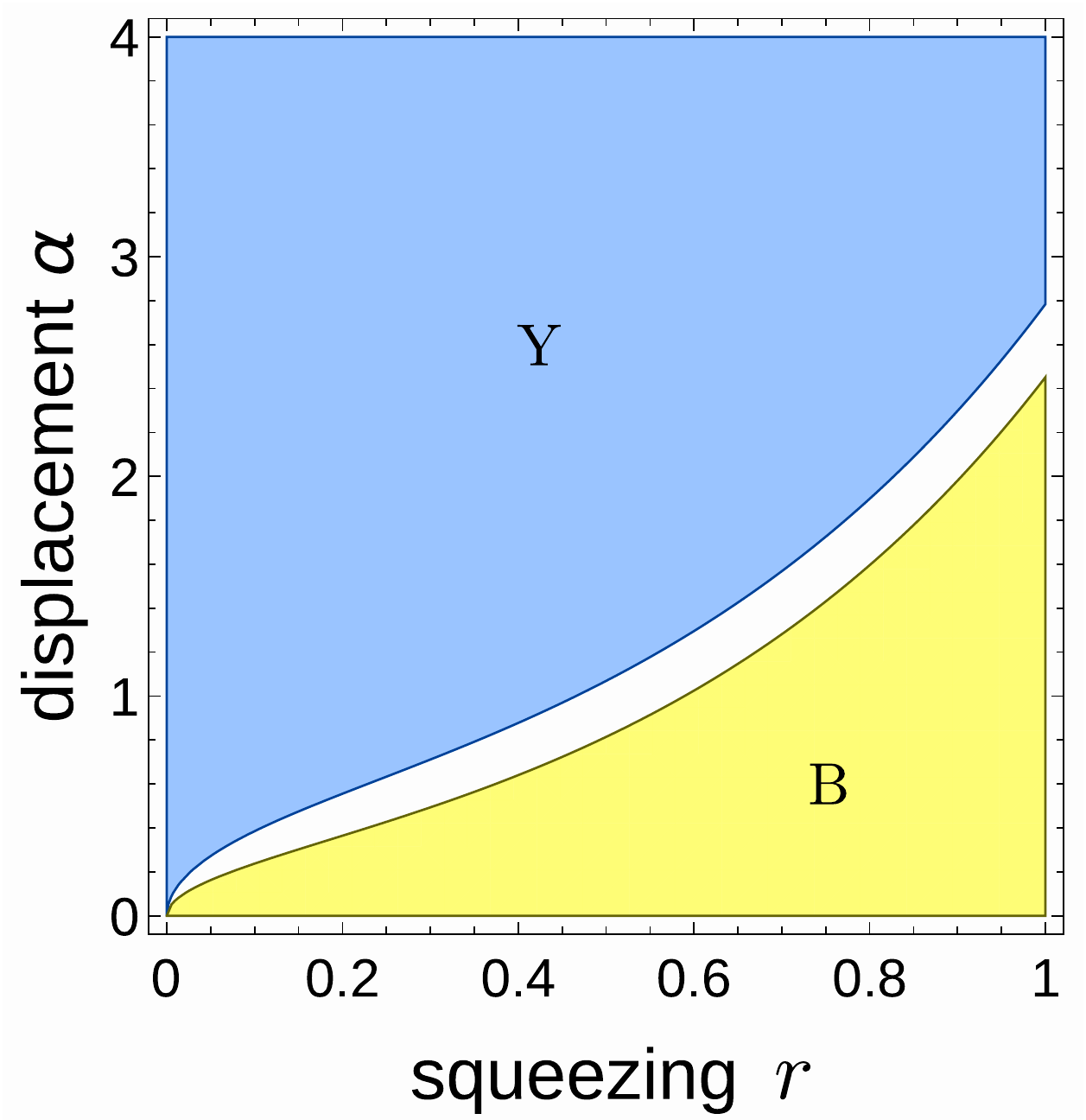}}\hspace*{-2pt}        
\subfloat[$(321)$ single-PB (type 1)]{\includegraphics[scale=0.35]{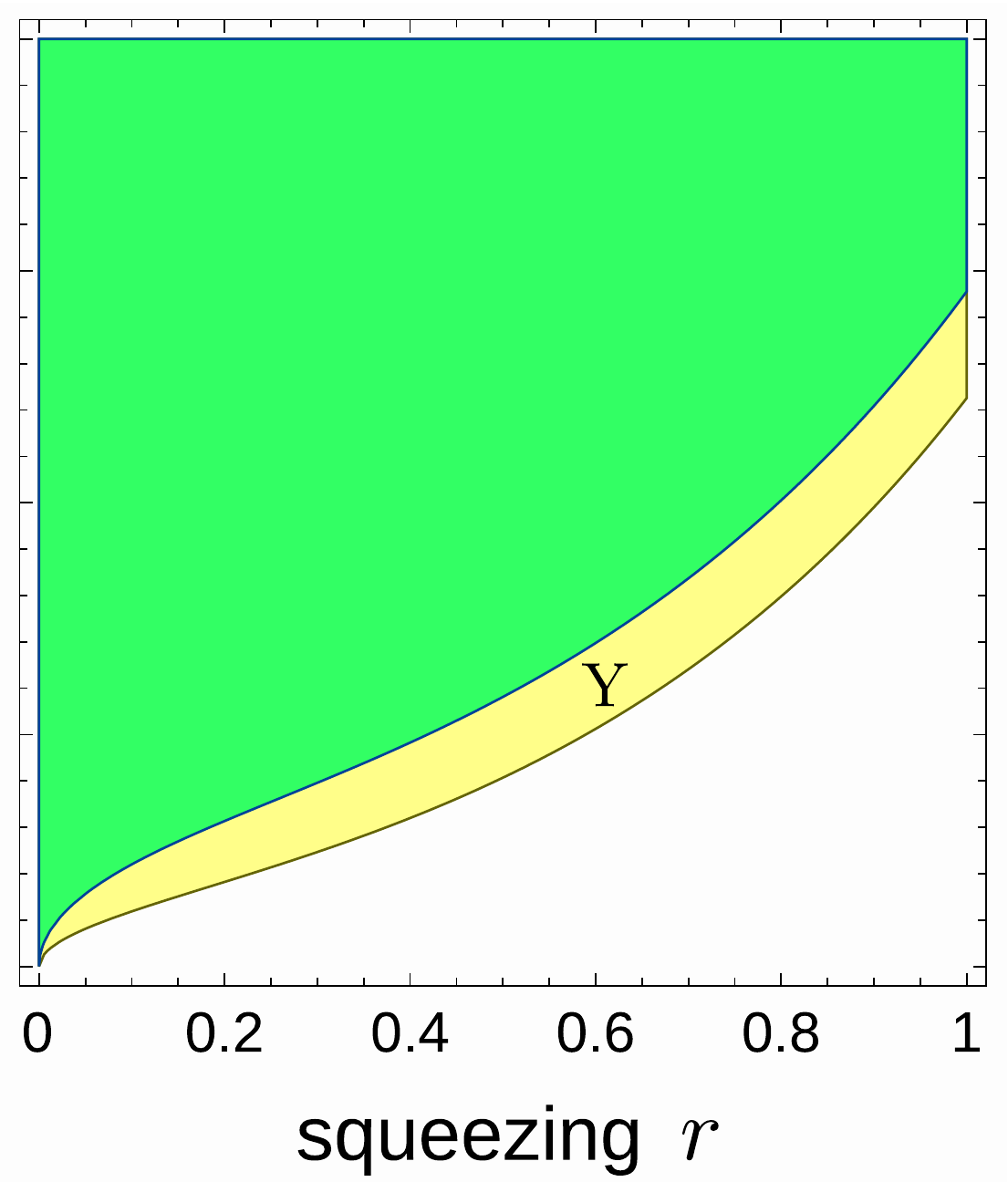}}                            
\end{center}
\caption{Photon-number correlations in the squeezed coherent
states showing the regions of the displacement ($\alpha$) and
squeezing ($r$) parameters for which the conditions in
Table~\ref{table2} are satisfied. This figure uses the same
notation, coloring, and carries a similar message as in
Fig.~\ref{fig03}. For example, the green region in figure (c)
(213) shows the ranges of parameters for which the conditions
$\cor{2}<1<\cor{3}$ are satisfied, as in Table~\ref{table2}(c).
The yellow (blue) region shows the parameter ranges satisfying
solely the condition $\cor{2}<1$ [$\cor{3}>1$]. {Yellow (blue) is
also indicated by ``Y'' (``B'')}.} \label{fig04}
\end{figure}
\begin{figure}[t]
\begin{center}
\subfloat[$(123)$ three-PT]{\includegraphics[scale=0.35]{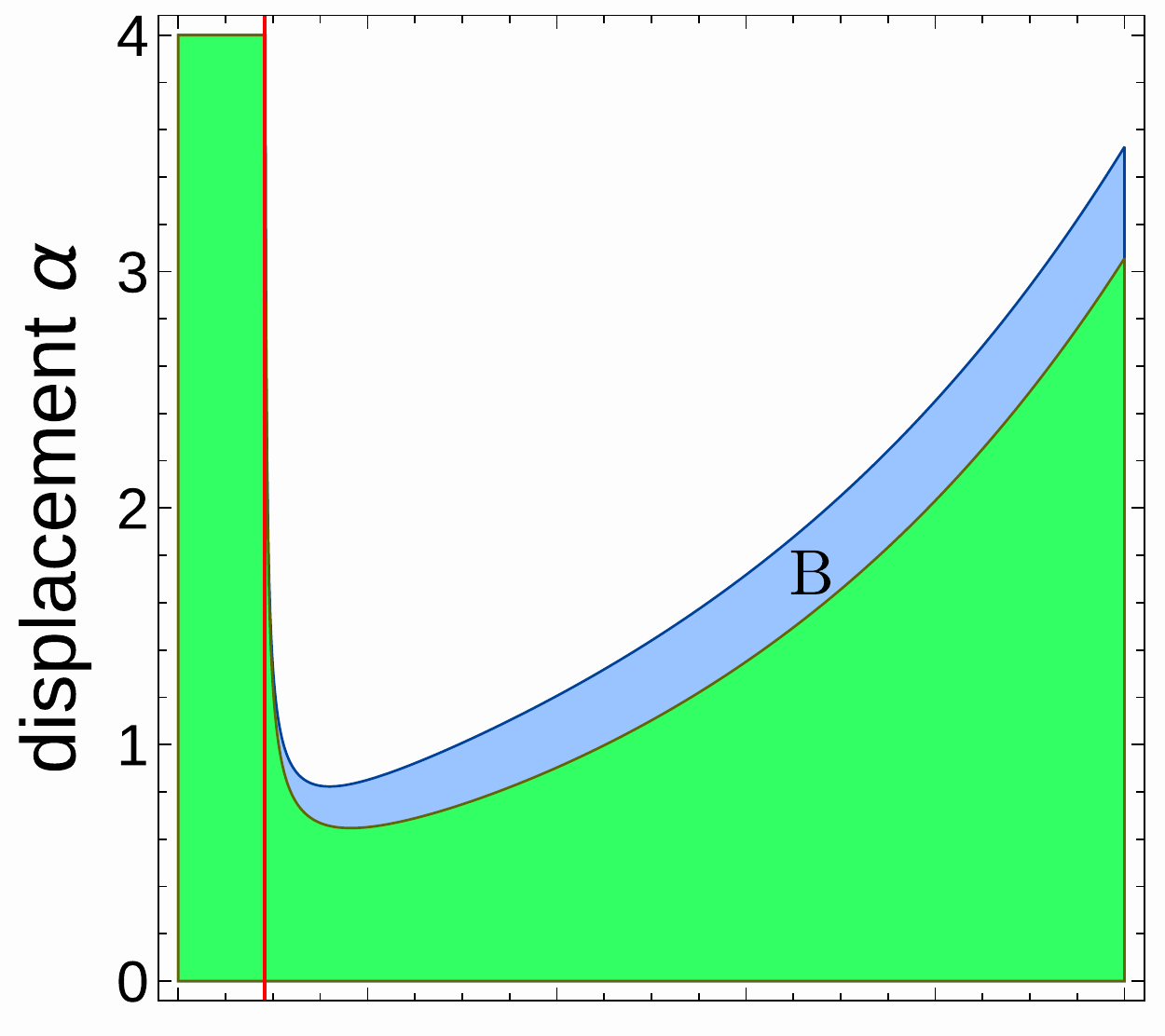}}\hspace*{1pt}                 
\subfloat[$(132)$ no two-PT]{\includegraphics[scale=0.35]{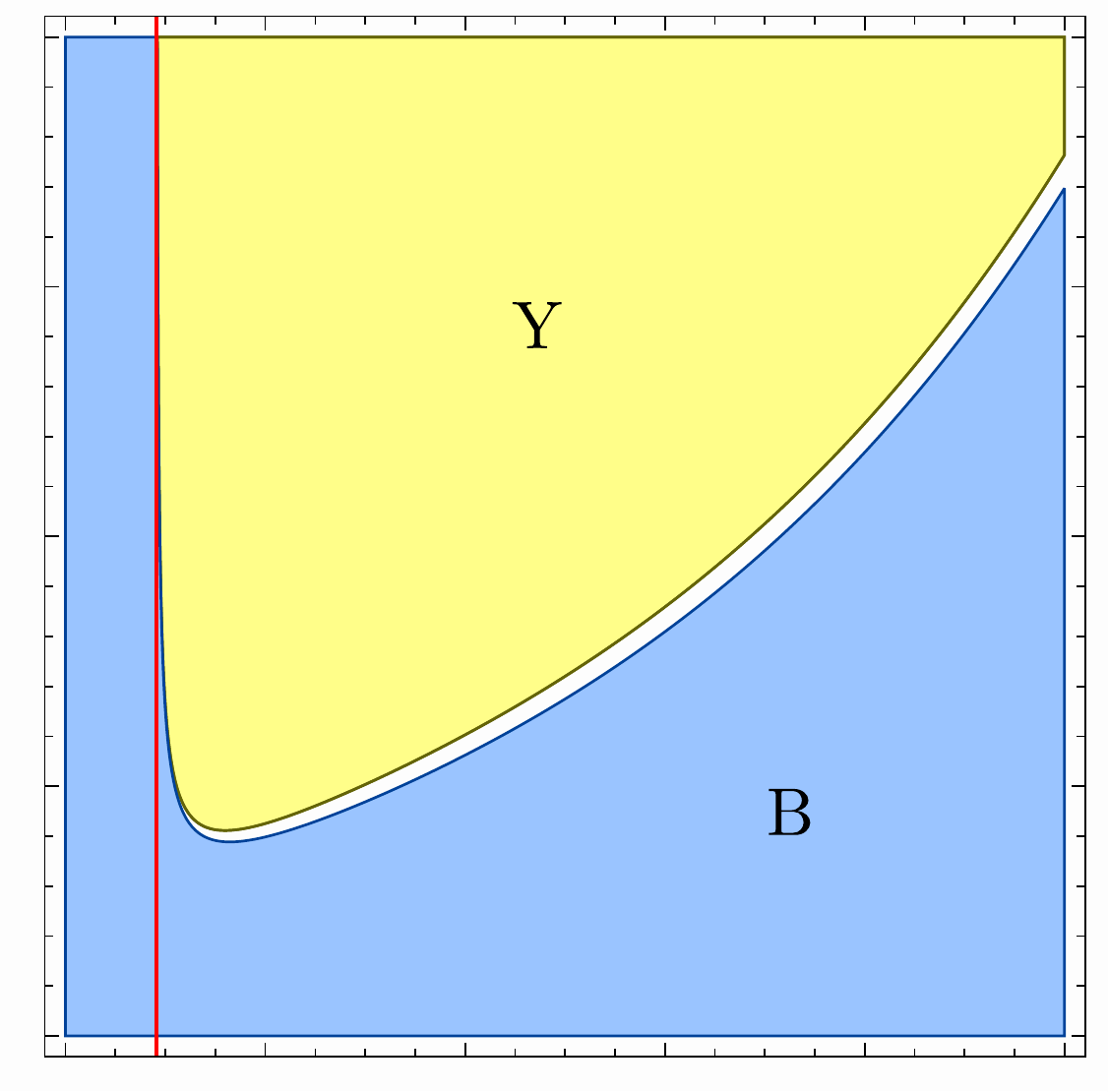}}\\               
\subfloat[$(213)$ single-PB (type 2)]{\includegraphics[scale=0.35]{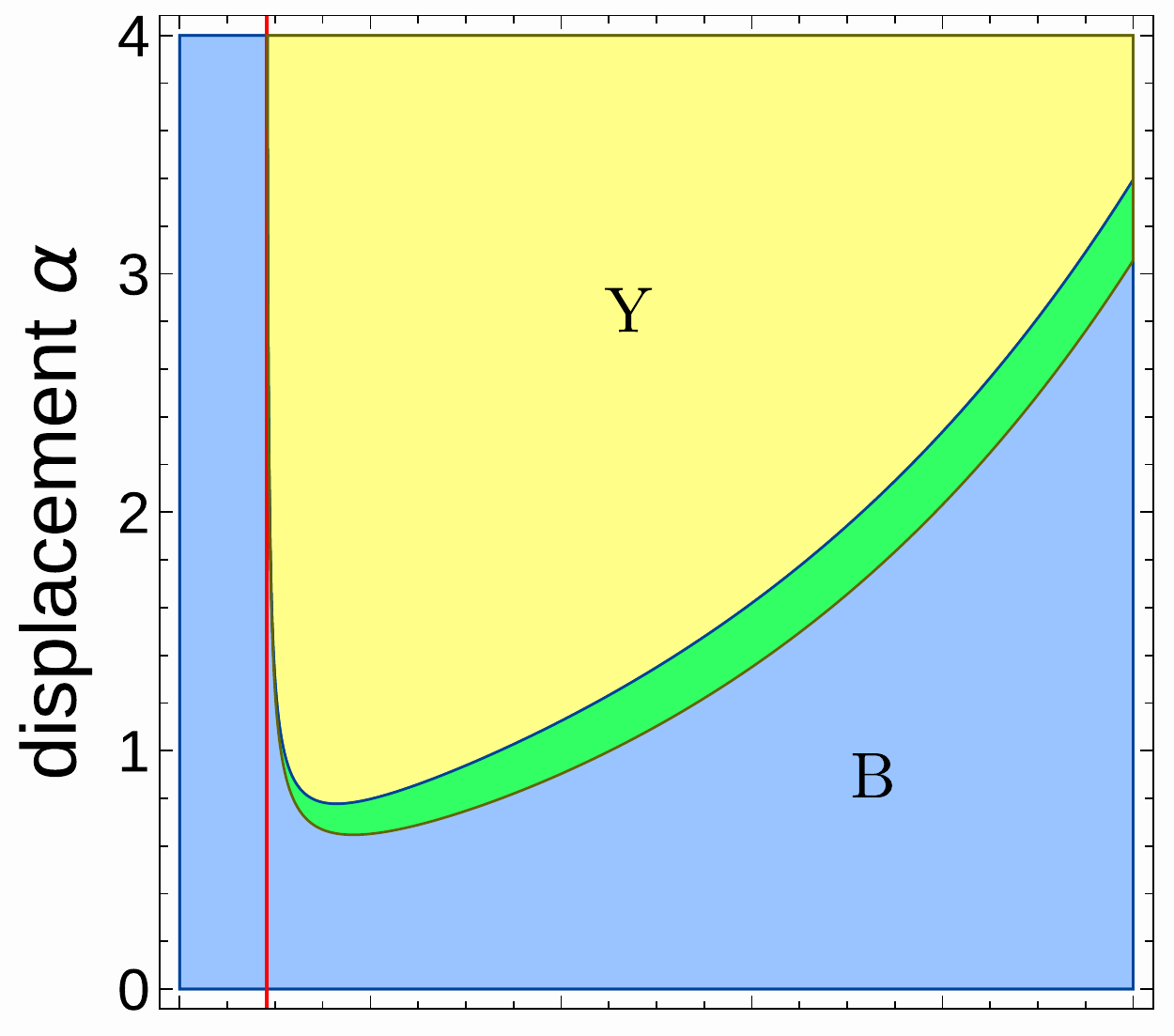}}\hspace*{3pt}     
\subfloat[$(231)$ single-PB (type 3)]{\includegraphics[scale=0.35]{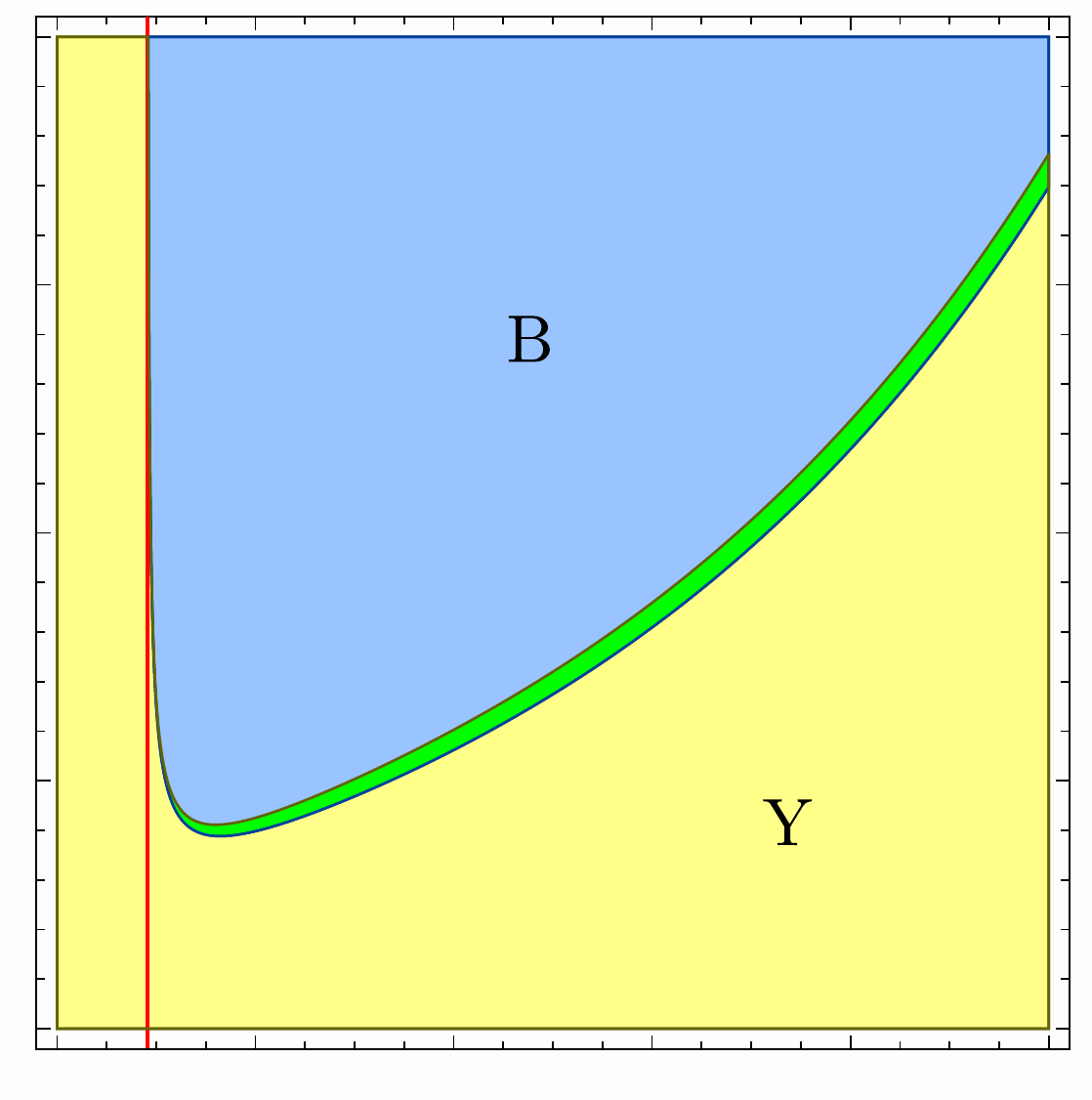}}\\                   
\subfloat[$(312)$ no two-PB]{\includegraphics[scale=0.35]{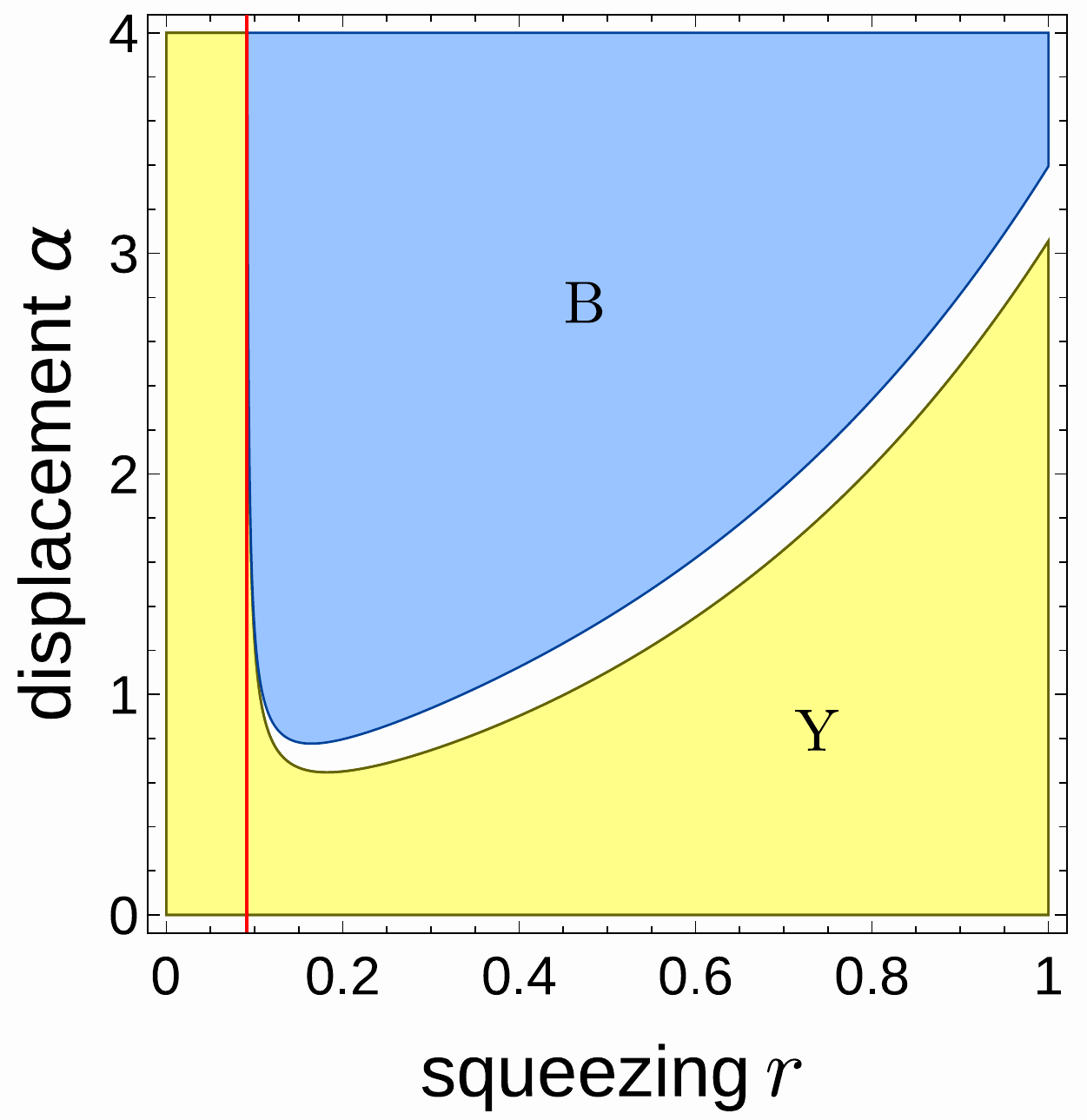}}\hspace*{-2pt}        
\subfloat[$(321)$ single-PB (type 1)]{\includegraphics[scale=0.35]{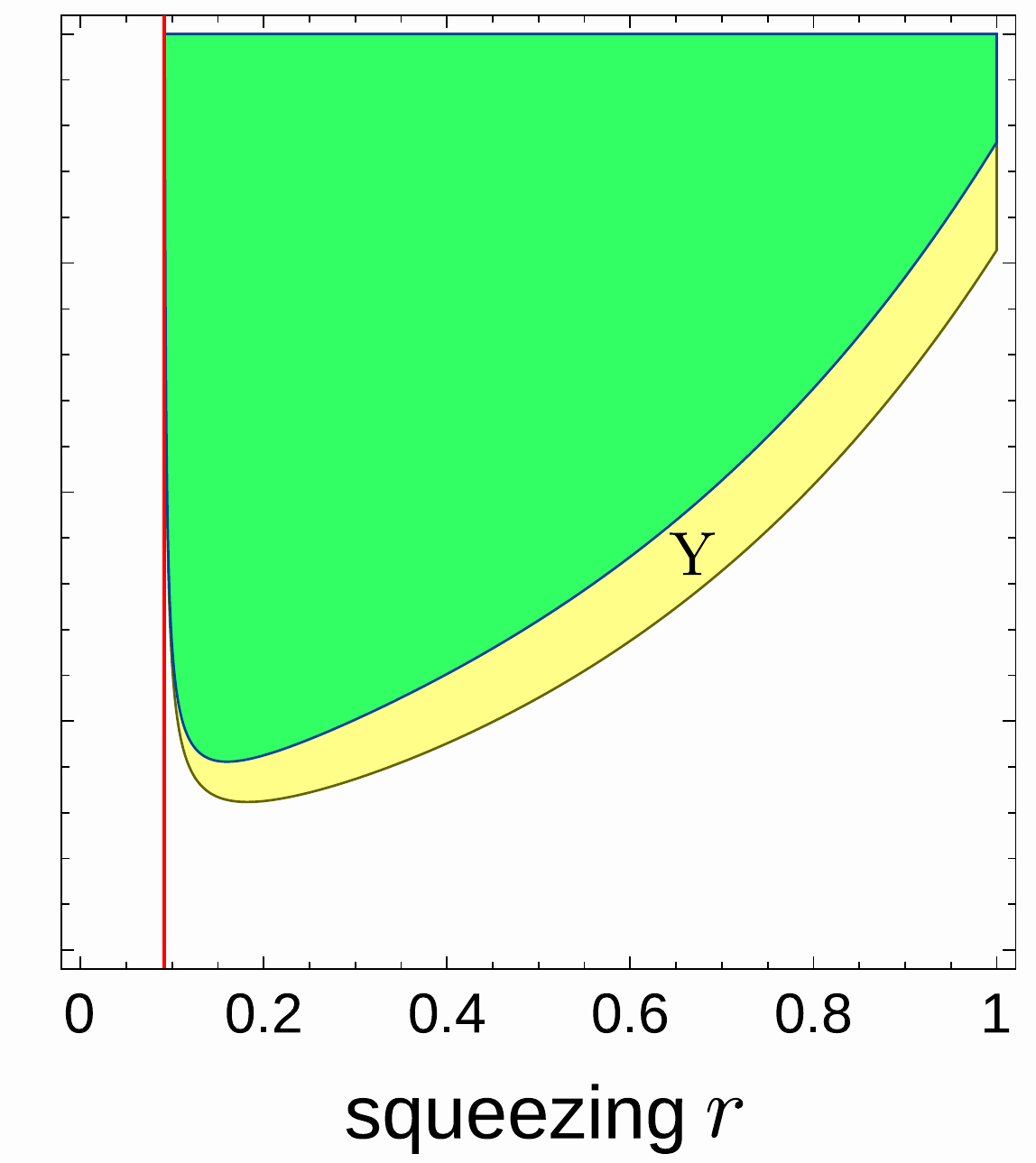}}                            
\end{center}
\caption{Photon-number correlations in the displaced squeezed
thermal states satisfying the inequalities in Table~\ref{table2}:
Same as in Fig.~\ref{fig04} but for the states defined in
Eq.~(\ref{DSTS}) with $\Nthermal=0.1$. The parameter region on the
left- (right-)hand side of the red vertical line in all the plots
corresponds to the classical (nonclassical) regimes of the states.
This red line is plotted at the critical squeezing parameter
$r_0=0.0912$, which is shown later in Fig.~\ref{fig12} by the
solid curve for $\Nthermal=0.1$ for the vanishing entanglement
potential, $\EP=0$.} \label{fig05}
\end{figure}
\subsection{Photon blockade and photon-induced tunneling via squeezing}
\label{sec1E}

It is known that SCS can exhibit the (second-order) sub-Poissonian
photon-number statistics (also referred to as single-time PAB).
This effect is also an important feature of light generated via
photon blockade.

The vast majority of previous works on PB assumed that dissipation
of a PB system can be modeled via its linear coupling to a
harmonic reservoir (a thermal bath). Only a few works, including
Refs.~\cite{Adam2014, Lemonde2014}, were analyzing PB in systems
coupled to nonlinear reservoirs. In such dissipative systems, PB
can result from: (i) a system nonlinearity, (ii) a reservoir
nonlinearity, or (iii) both of them. Single- and multi-PB effects
in a Kerr-nonlinear system coupled to a nonlinear (squeezed)
reservoir were analyzed in Ref.~\cite{Adam2014}. Shortly after
that publication, a single-PB effect generated solely by a
nonlinear (squeezed) reservoir was studied in a linear system in
Ref.~\cite{Lemonde2014}. Here we analyze various PB effects and
PIT in a harmonic cavity coupled to squeezed systems, as shown in
Fig.~\ref{fig01}(a). The two other common systems, which enable
the generation of conventional and unconventional PB are
schematically shown in Figs.~\ref{fig01}(b) and \ref{fig01}(c),
respectively. Note that some other schemes for PB can be obtained
by combing the three schemes shown in this figure.

The main objective of this paper is to analyze whether squeezing
plays an important role in generating various types of PB
(especially multiphoton effects). In other words, we address the
question whether PB can be observed in a driven harmonic resonator
without a strongly nonlinear medium [like in the standard PB setup
shown in Fig.~\ref{fig01}(b)] and without relying on multi-path
interference, as in the PB setup shown in Fig.~\ref{fig01}(c).

The paper is organized as follows: In Sec.~\ref{sec2}, we specify
the criteria of multi-PB and PIT. In Sec.~\ref{sec3}, we
numerically show that a two-photon decay process of light
generated in an optically linear system (a harmonic resonator) can
induce two-PB. Then, in Sec.~\ref{sec4}, we analytically study the
relations between $\cor{2}$ and the higher-order correlation
functions $\cor{k}$ for the squeezed coherent states and the
displaced squeezed thermal states, to demonstrate more explicitly
the possibility of generating two-PB, three-PT, and various types
of nonstandard single-PB via squeezing. The question of
nonclassicality of the studied effects and states is addressed in
Sec.~\ref{sec5} and Appendices~\ref{appC}, \ref{appD}, and
\ref{appE}. We also compare the proposed method for generating
multi-PB with the standard PB setups in Appendix~\ref{appA}.
Moreover, for pedagogical reasons, we present more details about
the master equation for a squeezed reservoir and recall its
relation to the standard master equation in Appendix~\ref{appB}.
We conclude in Sec.~\ref{sec6}.

In the main article, we use several abbreviations. We concisely
list them in Table~\ref{table1} to facilitate the following
exposition.

\section{Criteria for various types of photon blockade}
\label{sec2}

\subsection{Refined criteria for multi-photon blockade}
\label{sec2A}

The mechanisms of both conventional and unconventional single-PB
under proper resonance conditions can be generalized to generate
also two- and multi-PB, i.e., the generation of two or a larger
number of photons at the same instance of time.

Intuitively, $k$-PB can be understood as the generation of a state
$\hat\rho$ satisfying the conditions for the photon-number
probabilities $P_k=\langle k|\hat\rho|k\rangle$ as
follows~\cite{Adam2013, Hamsen2017}:
\begin{equation}
   P_{k+1}\approx 0 \AND P_{k}\gg P_{k+1}.
  \label{def1}
\end{equation}
However, in more realistic scenarios, the conditions in
Eq.~(\ref{def1}) are replaced by weaker criteria, where the
photon-number distribution $P_{k}$ of $\hat \rho$ is compared with
the Poissonian distribution $P^{\rm cs}_{k}$, describing the
photon-number statistics of a coherent state. Specifically:
\begin{equation}
   P_{k+1}< P^{\rm cs}_{k+1} \AND P_{k} \ge P^{\rm cs}_{k},
  \label{def2}
\end{equation}
where the probability $P^{\rm cs}_{k}=|\langle \alpha|k\rangle|^2$
is for a coherent state $\alpha$ with the same mean photon number
as that for $\hat \rho$, i.e., $\<\alpha|\hat
n|\alpha\>=|\alpha|^2=\tr(\hat \rho \hat n)$, where $\hat
n=\hat{a}^{\dagger}\hat{a}$ is the photon-number operator. The
conditions for the probabilities $P_k$ can be replaced by those
based on the experimentally-accessible $k$th-order correlation
function,
\begin{eqnarray}
 \cor{k}  = \frac{\mean{(\hat a^\dagger)^k\hat a^k}}{\mean{\hat n}^k} =
 \frac{\mean{\hat n^{[k]}}} {\mean{\hat n}^k} =
 \frac{\sum_{n=0}^{\infty} P_n n^{[k]}}{\mean{\hat n}^k}, \label{cor_k}
\end{eqnarray}
where, as usual, $\hat a$ ($\hat a^\dagger$) is the annihilation
(creation) operator, $\mean{\hat n^{[k]}}=\mean {(\hat
a^\dagger)^k\hat a^k}$, and $n^{[k]}=n(n-1)\cdots(n-k+1)$ is the
factorial power (also called the falling power). Thus, the
criteria for PB given in Eq.~(\ref{def1}) can be replaced by
\begin{equation}
   \cor{k+1}\approx 0 \AND \cor{k} \gg \cor{k+1},
  \label{def3}
\end{equation}
In this paper, we assume that $k$-PB is defined by the following
two criteria derived by Hamsen \etal~\cite{Hamsen2017}:
\begin{eqnarray}
{\bf criterion}~\#1:\quad&&   \cor{k+1} < A \equiv \exp(-\mean{\hat n}), \nonumber \\
{\bf criterion}~\#2:\quad&&   \cor{k} \ge B^{(k)} \equiv A+
\mean{\hat n} \cor{k+1},\quad\quad \label{C1}
  \label{refined_criteria}
\end{eqnarray}
which replace the criteria in Eq.~(\ref{def2}).

We note that the definition of multi-PB in
Eq.~(\ref{refined_criteria}) has some drawbacks and limitations.
Strictly speaking, the criteria in Eq.~(\ref{refined_criteria})
can only be considered a PB witness, i.e., necessary but not
sufficient conditions of PB. Note that second-order single-time
photon antibunching [$\cor{2}<1$] is the most common test of
single-PB, but it is also only a necessary but not sufficient
condition for PB. An intuitive ``orthodox'' interpretation of
single- and multi-PB effects can be given as follows: $k$-PB
($k=1,2,...$) corresponds to the effect, in which the photon
occupation of the first $k$ energy levels of a driven nonlinear
system blocks the generation of more photons in the system. In
other words, $k$-PB corresponds to an effective truncation of the
Hilbert space spanning a given state at the $k$-photon Fock state
$\ket{k}$ so the contributions of the Fock states $\ket{k+l}$ for
$l>0$ can be effectively ignored, which means that
$\<k|\hat\rho|k\>\gg \<k+l|\hat\rho|k+l\>$ or, alternatively,
$\cor{k}\gg \cor{k+l}$, for any $l>0$. However, the above
conditions are usually only checked for $l=1$, ignoring the
analysis of the cases for $l>1$. Such objection also applies to
many studies of single-PB based on requiring $\cor{2}<1$ and
ignoring the values of $\cor{3}$ and higher-order correlation
functions.

\subsection{Simplified criteria for multi-photon blockade}
\label{sec2B}

Note that if $\mean{\hat n}\ll 1$ then the refined conditions for
multi-PB, given in Eq.~(\ref{refined_criteria}), simplify to the
following familiar criteria for $\hat\rho$:
\begin{equation}
   \cor{k+1}< 1 \AND \cor{k}  \ge 1,
  \label{simplified_criteria}
\end{equation}
which mean that, in this small photon-number limit, the state
generated via $k$-PB exhibits (single-time) $(k+1)$-PAB, and
$k$-photon bunching if $\cor{k}>1$ or the so-called unbunching if
$\cor{k}=1$.

Thus, two-photon and three-PB effects can be given by the
following relations for the correlation functions:
\begin{eqnarray}
\cor{2}\ge 1 \;\;\;\;\;&\mbox{and}&\;\;\;\;\; \cor{3}<1;
\label{simplified_criteria2}\\
\cor{2},\; \cor{3}\ge1 \;\;\;\;\;&\mbox{and}&\;\;\;\;\;
\cor{4}<1,\label{simplified_criteria3}
\end{eqnarray}
respectively. Note that we have added an extra condition for
$\cor{2}$ in Eq.~(\ref{simplified_criteria3}), which is not
required in the criteria specified in
Eqs.~(\ref{refined_criteria}) and~(\ref{simplified_criteria}).
Moreover, in this simplified characterization of PB we ignore the
requirements on two-time correlation functions $g^{(k)}(\tau)$,
including $g^{(2)}(\tau)$.

Thus, in the case of two-PB, the three-photon probability has to
be suppressed and simultaneously the probability of observing two
photons should be enhanced. Analogously, the suppression of the
four-photon probability and the increase in the probabilities of a
lower number of photons would lead to three-PB.

Both types of PB, which are characterized by the simplified and
refined criteria, correspond to nonclassical effects, because they
require the sub-Poissonian photon-number statistics (of any given
order $k$), as described in greater detail in Appendix~\ref{appC}.

As mentioned above, the refined criteria in
Eq.~(\ref{refined_criteria}) reduce to the conditions in
Eq.~(\ref{simplified_criteria2}) for small photon numbers
$\mean{\hat n}\ll 1$. But, in principle, these simplified criteria
can be applied even if $\mean{\hat n}>1$, but then the predicted
PB can differ from that based on the refined criteria in
Eq.~(\ref{refined_criteria}). It might also be the case that a
given state exhibits, e.g., two-PB according to the refined
criteria, but not according to the simplified criteria. Actually,
we will show such cases in the following sections.

Now, we consider a simple example of such different predictions of
two-PB according to Eqs.~(\ref{refined_criteria}) and
(\ref{simplified_criteria}). Specifically, the two-photon Fock
state $\ket{2}$, for which $\cor{2}=1/2$ and $\cor{3}=0$, can be
considered a two-PB state according to the refined criteria in
Eq.~(\ref{refined_criteria}), because $\cor{2}>\exp(-2)\approx
0.135$ and $\cor{3}<\exp(-2)$. Note that the simplified criteria
in Eq.~(\ref{simplified_criteria}) can, in principle, be applied
to the two-photon Fock state $\ket{2}$. However, since
$A\equiv\exp(-\mean{\hat n})$ is not negligible, the predictions
of PB for $\ket{2}$ according to the refined and simplified
criteria are different. Indeed, the Fock state $\ket{2}$ is not
considered a two-PB state according to the simplified
criteria~(\ref{simplified_criteria}).

\subsection{Nonstandard types of photon blockade}
\label{sec2C}

As described in previous subsections, the simplified condition for
observing single PB corresponds to the requirement of single-time
PAB. If the following additional condition $\cor{3}< \cor{2}$ is
satisfied, as desirable for good single-photon sources, then we
refer to this effect as single-PB of type~1, which is
characterized by
\begin{equation}
  \cor{3}< \cor{2}< 1.
  \label{type1}
\end{equation}
Apart from this single-PB, there are other possibilities of
obtaining quantum photon-number statistics by specifying the
relations between higher-order single-time correlations $\cor{k}$
and/or the second-order two-time correlations $g^{(2)}(\tau)$.
These include the following types of PB:

(1) We recall that, in order to consider single-PB as a true
source of single photons, the generated light via PB should also
exhibit two-time PAB as given in Eq.~(\ref{type0}). Indeed, it is
known that the sub-Poissonian photon-number statistics (i.e.,
single-time PAB) of a field can be accompanied with both two-time
PAB and two-time photon bunching, and vice versa (see, e.g.,
Ref.~\cite{Adam2010} and references therein). Thus, if light
exhibits single-time PAB and two-time photon bunching [i.e., a
local maximum of $g^{(2)}(\tau)$ for small $\tau$], one can refer
to as nonstandard single-PB, because it is \emph{not}
characterized by Eq.~(\ref{type0}). Examples of this nonstandard
PB are analyzed in Sec.~\ref{sec3} and shown in Fig.~\ref{fig02}.
In the following we mainly analyze other types on nonstandard PB
based solely on single-time correlation functions.

(2) In greater detail we analyze a special kind of nonstandard PB
characterized by the single-time correlation functions satisfying
the conditions:
\begin{equation}
\cor{2} <1 < \cor{3},\label{type2}
\end{equation}
which was first studied in greater detail in
Ref.~\cite{Radulaski2017} under the name \emph{unconventional PB}.
However, in order to avoid confusion of this type of PB and
unconventional PB studied in
Refs.~\cite{Flayac2018review,Snijders2018,Vaneph2018}, we refer to
the effect characterized by Eq.~(\ref{type2}) as nonstandard PB of
type~2.

It is seen that this nonstandard PB occurs when the probability of
measuring two photons at the same time is suppressed and,
simultaneously, the probability of obtaining three photons is
enhanced. Note that this effect can be generated by different
physical mechanisms in different systems: (i) by using large
nonlinearities in conventional PB systems, as shown in
Fig.~\ref{fig01}(b), (ii) by small nonlinearities and multi-path
interference in unconventional PB systems, as shown in
Fig.~\ref{fig01}(c), or (iii) by exploiting squeezing in, e.g.,
linear systems coupled to a squeezed reservoir, as shown in
Fig.~\ref{fig01}(a) and studied here.

(3) One can modify the condition for $\cor{3}$ in
Eq.~(\ref{type2}) to consider another type (say type 3) of
single-PB, as characterized by:
\begin{equation}
    \cor{2}<\cor{3}<1.
  \label{type3}
\end{equation}

\begin{figure}[t]
\begin{center}
\includegraphics[width=\columnwidth]{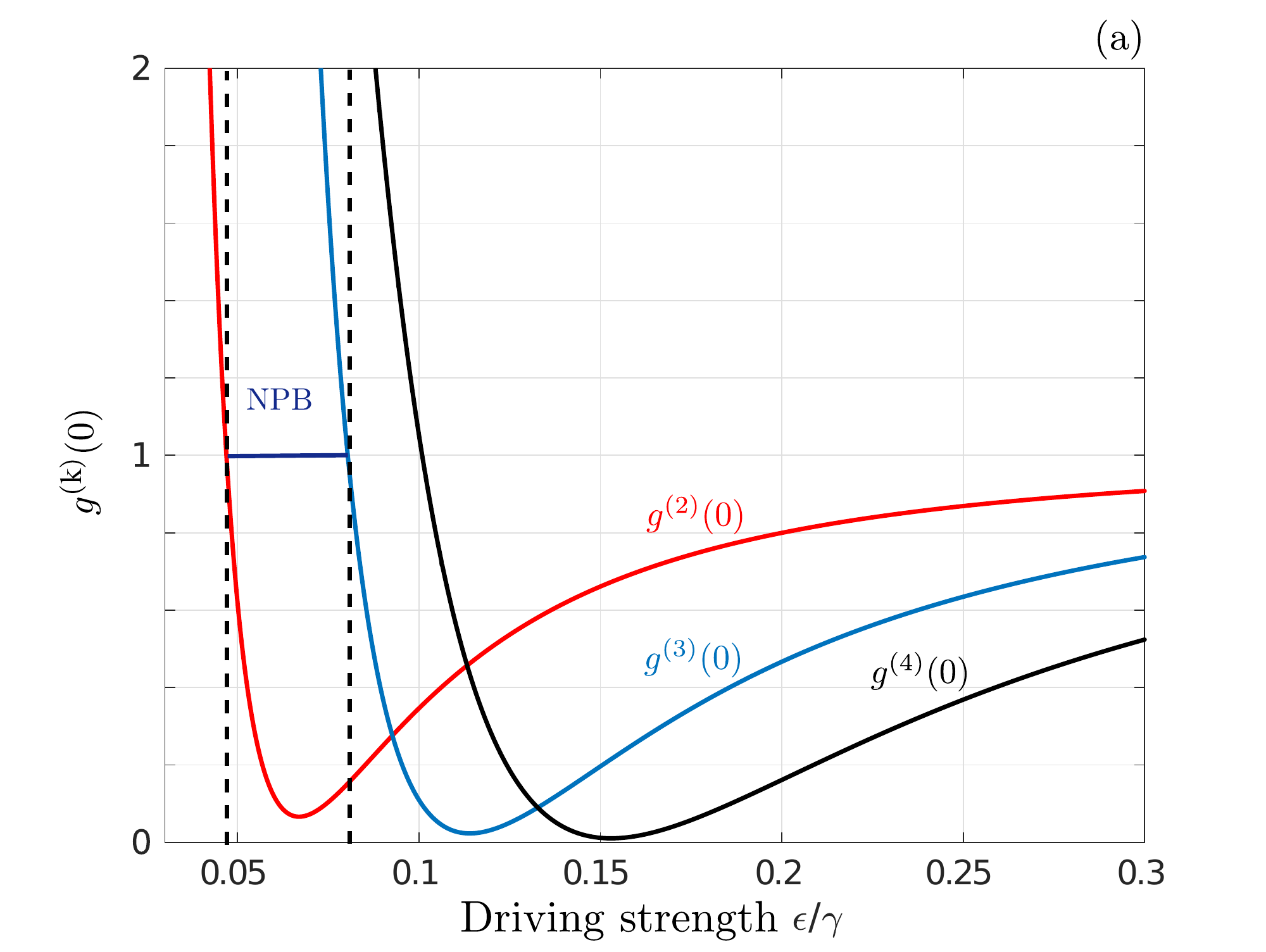}\\
\includegraphics[width=\columnwidth]{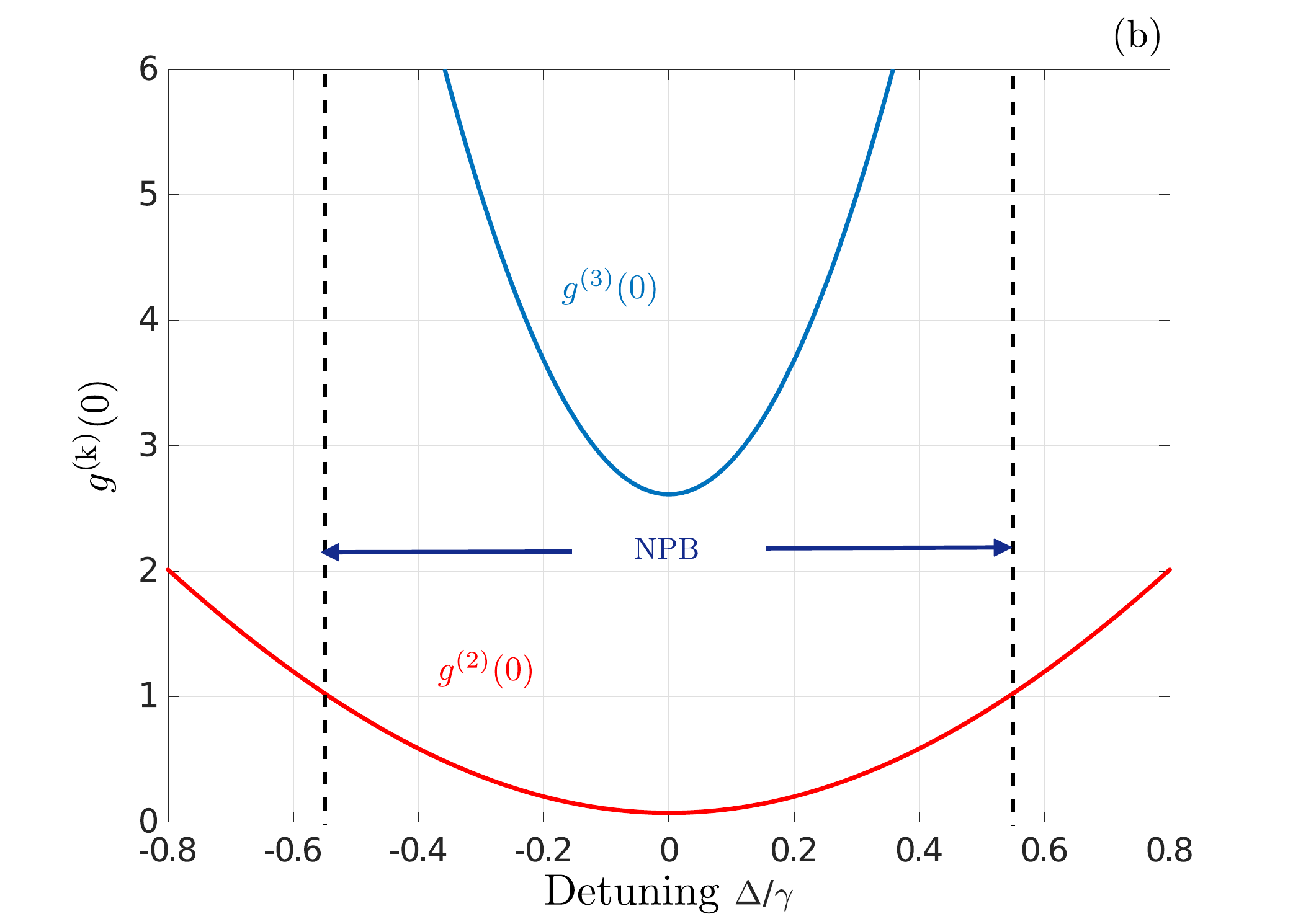}
\caption{Squeezed-reservoir model: Correlation functions $\cor{k}$
vs.: {(a) the driving strength $\varepsilon$ for fixed $\Delta=0$
and (b) the detuning $\Delta$ for the maximally squeezed reservoir
for $n=3\times 10^{-4}$, which corresponds to $M=0.017$, for fixed
$\varepsilon=0.07\gamma$}. All the parameters are scaled in
$\gamma$ units. The regions between the broken lines correspond to
nonstandard single-photon blockade (type~2).}
 \label{fig06}
 \end{center}
 \end{figure}

The latter two types of nonstandard single-PB are listed in
Table~II and a few examples of such effects generated via
squeezing are discussed in the following sections and shown in
Figs.~\ref{fig03}, \ref{fig04}, \ref{fig05}, and~\ref{fig06}.

Note that we found examples of nonstandard PB concerning unusual
properties of both single- and two-time correlation functions.
But, for brevity, we do not present such examples here.

We also note that nonclassical states often satisfy the conditions
$ g^{(2)} < g^{(3)}< \ldots < g^{(k)} < 1$, as those studied in
Refs.~\cite{PerinaJr2017c,PerinaJr2019a}, where the sub-Poissonian
statistics was resulting from postselection. Such states can also
be used for simulating nonstandard single-PB effects.

\section{Various types of photon blockade and tunneling
generated by squeezed reservoir}
 \label{sec3}

\subsection{Model}
\label{sec3A}

Here we will show that a squeezed reservoir can induce various
types of PB and PIT, including two-photon effects in a driven
\emph{harmonic} resonator.

Specifically, as an example of a physical system, in which
squeezing interactions induce PB, we use a single optical cavity
of a frequency $\omega_c$, which is externally driven by a laser
field of an amplitude $\varepsilon$ with a frequency $\omega_d$.
The cavity decays into a squeezed reservoir characterized by the
reservoir squeezing parameter $M$. The model is presented in
Fig.~\ref{fig01}(a). We will show that for such a linear optical
system, the two-photon dissipation process plays a crucial role in
obtaining single- and two-PB, as well as other nonstandard types
of nonclassical photon correlations.

The Hamiltonian of the system has the following form (hereafter we
set $\hbar=1$)
\begin{equation}
\hat{H}'=\omega_c\hat{a}^{\dagger}\hat{a}+\varepsilon\left(\hat{a}e^{i\omega_d
t}+\hat{a}^{\dagger}e^{-i\omega_d t}\right). \label{Hamiltonian1}
\end{equation}
After its transformation to the interaction picture to the frame
rotating with the driving frequency $\omega_d$, one obtains the
following effective Hamiltonian of the system
\begin{equation}
\hat{H}=\Delta\hat{a}^{\dagger}\hat{a}+\varepsilon
\left(\hat{a}^{\dagger}+\hat{a}\right),
\label{Hamiltonian}
\end{equation}
where $\Delta=\omega_c-\omega_d$ is the detuning between the
cavity and driving frequencies.

The evolution of the driven cavity interacting with a squeezed
reservoir is governed by the following master
equation~\cite{Agarwal1973,Perina1991,ScullyBook}:
\begin{eqnarray}
\frac{d\hat{\rho}}{dt}&=& -i[\hat H,\hat\rho]+
\frac{1}{2}\gamma(n+1)\left(2\hat{a}\hat{\rho}\hat{a}^{\dagger}-\hat{a}^{\dagger}\hat{a}\hat{\rho}-\hat{\rho}\hat{a}^{\dagger}
\hat{a}\right)
\nonumber\\
&&+\frac{1}{2}\gamma
n\left(2\hat{a}^{\dagger}\hat{\rho}\hat{a}-\hat{a}\hat{a}^{\dagger}\hat{\rho}-\hat{\rho}\hat{a}
\hat{a}^{\dagger}\right)
\nonumber\\
&&-\frac{1}{2}\gamma
M\left(2\hat{a}\hat{\rho}\hat{a}-\hat{a}\hat{a}\hat{\rho}-\hat{\rho}\hat{a}
\hat{a}\right)
\nonumber\\
&& -\frac{1}{2}\gamma M^{\star}\left(2\hat{a}^{\dagger}\hat{\rho}
\hat{a}^{\dagger}-\hat{a}^{\dagger}\hat{a}^{\dagger}\hat{\rho}-
\hat{\rho}\hat{a}^{\dagger}\hat{a}^{\dagger}\right). \label{ME1}
\end{eqnarray}
We refer to $M$ as a reservoir squeezing parameter and to $n$ as
the mean number of reservoir photons. These parameters satisfy the
inequality $|M|\leq\sqrt{n(n+1)}$. For the squeezed-vacuum
reservoir, these parameters are given by $n=\sinh^2(r)$ and
$M=\cosh(r)\sinh(r)\exp(-i\theta)$, implying the equality
$|M|=\sqrt{n(n+1)}$, where $r$ and $\theta$ correspond,
respectively, to the amplitude and phase of the squeezing
parameter $\xi=r\exp(i\theta)$ (see Appendix~\ref{appB} for more
details). Apart from the standard parts in Eq.~(\ref{ME1}), which
describe a thermal-like Markovian reservoir with the mean photon
number $n$ allowing for single-photon dissipation, this master
equation includes also two-photon decay processes. Indeed,
Eq.~(\ref{ME1}) reduces to the standard master equation for the
thermal reservoir by setting $M\rightarrow 0$ and $n\rightarrow
\Nthermal= \{\exp[\hbar\omega/(k_B T)]-1\}^{-1}$, which becomes
the mean number of thermal photons at the frequency $\omega$ and
temperature $T$, where $k_B$ is the Boltzmann constant.

 \begin{figure}[t]
 \includegraphics[width=\columnwidth]{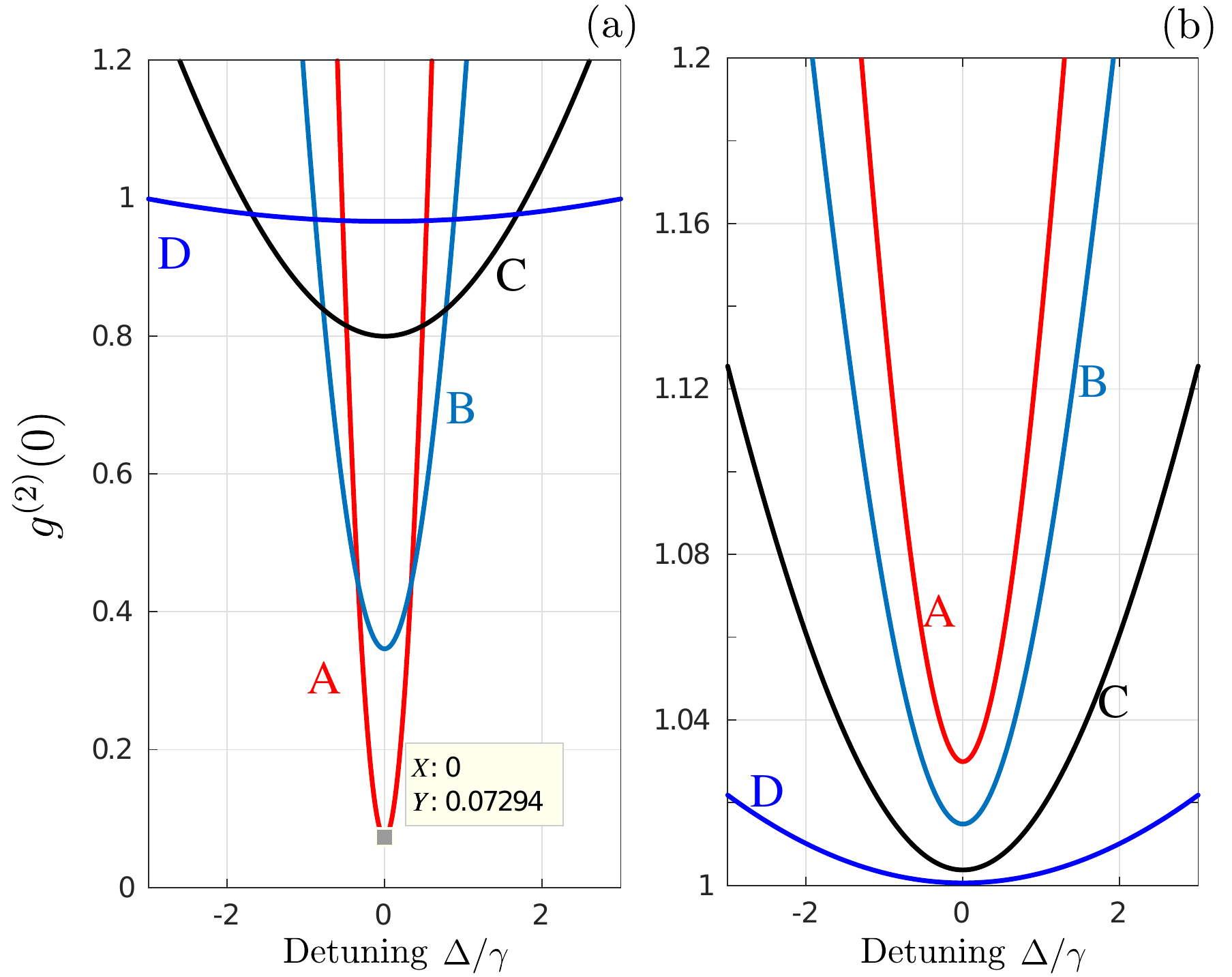}
\caption{Squeezed-reservoir model: Single-photon blockade in a
driven harmonic resonator coupled to a squeezed reservoir.
Specifically, the steady-state second-order correlation function
$\cor{2}$ vs. the detuning $\Delta$ between the cavity and driving
frequencies for various values of the external field strength
$\varepsilon$, assuming: (a) the squeezed-vacuum reservoir (see
Appendix~\ref{appB}) with $M=\sqrt{n(n+1)}$ and (b) no squeezing
($M=0$) of the reservoir. We set the reservoir mean photon number
as $n=3\times 10^{-4}$, {and $\varepsilon/\gamma=0.07$ (curve A),
0.1 (B), 0.2 (C), and 0.5 (D).} All the parameters are scaled in
$\gamma=1$ units. Panel (a) shows strong single-time photon
antibunching, especially for $\varepsilon=0.07\gamma$ and
$\Delta=0$, which characterizes single-photon blockade. Panel (b)
shows single-time photon bunching, which confirms that the
single-photon blockade in panel (a) results from the squeezed
reservoir.}
 \label{fig07}
 \end{figure}
\begin{figure}[t]
\begin{center}
 \includegraphics[width=0.8\columnwidth]{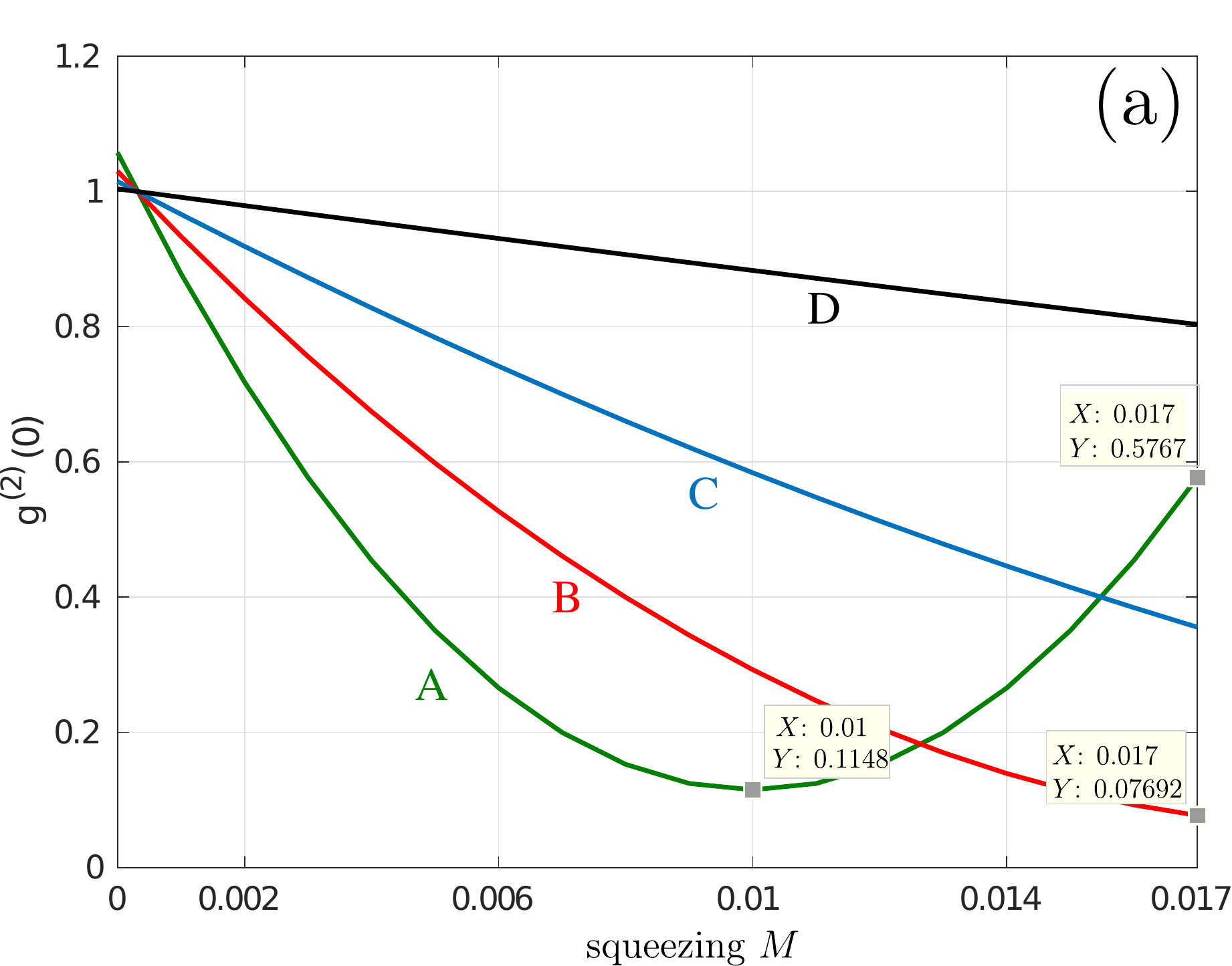}\\
 \includegraphics[width=0.8\columnwidth]{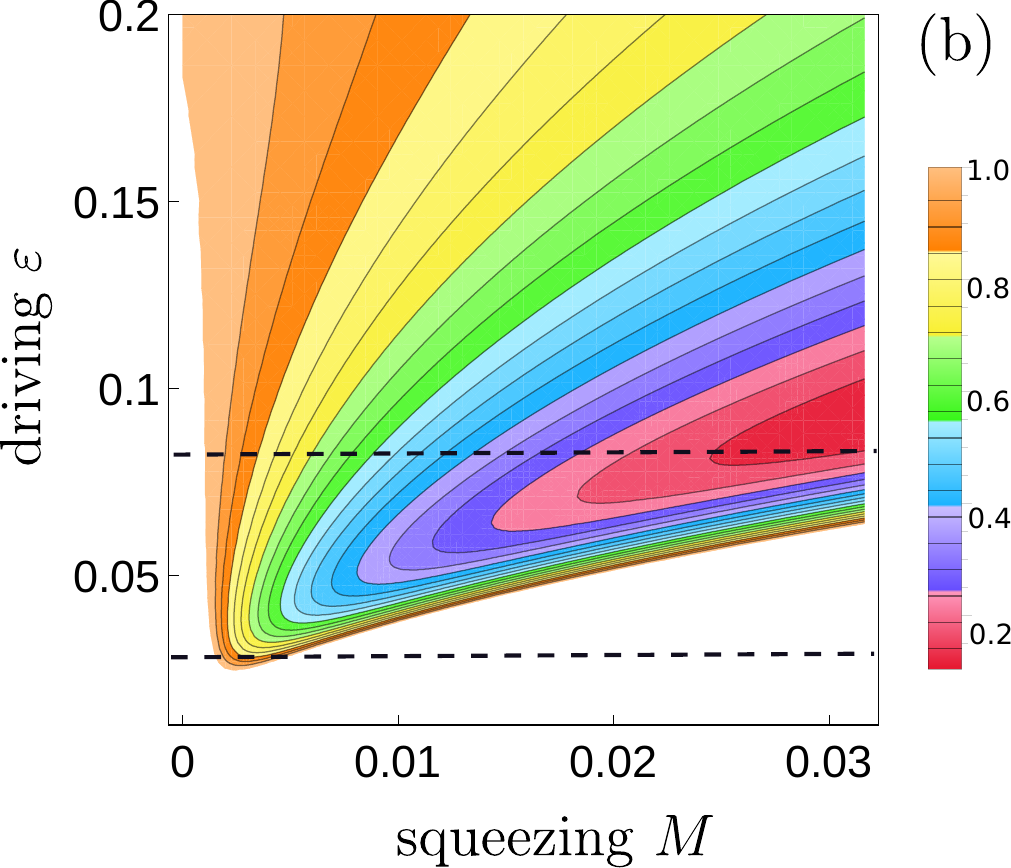}
\caption{Squeezed-reservoir model: Steady-state correlation
function $\cor{2}$ vs. the reservoir squeezing parameter $M$ and
the driving strength $\varepsilon$ in a driven harmonic resonator
coupled to a squeezed reservoir. {We set $\varepsilon/\gamma=$0.05
(curve A), 0.07 (B), 0.1 (C) and 0.2 (D). Moreover,} we assume
resonance between the cavity and external fields, $\Delta=0$, and
the mean photon number of the squeezed reservoir is $n=3\times
10^{-4}$. All the parameters are scaled in $\gamma$ units. It is
seen that, usually, a larger reservoir squeezing parameter $M$
implies stronger single-time PAB, reaching the smallest value of
$\cor{2}$ for the squeezed-vacuum reservoir with $M=\sqrt{n(n+1)}$
(see Appendix~\ref{appB}). However, this is not the case for,
e.g., $\varepsilon=0.05\gamma$ in panel (a), when there is an
optimal value of $M\ll \sqrt{n(n+1)}$, which results in the
strongest single-time PAB. The same surprising result is shown in
panel (b) in the area between the dashed lines indicating the
range of the external field strength $\varepsilon$ for which
$\cor{2}$ has a minimum for $M<\sqrt{n(n+1)}$.}
 \label{fig08}
 \end{center}
 \end{figure}
\subsection{Standard single-photon blockade}
\label{sec3B}

As mentioned above, the standard indicator of single-PB is the
condition $\cor{2}<1$ showing the decreased probability of
measuring simultaneously two photons  during the process of the
cavity-field dissipation.

In Figs.~\ref{fig07}(a) and~\ref{fig07}(b), we have shown the
dependence of single-time steady-state second-order correlation
function $\cor{2}$ vs. the detuning $\Delta$ for the harmonic
cavity field decaying, respectively, into (a) the squeezed-vacuum
reservoir (i.e., the maximally squeezed reservoir with
$M=\sqrt{n(n+1)}$) and (b) the standard thermal reservoir ($M=0$)
with the same mean number $n=0.003$ of reservoir photons. Various
external-driving-field strengths are considered. Our first
conclusion is that the squeezing of the field in the reservoir is
responsible for generating single-PB of the linear-cavity field,
as described by the sub-Poissonian photon-number statistics shown
in Fig.~\ref{fig07}(a); while the interaction with the thermal
field of the environment inevitably leads to the super-Poissonian
photon-number statistics of the cavity field shown in
Fig.~\ref{fig07}(b). This effect can be interpreted as PIT. In all
of these cases, by tuning the frequency of the external excitation
with the cavity frequency, one can assure the lowest possible
value of $\cor{2}$. Additionally, a weaker external driving is
preferable to obtain lower values of $\cor{2}$. For the parameters
presented in Fig.~\ref{fig07}(a), the lowest value of $\cor{2}$ is
$0.0729$. By decreasing the mean photon number inside the squeezed
reservoir, or by applying a weaker external field, one can obtain
even smaller values of $\cor{2}$ under the exact resonance
condition $\Delta=0$.

Figures~\ref{fig08}(a) and \ref{fig08}(b) show the dependence of
the steady-state single-time second-order correlation $\cor{2}$ on
the reservoir squeezing parameter $M$ and the driving strength
$\varepsilon$. Usually, the minimal possible values of $\cor{2}$
are obtained when the field inside the reservoir is maximally
squeezed, i.e., for the squeezed-vacuum reservoir satisfying
$M=\sqrt{n(n+1)}$. However, for very weak excitations, the
dependence $\cor{2}$ vs. $M$ has a minimum for $M<\sqrt{n(n+1)}$.
Thus, it is worth stressing that it is possible to use a
non-maximally squeezed reservoir, which still enables strong
single-time PAB for very weak excitations, as shown in
Fig.~\ref{fig08}(a).

\subsection{Nonstandard single-photon blockade with two-time photon bunching}
\label{sec3C}

Here we discuss whether a squeezed reservoir can generate
nonstandard PB exhibiting two-time photon bunching and single-time
photon antibunching. Three examples of this PB are shown in
Figs.~\ref{fig02}(a) and~\ref{fig02}(c), as indicated by symbols
A, B, and 4. These examples should be compared with the examples
of true single-PB indicated there by arrows 2 and 3.

More specifically, in Fig.~\ref{fig02}(a), the steady-state
two-time second-order correlation function $g^{(2)}(\tau)$ is
shown vs. the rescaled delay time $\gamma\tau$ for the same values
of the parameters as those in Fig.~\ref{fig08}(a). We assumed here
the maximal squeezing of the field in the reservoir, i.e.,
$M=\sqrt{n(n+1)}$. For each of the considered cases, having the
minimum of $\cor{2}<1$, the cavity field clearly exhibits two-time
PAB, $g^{(2)}(\tau)>\cor{2}$. When $\varepsilon$ takes such a
value, which results in the minimal value of $\cor{2}$ for a
non-maximally squeezed reservoir field, the cavity field exhibits
two-time bunching of photons for short delay times. PAB appears
for longer delay times. In Figs.~\ref{fig02}(b) and
\ref{fig02}(c), this behavior is studied in more details. It
appears that, depending on the reservoir squeezing degree $M$ of
the reservoir, both two-time photon bunching and antibunching are
possible. But bunching for short delay times is possible only for
such values of $M$, which result in decreasing $\cor{2}$ for
increasing $M$.

\subsection{Nonstandard single-photon blockade of types 2 and 3}
\label{sec3D}

We will show now the possibility of generating nonstandard
single-PB of the second and third types in the system considered
here.

In Fig.~\ref{fig06}(a), the correlation functions of $\cor{2}$,
$\cor{3}$, and $\cor{4}$ are shown in their dependence on the
external excitation strength $\varepsilon$ for a specified mean
number of photons in the squeezed reservoir; while
Fig.~\ref{fig06}(b) shows $\cor{2}$ and $\cor{3}$ as a function of
the detuning $\Delta$. As one can see, there are ranges of the
excitation strengths $\varepsilon$ and the detuning $\Delta$ for
which $\cor{2}<1$ is accompanied by the additional condition for
$\cor{3}>1$, which implies the occurrence of NPB of type 2. For
these regions, the condition $\cor{2}<1$ is not sufficient for
identifying ``true'' single-PB, because there still exists a
nonzero probability of measuring more than two photons at the same
time. Only the two-photon statistics is suppressed and that can be
also achieved when the external driving field is off resonance
with the cavity frequency. Larger values of $\varepsilon$ are
related to the simultaneous suppression of the higher-order
correlations. Although the values of $\cor{2}$ are increasing,
still we can decrease the higher-order correlations below the
value of $\cor{2}$, as shown in Fig.~\ref{fig06}(a).

Moreover, in Fig.~\ref{fig04}(d), we show the ranges of the
squeezing $r$ and displacement $\alpha$ parameters, for which
another type of nonstandard single-PB (i.e., type~3) can be
observed. This NPB in Fig.~\ref{fig04}(d) is shown in addition to
the NPB of type~2 presented in Fig.~\ref{fig04}(c).

\subsection{Two-photon blockade}
\label{sec3E}

As shown in Fig.~\ref{fig03}, various types of single-PB can be
generated via dissipation of a linearly driven optical cavity
field into a squeezed environment. However, two-PB, according to
the \emph{simplified} criteria in
Eq.~(\ref{simplified_criteria2}), is not observed in this model,
which is demonstrated in Fig.~\ref{fig03}(e) for a specific choice
of $n$. Also our numerical calculations show that it is very
unlikely to generate three-PB according to the \emph{simplified}
criteria Eq.~(\ref{simplified_criteria3}) for an arbitrary value
of $n$.

\begin{figure}[t]
\begin{center}
\subfloat[$M\leq\sqrt{n(n+1)}$,
$n=0.03$]{\includegraphics[scale=.5]{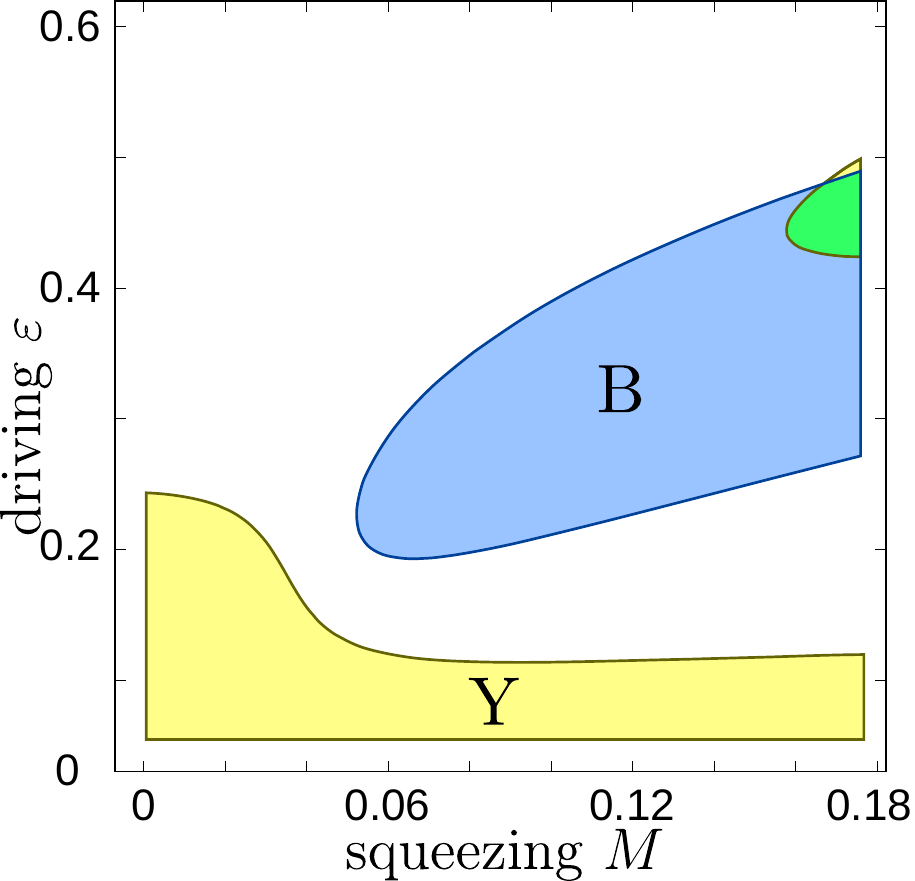}}
\subfloat[$M=\sqrt{n(n+1)}$]{\includegraphics[scale=.5]{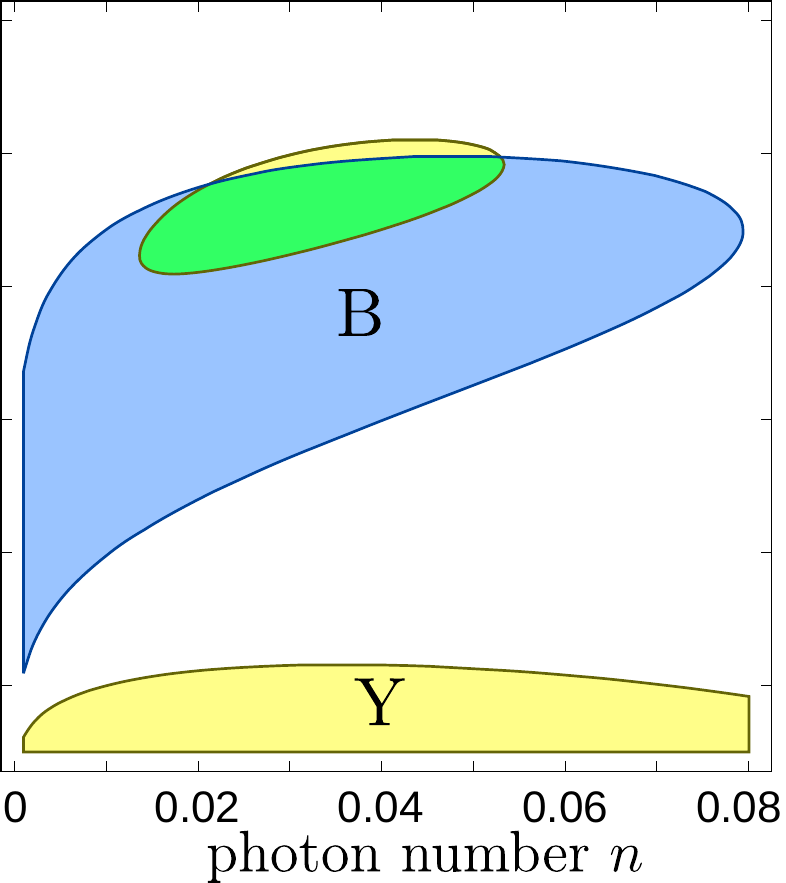}}
\end{center}
\caption{Squeezed-reservoir model: Two-photon blockade generated
in a driven harmonic cavity coupled to a squeezed reservoir
according to the refined criteria in Eq.~(\ref{refined_criteria})
for $n=0.03$ and (a) $M\leq\sqrt{n(n+1)}$ and (b)
$M=\sqrt{n(n+1)}$ corresponding to the squeezed-vacuum reservoir
(see Appendix~\ref{appB}), assuming $\Delta=0$. Specifically, by
changing the driving strength $\varepsilon$ vs. (a) the reservoir
squeezing parameter $M$ and (b) the reservoir mean photon number
$n$, we show the regions in which the criteria \#1 and \#2 are
satisfied, as indicated in yellow and blue, respectively.
Two-photon blockade occurs when both criteria \#1 and \#2 are
satisfied, which corresponds to the green regions. }\label{fig09}
\end{figure}

However, two-PB, according to the \emph{refined} criteria in
Eq.~(\ref{refined_criteria}), can be observed in this model.
Indeed, the green regions in Fig.~\ref{fig09} show the ranges of
the parameters $M$, $\epsilon$, and $n$ for which two-PB can be
observed.

\section{Simulating various types of photon blockade and tunneling with squeezed coherent states}
\label{sec4}

\subsection{Squeezed coherent states}
\label{sec4A}

Ideal SCS, or more precisely the displaced squeezed vacuum, can be
obtained by applying the squeezing and displacement operators to
the vacuum state as follows:
\begin{equation}
|\alpha,\xi\rangle= \hat D(\alpha) \hat
S(\xi)|0\rangle,\label{SCS}
\end{equation}
where
\begin{equation}
  \hat S(\xi)=\exp\left[\tfrac{1}{2}\left(\xi^{\star}\hat{a}^2-\xi\hat{a}^{\dagger
2}\right)\right] \label{SqueezingOp}
\end{equation}
is the squeezing operator with a complex squeezing parameter
$\xi=r\exp(i\theta)$ and
$D(\alpha)=\exp(\alpha\hat{a}^{\dagger}-\alpha^{\star}\hat{a})$ is
the displacement operator with $\alpha=\bar\alpha \exp(i\phi)$,
for arbitrary phases $\theta,\phi\in[0,2\pi]$ and amplitudes
$\bar\alpha,r\ge 0$.

The second-order correlation function $\cor{2}$ for the SCS with
arbitrary values of $\theta$ and $\phi$ is given by
\begin{equation}
\cor{2}=3+2\left(1-2\bar{\alpha}^2\right)\bar
N^{-1}-\bar{\alpha}^2\left[1 +C\right]\bar N^{-2}, \label{g2_n0}
\end{equation}
where the mean photon number is
\begin{eqnarray}
  \bar N\equiv \langle \hat a^\dagger \hat a \rangle
=\tfrac12[2\bar{\alpha}^2+\cosh\left(2r)-1\right], \label{N_scs}
\end{eqnarray}
and $C=\cos\left(2\phi-\theta\right)\sinh \left(2r\right)$. For a
special case with the optimally squeezed amplitude quadrature
($\theta=2\phi$), Eq.~(\ref{g2_n0}) simplifies to the formula
given in Ref.~\cite{Lemonde2014}. Note that such phase
optimization corresponds to the so-called principal
squeezing~\cite{Luks1988,Loudon1989,Adam2010}.

Our main objective is to determine whether two-PB (2PB) and
three-PB (3PB), as well as various types of nonstandard single-PB
(NPB) and other phenomena such as PIT, can be generated or
simulated with squeezed states. Thus,  we have to examine
higher-order correlation functions, namely $\cor{3}$ and
$\cor{4}$. We find that the third-order correlation function for
the SCS with arbitrary angles $\theta$ and $\phi$ is
\begin{eqnarray}
\cor{3}&=&15+9\left(1-3\bar{\alpha}^2\right)\bar N^{-1}
-9\bar{\alpha}^2(1+B)\bar
N^{-2}\nonumber\\
&&+2\bar{\alpha}^2\left(2\bar{\alpha}^2+3C\right) \bar N^{-3},
\label{g3_n0}
\end{eqnarray}
which considerably simplifies for the optimally squeezed amplitude
quadratures ($\theta=2\phi$).

The analytical solution of the simplified criteria in
Eq.~(\ref{simplified_criteria2}) can be obtained for the optimally
squeezed state holding the relation of $\theta=2\phi$.
Additionally, analytical solutions can also be found whenever one
of the phases is fixed and the other takes any value from the
range $[0,2\pi]$. Our numerical and analytical results show that
it is very unlikely to obtain the simplified conditions in
Eq.~(\ref{simplified_criteria2}) for two- and three-PB for the SCS
having the optimally squeezed amplitude quadratures. The same
conclusion holds for the SCS with one of the phases fixed and for
any values of: the other phase, $\alpha$, and $r$. This conclusion
has been confirmed numerically for $10^6$ randomly generated SCS
without fixing any parameters.

Thus, we have shown that multi-PB, according to the simplified
criteria in Eq.~(\ref{simplified_criteria2}), are very unlikely
for any choice of the parameters of the SCS. This suggests that,
by having a physical system evolving into a squeezed state, one
can expect the possibility of generating single-PB but standard
squeezing does \emph{not} lead to the generation of this type of
multi-PB.

In contrast to this, we find that two-PB is still possible, but
according to the refined criteria in Eq.~(\ref{refined_criteria}).
Indeed, for properly chosen parameters $M$ and $\varepsilon$ of
the SCS, two-PB can be observed as shown by the green regions in
Fig.~\ref{fig09}.

Nonstandard single-PB (of type 2) can occur for the SCS. Indeed,
we have found analytical solutions satisfying both criteria in
Eq.~(\ref{type2}). Such solutions exist only for some relations
between the phases of the displacement and squeezing operators.
The ranges of these phases are collected in Table~\ref{table3}.
The nonstandard PB effect cannot be observed for other phase
relations.

\begin{table}[t]
\begin{tabular}{c c c}
\hline %
 $\theta$ & $\phi$ & NPB\\[2pt] \hline
 0 & $(-\pi/4;\pi/4)$ \& $(3\pi/4;5\pi/4)$ & yes\\[2pt]
   & $[\pi/4;3\pi/4]$ \& $[5\pi/4;7\pi/4]$ &  no\\[5pt]
$\pi$  & $(\pi/4;3\pi/4)$ \& $(5\pi/4;7\pi/4)$ & yes\\[2pt]
       & $[ -\pi/4;\pi/4]$  \& $[ 3\pi/4;5\pi/4]$ & no\\[2pt]
\hline
$(-\pi/2;\pi/2)$& $0,\pi$ & yes\\[2pt]
 $[\pi/2;3\pi/2]$   &  & no\\[5pt]
$[-\pi/2;\pi/2]$ & $\pi/2$ & no\\[2pt]
    $(\pi/2;3\pi/2)$    &   & yes\\[2pt] \hline
\end{tabular}
\caption{Squeezed coherent states simulating nonstandard photon
blockade (of type 2), for which $\cor{2}<1$ and $\cor{3}>1$ hold,
vs. the phase $\phi=\Arg\alpha$ of the displacement operator, and
the phase $\theta=\Arg\xi$ of the squeezing parameter. }
\label{table3}
\end{table}

\subsection{Displaced squeezed thermal states}
\label{sec4B}

In addition to the SCS, we also analyze the displaced squeezed
thermal states (DSTS), which can be obtained by applying the
displacement $\hat D(\alpha)$ and squeezing $\hat S(\xi)$
operators to a thermal state $\hat{\rho}_{\rm th}(\Nthermal)$,
i.e.:
\begin{equation}
\hat{\rho}(\alpha,\xi,\Nthermal)=\hat D(\alpha)\hat
S(\xi)\hat{\rho}_{\rm th}(\Nthermal)\hat S^{\dagger}(\xi)\hat
D^{\dagger}(\alpha).\label{DSTS}
\end{equation}
The thermal state is characterized by the density matrix
$\hat{\rho}_{\rm th}(\Nthermal)=\sum\limits_n P_n|n\rangle\langle
n|,$ where $P_n=n_{\rm th}^n/(1+\Nthermal)^{n+1}$ is the
probability of finding $n$ thermal photons in a thermally-excited
mode having a geometric probability distribution, and $\Nthermal$
is the mean number of thermal photons.

In Appendix~\ref{appE} we show explicitly that the DSTS
$\hat{\rho}(\alpha,\xi,\Nthermal)$ are nonclassical if and only if
the squeezing parameter $r\equiv |\xi|$ is greater than the
critical value $r_0$:
\begin{equation}
  r > r_0\equiv \tfrac 12 \ln(1+2\Nthermal).\label{r_0}
\end{equation}
These states are nonclassical, independent of the displacement
parameter $\alpha$, because they are described by a
non-positive-semidefinite Glauber-Sudarshan $P$-function. This is
demonstrated in Appendix~\ref{appE} without recalling the explicit
form of the $P$-function for the DSTS. Further discussion of the
nonclassical ($r > r_0$) and classical ($r \le r_0$) regimes of
the DSTS in relation to their simulation of PIT is presented in
Sec.~\ref{sec5}.

Applying the definition of the $k$th-order correlation functions
$\cor{k},$ we can easily obtain the following relations describing
the second- and third-order equal-time correlation functions:
\begin{eqnarray}
\cor{2}&=&3+\left(1-2\bar{\alpha}^2\right)\bar N^{-1}-h^{-},\label{g2_term}\\
\cor{3}&=&15+9\left(1-3\bar{\alpha}^2\right)\bar N^{-1}-9h^{+}
\nonumber
\\ &&+2\bar{\alpha}^2\left[2\bar{\alpha}^2+3(2\Nthermal+1)B\right]\bar
N^{-3}, \label{g3_term}
\end{eqnarray}
where the mean photon number is
\begin{equation}
  \bar N\equiv \langle \hat a^\dagger \hat a \rangle
=\tfrac12 [2\bar{\alpha}^2+\left(1+2\Nthermal\right)\cosh(2r)-1],
\label{N_DSTS}
\end{equation}
and the auxiliary functions are
\begin{equation}
  h^{\pm}=\left\lbrace \Nthermal(1+\Nthermal)+\bar{\alpha}^2\left[1\pm (2\Nthermal
+1)C\right]\right\rbrace \bar N^{-2},\quad \label{h_pm}
\end{equation}
where $C$ is defined below Eq.~(\ref{N_scs}). For $\theta=2\phi$,
Eqs.~(\ref{g2_term}) and~(\ref{g3_term}) considerably simplify. In
this special case, Eq.~(\ref{g2_term}) reduces to the
corresponding formula given in Ref.~\cite{Lemonde2014}.

\begin{figure}[t]
\begin{center}
\subfloat[$\theta=\pi$, $\phi=0$]
{\includegraphics[scale=0.30]{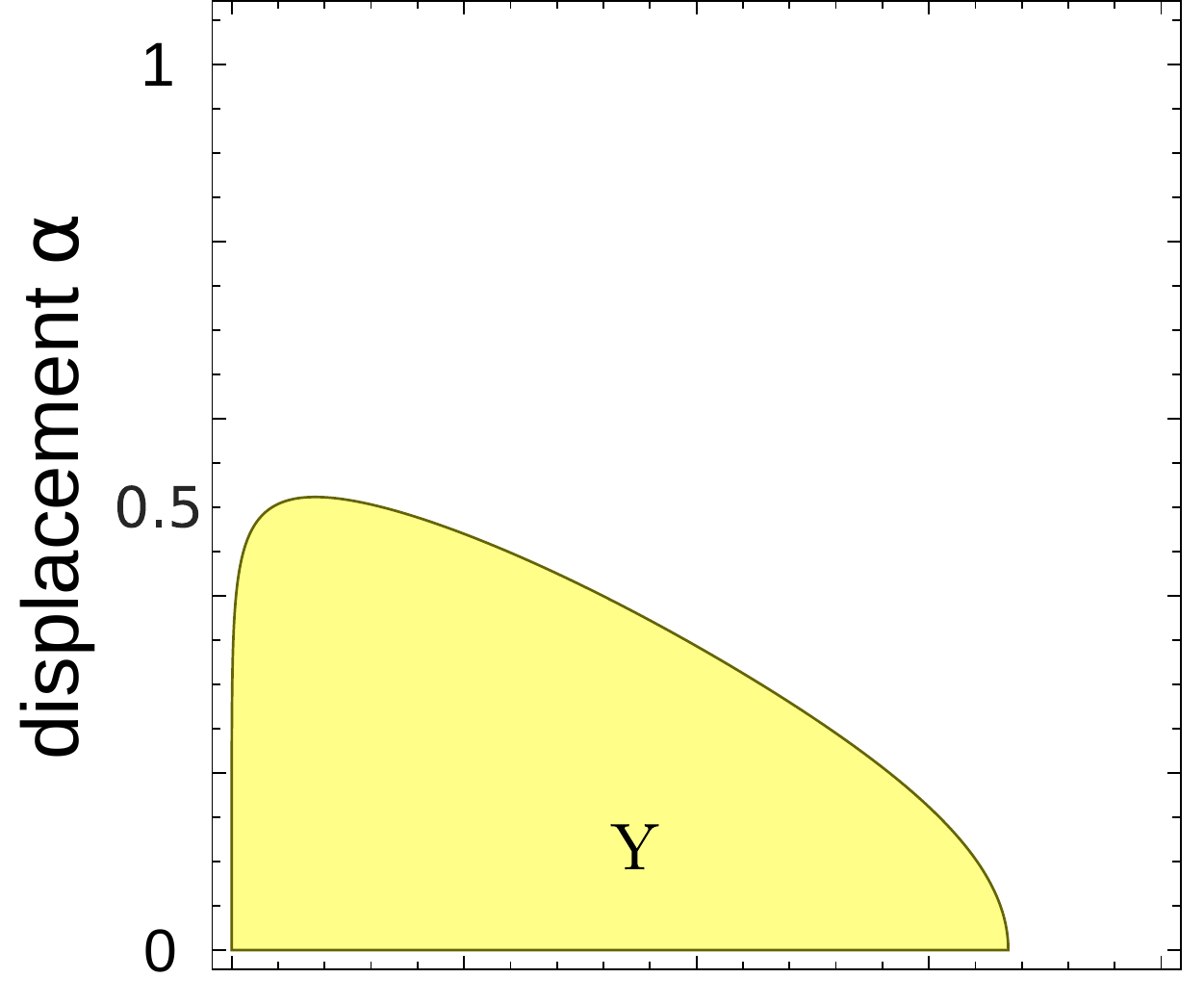}}\hspace*{2pt}
\subfloat[$\theta=\pi$, $\phi=3\pi/8$]{\includegraphics[scale=0.30]{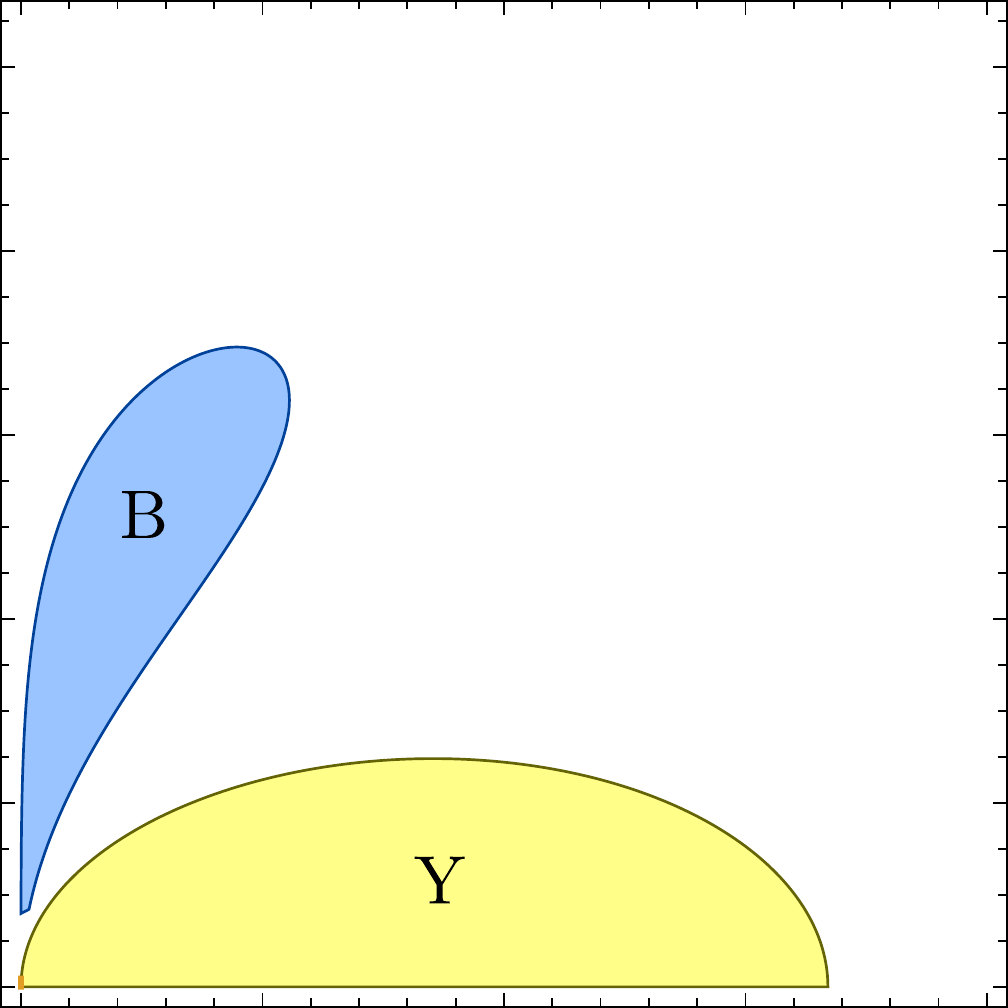}}\\
\subfloat[$\theta=\pi$,
$\phi=4\pi/8$]{\includegraphics[scale=0.30]{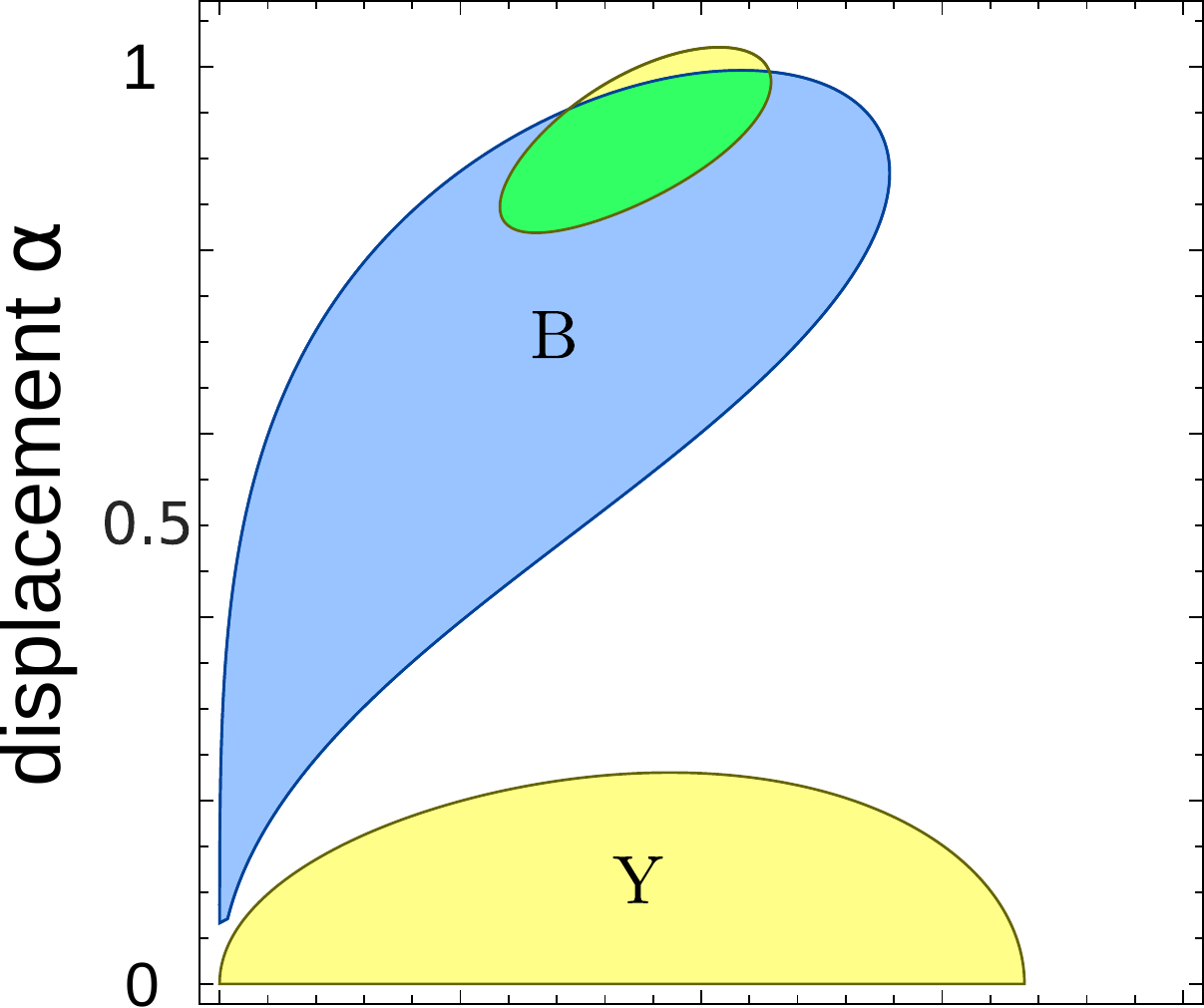}}\hspace*{2pt}
\subfloat[$\theta=\pi$, $\phi=0$]     {\includegraphics[scale=0.30]{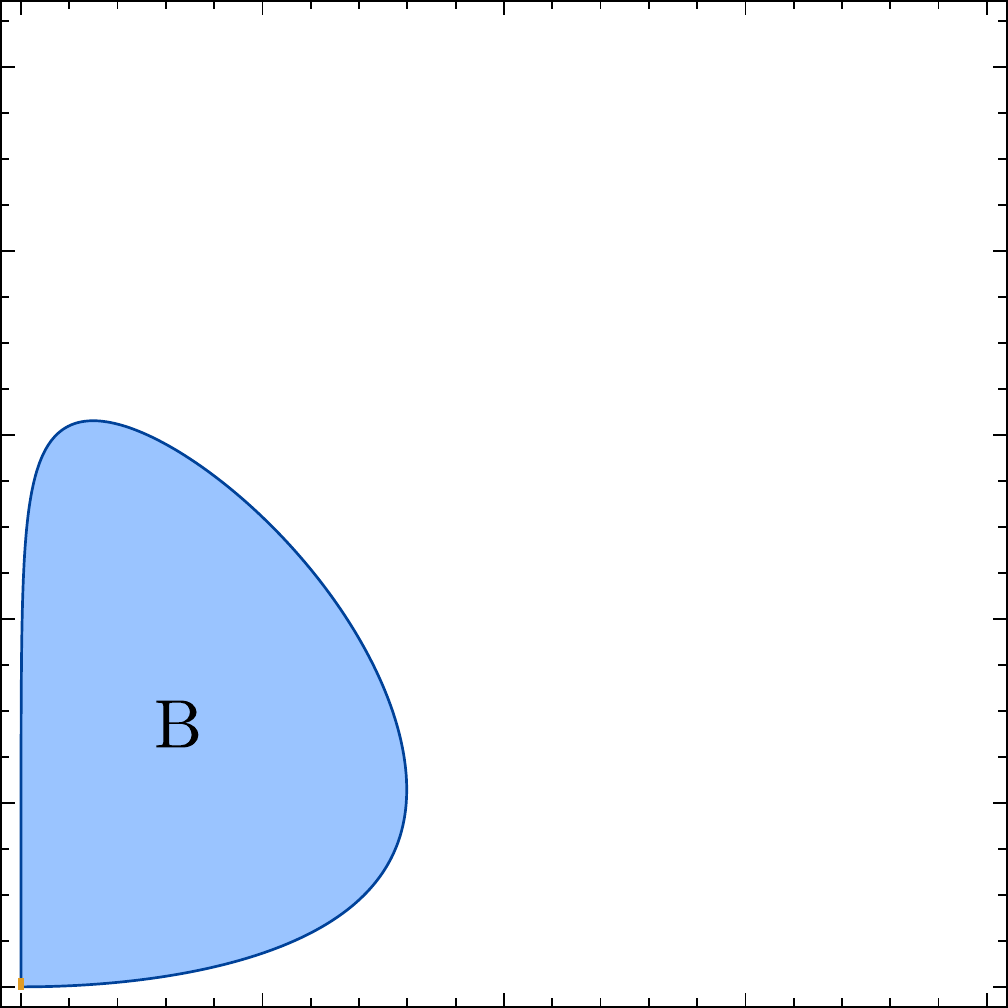}}\\
\subfloat[$\theta=\pi$,
$\phi=3\pi/8$]{\includegraphics[scale=0.30]{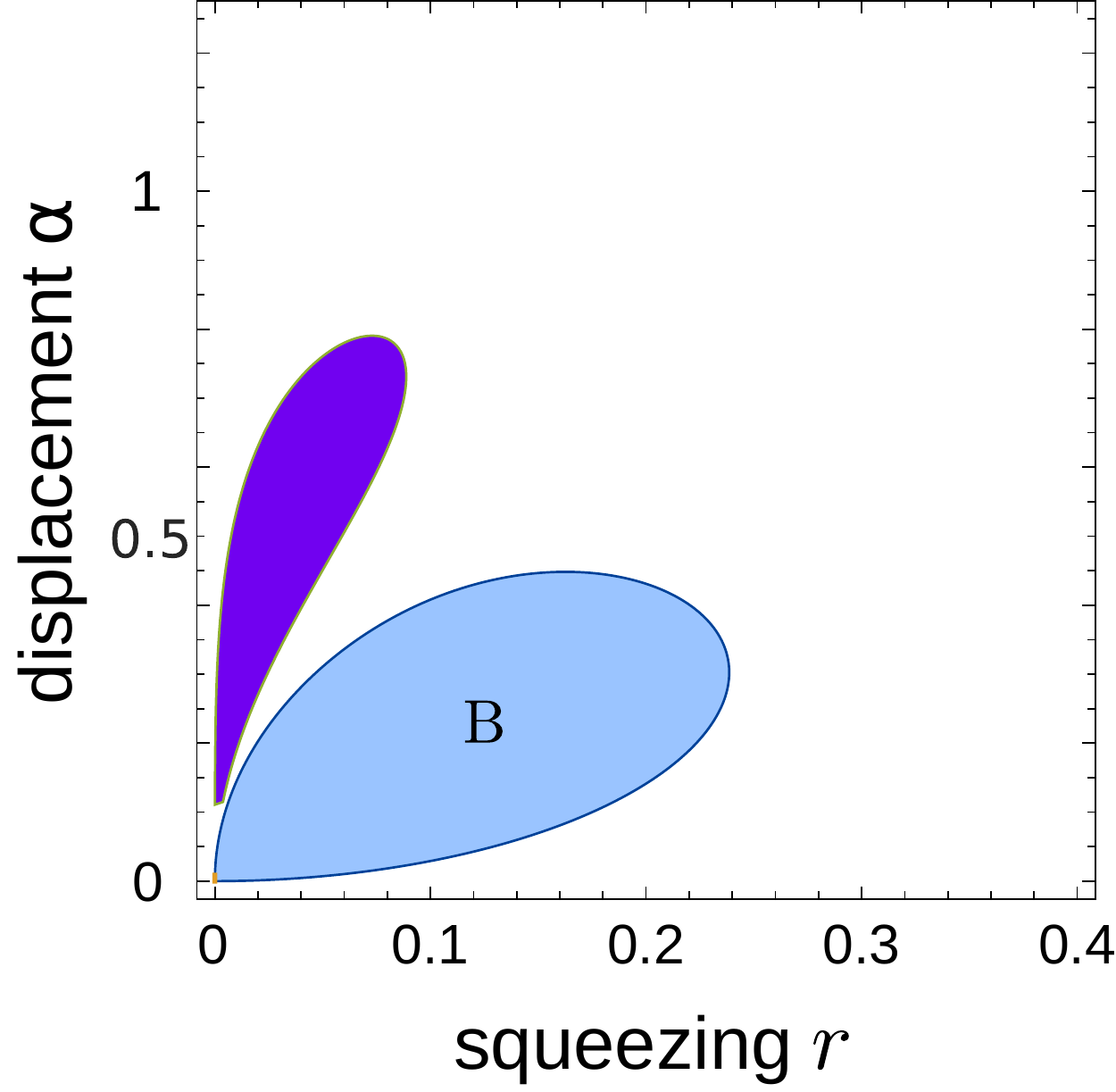}}\hspace*{1pt}
\subfloat[$\theta=\pi$,
$\phi=4\pi/8$]{\includegraphics[scale=0.30]{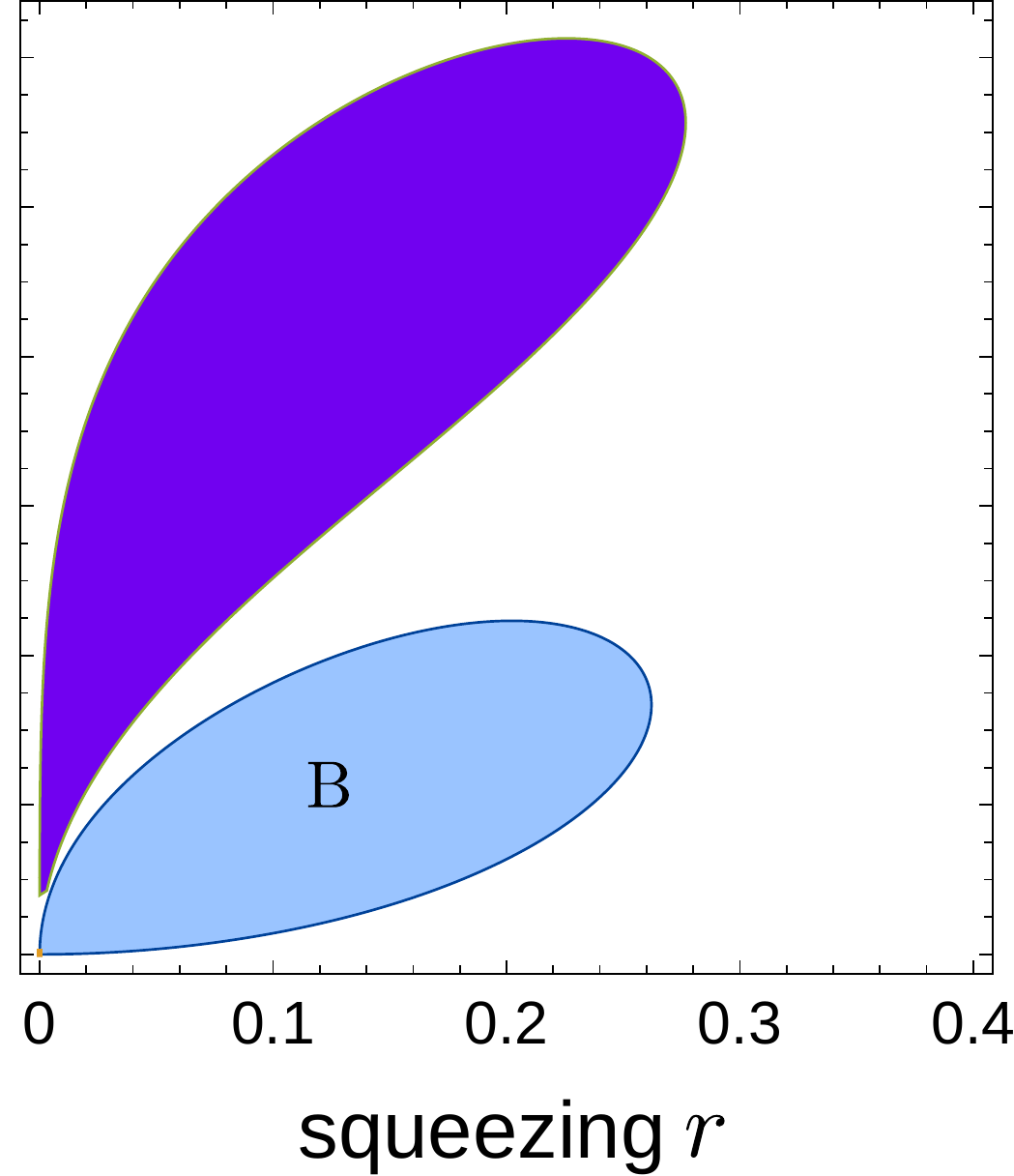}}
\end{center}
\caption{Squeezed coherent states: Regions of the displacement
($\alpha$) and squeezing ($r$) parameters for which the
\emph{refined} photon blockade criteria in
Eq.~(\ref{refined_criteria}) are satisfied for two-photon (a,b,c)
and three-photon (d,e,f) blockades. Specifically, the regions of
the SCS parameters, where the criteria \#1 and \#2, are satisfied
for $\cor{2}$, $\cor{3}$, and $\cor{4}$, are shown in yellow,
blue, and violet (or navy blue), respectively. {In grayscale,
yellow (indicated by ``Y'') is the brightest, and violet is the
darkest color. Blue is marked by ``B''.} Two-photon blockade
occurs if the criteria for $\cor{2}$ and $\cor{3}$ are both
satisfied, which corresponds to the green region in panel (c).
Three-photon blockade does not occur as the regions for $\cor{3}$
and $\cor{4}$ do not overlap.} \label{fig10}
\end{figure}
\subsection{Photon correlations in squeezed coherent states}
\label{sec4C}

\begin{figure}[t]
\begin{center}
\subfloat[two-PB for $\Nthermal=0.005$]
{\includegraphics[scale=0.35]{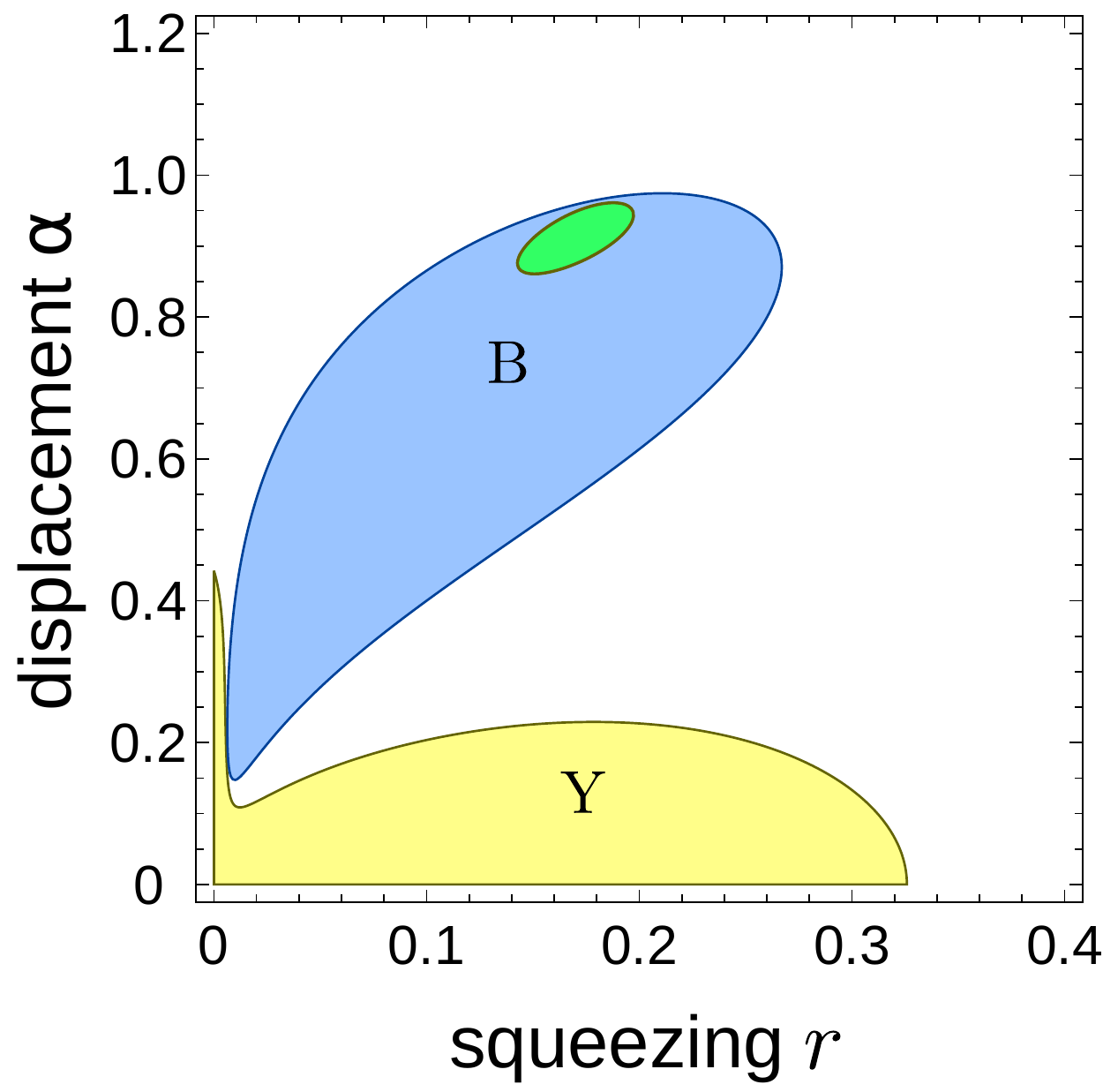}}\hspace*{2pt}
\subfloat[no two-PB for $\Nthermal=0.01$]
{\includegraphics[scale=0.35]{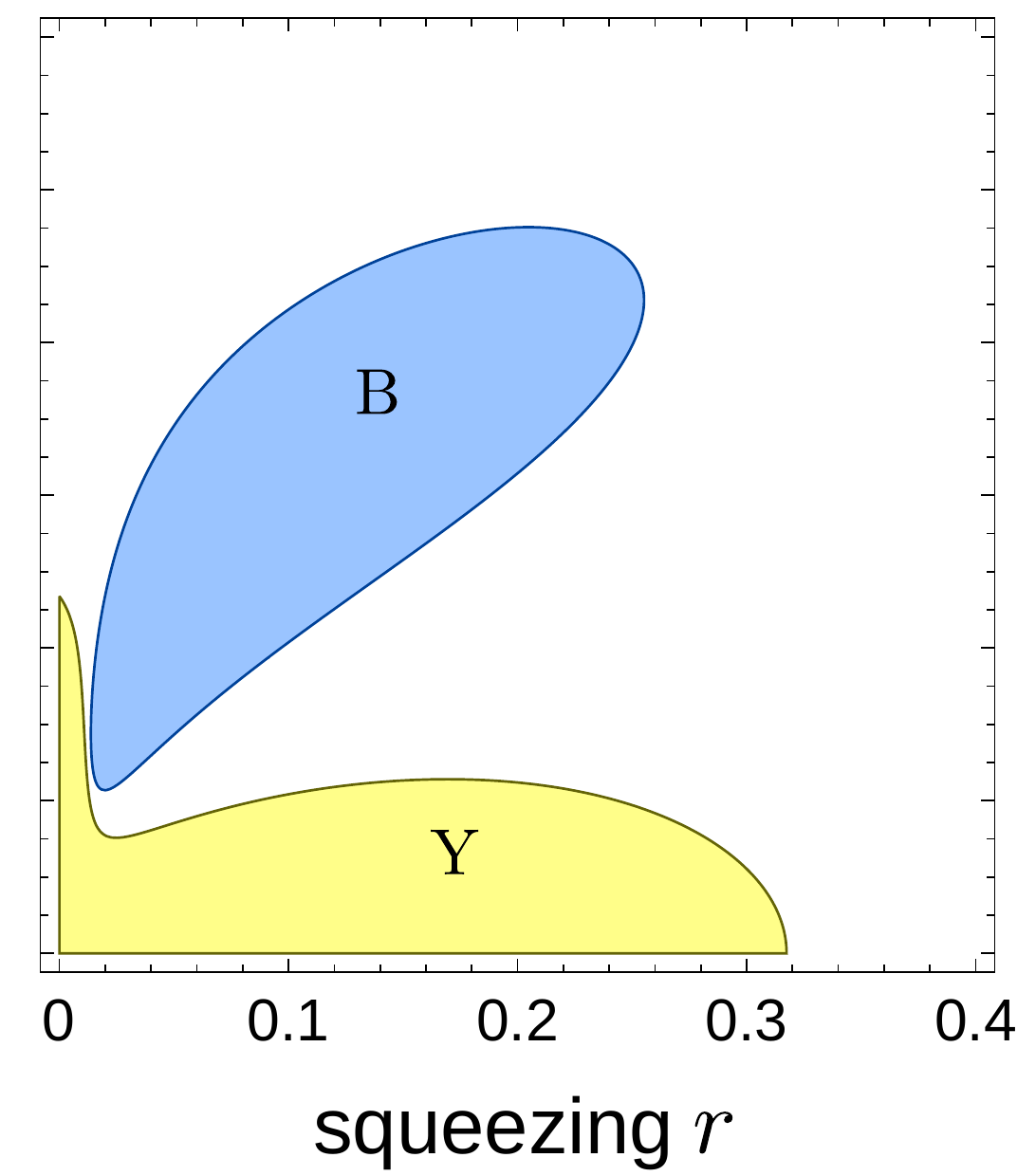}}
\end{center}
\caption{Simulation of two-photon blockade with the displaced
squeezed thermal states according to the refined criteria \#1 and
\#2. Same as in Fig.~\ref{fig10} but for the DSTS with (a)
$\Nthermal=0.005$ and (b) $n_{\rm th}=0.01$. We set here
$\theta\equiv\Arg\xi=\pi$ and $\phi\equiv\Arg\alpha=4\pi/8$.
Two-photon blockade occurs only in panel (a) (in the green
region). This green region is much larger for $\Nthermal=0$ as
shown in Fig.~\ref{fig10}(c). It is seen that even a very small
number of thermal photons severely shrinks the range of the
parameters allowing for the observation of two-photon blockade. }
\label{fig11}
\end{figure}

\begin{figure}[t]
\includegraphics[width=0.95\columnwidth]{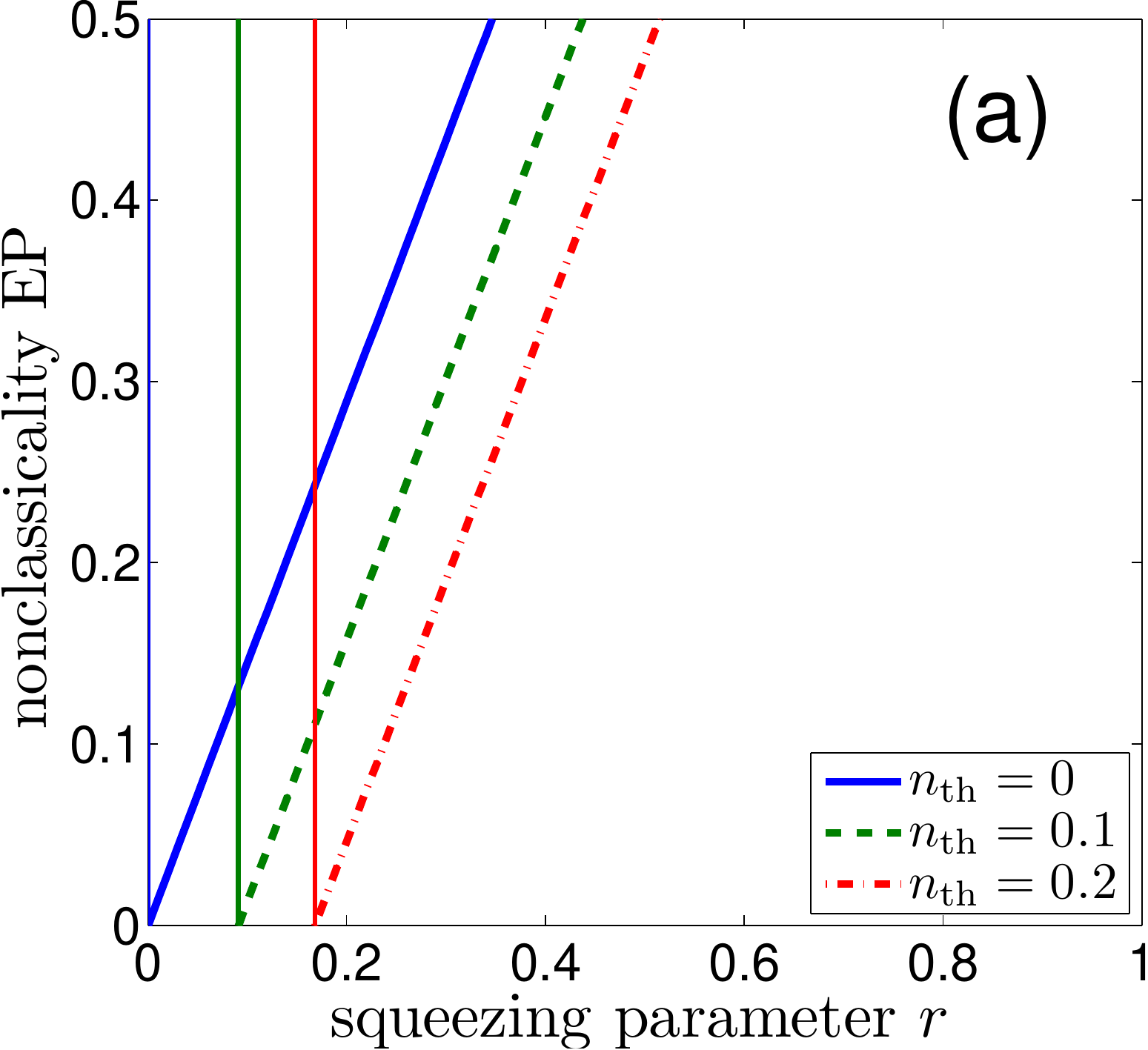}
\includegraphics[width=0.95\columnwidth]{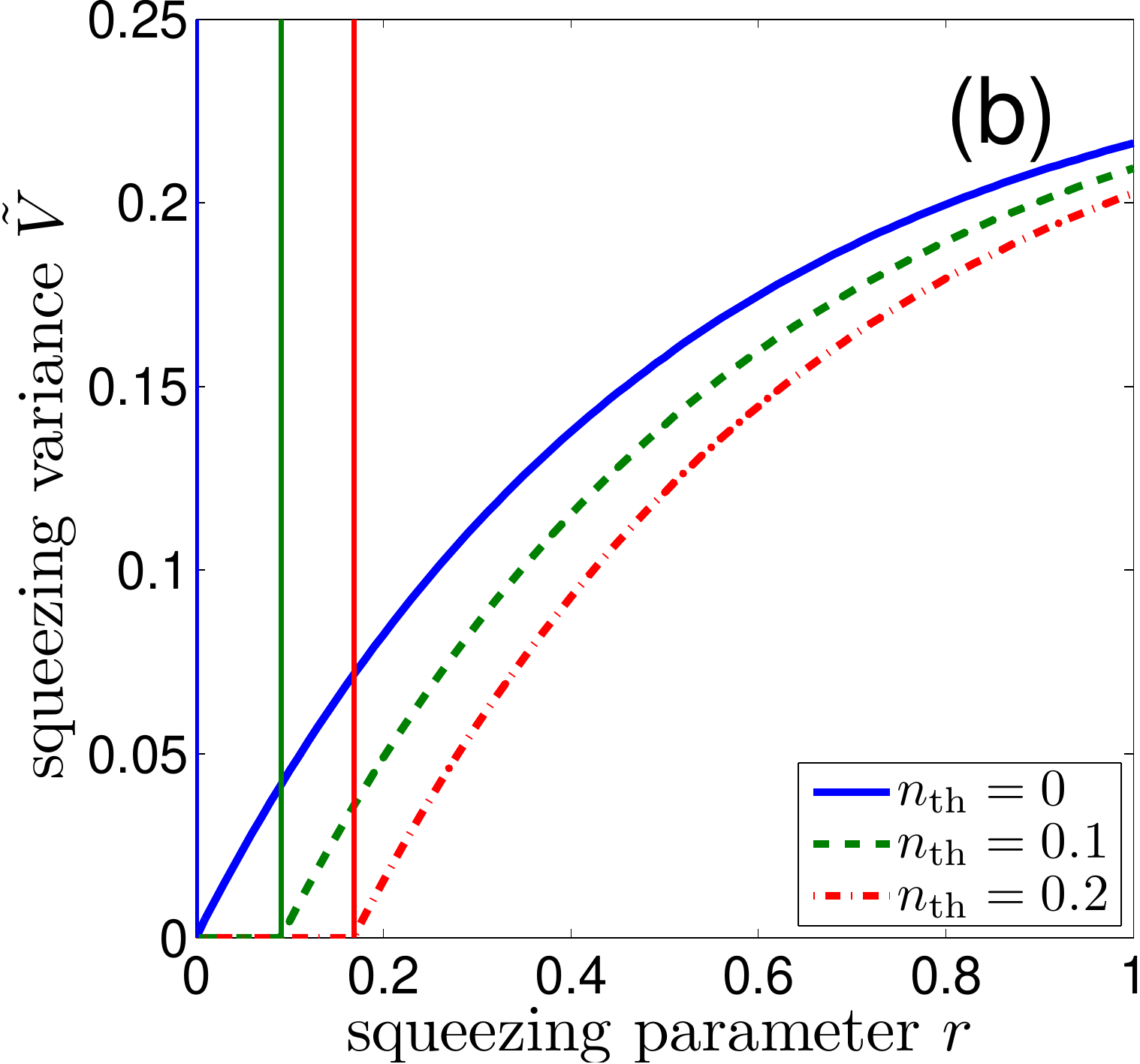}
\caption{Nonclassicality measures of the displaced squeezed
thermal states: (a) Entanglement potential, EP, of
$\hat{\rho}(\alpha,\xi,\Nthermal)$ and (b) the truncated squeezing
variance $\tilde V$, defined in Eq.~(\ref{truncated_var}), vs. the
squeezing parameter $r=|\xi|$ for different values the mean
numbers of thermal photons $\Nthermal$. The EP and squeezing
variance are independent of the displacement parameter $\alpha$
and the squeezing phase $\theta$. It is seen that the critical
values $r_0$ (specifically, $r_0=0,0.0912,0.1682$) of the
squeezing parameter $r$ increase with increasing $\Nthermal$
(i.e., $\Nthermal=0,0.1,0.2$), according to Eq.~(\ref{r_0}) as
indicated by the vertical thin solid lines.} \label{fig12}
\end{figure}

Here we analyze different kinds of PB and PIT effects as listed in
Table~\ref{table2} and shown in Figs.~\ref{fig03}-\ref{fig12}.

(i) Three-PT occurs when $1<\cor{2}<\cor{3}$. We find that these
conditions are satisfied for the SCS if $r>0$ and $\alpha$ is
smaller than a critical parameter $\alpha_0$, i.e.,
\begin{eqnarray}
  0<\alpha<\alpha_0\equiv
  \tfrac1{\sqrt{2}}\sqrt{1+c^4+c(2c+s)(cs-1)},
\label{alpha0}
\end{eqnarray}
where hereafter $c=\cosh(r)$ and $s=\sinh(r)$.

(ii) Nonstandard single-PB of type~2, which is also referred to as
unconventional PB in Ref.~\cite{Radulaski2017}, occurs if
$\cor{2}<1<\cor{3}$, which can be observed for the SCS if
$\alpha\in(\alpha_0,\alpha_1)$ for $r>0$, where the critical
parameter $\alpha_0$ is defined in Eq.~(\ref{alpha0}) and another
critical value of $\alpha$ is
\begin{eqnarray}
\alpha_1=\tfrac{1}{4 \sqrt{6s}}\sqrt{3f_7-4(3c-21s)+8
\sqrt{3\beta_1}(c+ s)s^2},\quad \label{alpha1}
\end{eqnarray}
given in terms of the auxiliary functions:
\begin{eqnarray}
  f_x&=&x\exp(-3r)-4\exp(3r)+\exp(5r),\\
  \beta_1&=&35+94 c^4+2 cs (70+47 cs)-2 c^2 (51+88 cs).\nonumber \label{beta1}
\end{eqnarray}

(iii) Nonstandard three-PT occurs when $\cor{2}<\cor{3}<1$. This
effect can be observed for the SCS if
$\alpha\in(\alpha_1,\alpha_2)$ for $r>0$, where $\alpha_1$ is
defined in Eq.~(\ref{alpha1}) and
\begin{eqnarray}
\alpha_2&=&\tfrac{1}{4 \sqrt{2r}}\sqrt{f_9-2(3c-17s)+8
\sqrt{2\beta_2}(c+s)s^2}, \label{alpha2}\\
\beta_2&=&10+29c^4+cs (40+29cs)-c^2 (31+56cs).\nonumber
\end{eqnarray}

(iv) Single-PB is usually verified by the simplified condition
$\cor{2}<1$. Stricter conditions for single-PB can be given as
$\cor{3}<\cor{2}<1$. These conditions are satisfied for the SCS if
$\alpha>\alpha_2$ and $r>0$.

(v) Our numerical and analytical calculations show that there are
no solutions for $\alpha$ satisfying the conditions
$1<\cor{3}<\cor{2}$ for two-PT.

(vi) Two-PB can indeed be observed according to the refined
conditions given in Eq.~(\ref{refined_criteria}) for both SCS and
DSTS, as shown by the green regions in Figs.~\ref{fig10}(c) and
\ref{fig11}(a), respectively, for specific choices of the
squeezing phase $\theta=\pi$ and the displacement phase
$\phi=4\pi/8$. It is seen in Figs.~\ref{fig10}(a) and
\ref{fig10}(b) that two-PB cannot be observed for the phases
$\phi=0,3\pi/8$. Figures~\ref{fig10}(c) and~\ref{fig11} show the
destructive role of thermal photons $\Nthermal$ for the generation
of two-PB. Indeed, the green region in Figs.~\ref{fig10}(c)
and~\ref{fig11} decreases with increasing $\Nthermal$, and it is
not seen any more for $\Nthermal=0.01$ in Fig.~\ref{fig11}(b).

In contrast to this refined two-PB, our analytical and numerical
calculations show that the simplified criteria in
Eq.~(\ref{simplified_criteria2}) for two-PB are very unlikely to
be satisfied as graphically explained in Fig.~\ref{fig04}(e) for
the SCS and Fig.~\ref{fig05}(e) for the DSTS.

Moreover, our both numerical and analytical results show that
three-PB can be simulated by neither SCS nor DSTS according to the
refined and simplified criteria of PB, given in
Eqs.~(\ref{refined_criteria}) and~(\ref{simplified_criteria}),
respectively. Indeed, the criteria \#1 and \#2 can be satisfied
separately, as shown by the violet and blue regions in
Figs.~\ref{fig10}(d)--(e), but they cannot be satisfied
simultaneously for the same values of the squeezing parameter $r$
and the displacement parameter $\alpha$. This result implies that
the colored regions in these figures do not overlap.

\section{Nonclassical and classical effects and states}
\label{sec5}

Now we address the question whether the analyzed effects and
states are nonclassical or not.

We apply here the standard quantum-optical definition (or
criterion) of the nonclassicality of a single-mode bosonic state
$\hat\rho$ in terms of the Glauber-Sudarshan
$P$-function~\cite{VogelBook}:
\begin{eqnarray}
  \hat\rho &=& \int {\rm d}^2{\beta}\; P(\beta,\beta^*)|\beta\rangle\langle \beta|,
 \label{Pfunction}
\end{eqnarray}
where $|\beta\rangle$ is a coherent state with a complex amplitude
$\beta$. According to this common definition (see, e.g.,
Refs.~\cite{VogelBook, Adam2010}): a given state $\hat\rho$ is
referred to as classical, if it is described by a classical-like
(i.e., non-negative) $P$-function. Otherwise, a state $\hat\rho$
is considered nonclassical (or quantum), i.e., when it is
described by a negative (or more precisely non-positive or
non-positive semi-definite) $P$-function. Thus, according to this
definition, \emph{only} coherent states and their mixtures
(including thermal states) can be considered classical, while all
other mixed and pure states (including squeezed states) are
nonclassical.

Single- and multi-PB effects are indeed purely nonclassical as
shown explicitly in Appendix~\ref{appC}.

PIT is usually also considered a quantum effect (as emphasized in,
e.g.,~\cite{Faraon2008,Huang2018}), even if it is characterized by
a classical-like property of the photon-number distributions,
i.e., the second-order or higher-order super-Poissonian
photon-number statistics (i.e., single-time photon bunching). We
note that $\cor{2}\ge 1$ is usually regarded as ``a general
property of all kinds of classical light''~\cite{LoudonBook}.
Indeed, thermal states, which are classical as given by the
mixtures of coherent states, can simulate PIT as shown in
Appendix~\ref{appD}.

A number of nonclassicality measures of bosonic fields have been
proposed, which include nonclassical depth~\cite{Lee1991},
nonclassical distance~\cite{Hillery1987}, and the nonclassicality
volume~\cite{Kenfack2004}, which corresponds to the volume of the
negative part of the Wigner function (see, e.g.,~\cite{Adam2015a}
and references therein). Here, we apply an entanglement potential
introduced by Asb\'oth \etal~\cite{Asboth2005}. Entanglement
potentials are, in general, numerically and experimentally simpler
than other formally-defined nonclassicality measures, including
the nonclassical depth and distance. Moreover, entanglement
potentials are much more sensitive in detecting nonclassicality
compared to the nonclassicality volume. Indeed, the
nonclassicality volume of the SCS studied here is exactly zero,
although the states are nonclassical according entanglement
potentials.

The basic idea of entanglement potentials is physically quite
simple: By combining a classical single-mode light with the vacuum
on a beam splitter, then the output state is separable. In
contrast to this, if the input light is nonclassical then the
output light from a lossless beam splitter is entangled. Moreover,
the degree of nonclassicality is not changed by lossless
linear-optical transformations (including beam splitters). Thus,
the degree of nonclassicality of the input state can be measured
by the output-state entanglement by applying standard entanglement
measures~\cite{Horodecki09review}, e.g., the negativity, the
concurrence, or the relative entropy of
entanglement~\cite{Asboth2005,Adam2015b}.

To be more specific, the nonclassicality of a single-mode state
$\hat\rho\equiv \hat\rho_{\rm in}$ can be quantified, according to
Ref.~\cite{Asboth2005}, by the entanglement of the output state
$\rho_{\rm out}$ of an auxiliary lossless balanced BS with the
state $\hat\rho$ and the vacuum $\ket{0}$ at the inputs, i.e.,
\begin{equation}
  \hat\rho_{\rm out} = \hat U_{\rm BS} (\hat\rho_{\rm in}\otimes \ket{0}\bra{0} )
  \hat U^\dagger_{\rm
  BS},
\label{rho_BS}
\end{equation}
where $ \hat U_{\rm BS}$ is the unitary transformation of a
balanced (50:50) lossless beam splitter,
\begin{eqnarray}
  \hat U_{\rm BS}=\exp\left[-i \tfrac{\pi}{2} (\hat a_{1}^\dagger \hat a_{2}+\hat a_{1} \hat a_{2}^\dagger)
  \right],
\label{U_BS}
\end{eqnarray}
and $\hat a_{1,2}$ ($\hat a_{1,2}^\dagger$) are the annihilation
(creation) operators of the input modes. We apply here the
entanglement potential (EP) based on the negativity
($N$)~\cite{Asboth2005,Horodecki09review}:
\begin{eqnarray}
 \EP(\hat\rho_{\rm in})&\equiv& E_N(\hat\rho_{\rm
 out})
 =\log_2[N(\hat\rho_{\rm out})+1]
 \nonumber \\
 &=& \log_2 \vert\vert\hat\rho_{\rm out}^{{\Gamma}}\vert\vert_{1},
 \label{EP}
\end{eqnarray}
which is given in terms of the trace norm $\vert\vert\hat
\rho^{{\Gamma}}\vert\vert_{1}$ of the partially-transposed
statistical operator $\hat \rho^{{\Gamma}}$, and the logarithmic
negativity $E_N$. We note that the negativity and, thus, the
corresponding entanglement potential determine, e.g.: (i) the
entanglement cost $E_{\rm cost}\equiv E_N$ under operations
preserving the positivity of the partial transpose (at least for
single-PB entangled states)~\cite{Horodecki09review} and (ii) the
dimensionality of entanglement, which is the number of the degrees
of freedom of entangled beams~\cite{Eltschka2013,Adam2016}.

The entanglement potential, defined in Eq.~(\ref{EP}), for the
DSTS is given by the following simple formula~\cite{Asboth2005}:
\begin{equation}
  \EP[\hat{\rho}(\alpha,\xi,\Nthermal)]=\frac{r-r_0}{\ln2}, \label{EP_DSTS}
\end{equation}
where the critical parameter $r_0$ is given in Eq.~(\ref{r_0}).
This entanglement potential is plotted in Fig.~\ref{fig12}(a)
together with the squeezing variance, which is another
nonclassicality measure of the DSTS. Indeed, in
Fig.~\ref{fig12}(b), we plotted the truncated squeezing variance
defined as~\cite{Bartkowiak2011}:
\begin{eqnarray}
  \tilde V &\equiv& \min[0,-\mean{:(\Delta \hat
  X_{\varphi_0})^2:}],
  \label{truncated_var}
\end{eqnarray}
where the squeezing variance for the DSTS is (see
Appendix~\ref{appE}):
\begin{equation}
  \mean{(\Delta \hat  X_{\varphi_0})^2}
  = \tfrac12 \big(\tfrac12+\Nthermal\big) \exp(-2r)
  = \tfrac14 \exp[-2(r-r_0)] ,
  \label{var_DSTS1}
\end{equation}
and $\mean{:(\Delta \hat  X_{\varphi_0})^2:}=\mean{(\Delta \hat
X_{\varphi_0})^2}-1/4$. We note that, in general, squeezing for an
optimal phase $\varphi_0$ is referred to as principal
squeezing~\cite{Luks1988,Adam2010} and its geometrical
interpretation can be provided by Booth's elliptical
lemniscates~\cite{Loudon1989}. Figure~\ref{fig12}(b) clearly shows
the same nonclassical and classical regimes of the DSTS, as those
in Fig.~\ref{fig12}(a) for the entanglement potential, as
explained in greater detail in Appendix~\ref{appE}.

By comparing Eqs.~(\ref{EP_DSTS}) and (\ref{var_DSTS1}), it is
seen that the EP is a monotonic function of the squeezing variance
for the DSTS, i.e.:
\begin{equation}
  \EP=-\tfrac{1}{2}\log_2\mean{(\Delta \hat  X_{\varphi_0})^2}-1.
  \label{EPV}
\end{equation}
Note that these quantities are also monotonically related to the
nonclassical depth for the DSTS~\cite{Lee1991}.

The nonclassicality of an arbitrary two-mode Gaussian state
$\hat\rho_{\rm out}$ (which includes an arbitrary single-mode
state $\hat\rho_{\rm in}$ studied in this paper) can also be
analyzed by applying a numerically-efficient nonclassicality
invariant proposed in Ref.~\cite{Arkhipov2016}. That quantifier is
invariant under any global unitary photon-number-preserving
transformations of the covariance matrix of a Gaussian state.

Thus, we have shown that not all our numerical predictions of PIT
and other photon-number correlations correspond to quantum states,
but only those for $r>r_0$ are nonclassical for the DSTS. To make
this distinction clearer, we plotted in Figs.~\ref{fig05}
and~\ref{fig12} the borderline at $r=r_0$ between the classical
and nonclassical regimes of the DSTS. We emphasize that all SCS
with nonzero squeezing parameter $r$ are nonclassical, which is a
special case of the DSTS for $\Nthermal=0$. Thus, all our
numerical predictions shown in Figs.~\ref{fig04} and~\ref{fig10}
correspond to nonclassical states.

\section{Conclusions}
\label{sec6}

Single- and two-photon blockades have been usually studied in a
driven nonlinear cavity [see Fig.~\ref{fig01}(b)] or cavities [see
Fig.~\ref{fig01}(c)] coupled to a harmonic reservoir (a thermal
bath). Only a few works (including Refs.~\cite{Adam2014,
Lemonde2014}) were devoted to the analysis of single-photon
blockade via quantum nonlinear reservoir engineering.

In this paper we showed that a driven harmonic cavity coupled to a
squeezed reservoir, as schematically shown in Fig.~\ref{fig01}(a),
can generate light exhibiting various types of photon blockade and
related phenomena. These include: two-photon blockade (as defined
in Sec.~\ref{sec2A}), three-photon tunneling (defined in
Sec.~\ref{sec1D}), and three nonstandard types of single-photon
blockade (defined in Sec.~\ref{sec2C}), in addition to standard
single-photon blockade. Our theoretical interest in studying
two-photon blockade~\cite{Adam2013} has been stimulated by a
recent experiment of the Rempe group~\cite{Hamsen2017}.

As shown in Refs.~\cite{Bartolo2016,Minganti2016}, the roles of
the Kerr nonlinear interaction and two-photon dissipation can be
interchanged in the steady states of the systems undergoing these
processes. This might explain why the linear system shown in
Fig.~\ref{fig01}(a), being coupled to a squeezed reservoir,
enables the generation of photon blockade analogously to the
standard Kerr nonlinear systems shown in Figs.~\ref{fig01}(b) and
\ref{fig01}(c) in the dispersive limit. Indeed, a squeezed
reservoir allows for two-photon dissipation.

We considered various types of nonstandard
photon-number-correlation effects by analyzing different
properties of second- and third-order single-time correlation
functions (as listed in Table~\ref{table2}), and two-time
correlations described by $g^{(2)}(\tau)$.

We also simulated these multi-photon effects with squeezed
coherent states and displaced squeezed thermal states, inspired by
the prediction~\cite{Lemonde2014} of single-photon blockade in a
linear system with nonlinear damping. The relation between the
squeezed-state simulations of these effects and their generation
via squeezed-reservoir is explained in Appendix~\ref{appB}.

Photon blockade in nonlinear systems coupled to thermal reservoirs
has already attracted considerable interest, as confirmed by a
number of experimental demonstrations~\cite{Birnbaum2005,
Faraon2008, Lang2011, Hoffman2011, Reinhard2011, Peyronel2012,
Muller2015, Hamsen2017, Snijders2018, Vaneph2018}. Thus, we hope
that the described method of quantum reservoir engineering, which
enables the generation of multi-photon blockade, photon-induced
tunneling, and related phenomena, can also stimulate further
theoretical and experimental research in optical and microwave
photonics~\cite{Gu2017}.

\section*{Acknowledgments}

The authors kindly acknowledge insightful discussions with
Fabrizio Minganti and Wei Qin. J.P. is supported by the GA \v{C}R
project 18-22102S. F.N. is supported in part by: the MURI Center
for Dynamic Magneto-Optics via the Air Force Office of Scientific
Research (AFOSR) (FA9550-14-1-0040), Asian Office of Aerospace
Research and Development (AOARD) (Grant No. FA2386-18-1-4045),
Japan Science and Technology Agency (JST) (via the Q-LEAP program,
and the CREST Grant No. JPMJCR1676), Japan Society for the
Promotion of Science (JSPS) (JSPS-RFBR Grant No. 17-52-50023, and
JSPS-FWO Grant No. VS.059.18N), the FQXi, the NTT PHi Lab and the
RIKEN-AIST Challenge Research Fund. Moreover, F.N. and A.M. are
supported by the Army Research Office (ARO) (Grant No.
W911NF-18-1-0358).

\section*{Appendices}

\appendix

\section{Standard systems for studying conventional and unconventional photon blockade}
\label{appA}

For a better comparison of  the proposed PB system shown in
Fig.~\ref{fig01}(a), we briefly recall here the prototype systems
for generating conventional [see Fig.~\ref{fig01}(b)] and
unconventional [see Fig.~\ref{fig01}(c)] PB effects.

(1) Conventional PB is usually studied in a driven Kerr nonlinear
system described by the Hamiltonian
\begin{equation}
  \hat{H}_a=\omega_c\hat{a}^{\dagger}\hat{a}+\varepsilon\left(\hat{a}e^{i\omega_d
t}+\hat{a}^{\dagger}e^{-i\omega_d
t}\right)+\chi\hat{a}^{\dagger}\hat{a}^{\dagger}\hat{a}\hat{a},
  \label{Ha}
\end{equation}
where $\chi$ is a Kerr nonlinearity proportional to the
third-order susceptibility $\chi^{(3)}$, and the other terms of
this Hamiltonian are the same as in Eq.~(\ref{Hamiltonian1}). The
Hamiltonian~(\ref{Ha}) can be effectively derived (see, e.g.,
Ref.~\cite{Adam2013} and references therein) from the
Jaynes-Cummings model in the dispersive limit (i.e., far off
resonance) describing a driven cavity interacting with a two-level
system (qubit) under the rotating wave approximation. Thus, the
system shown schematically in Fig.~\ref{fig01}(b), can be given by
the Hamiltonian:
\begin{eqnarray}
\hat H^q_a &=& \frac12
\omega\hat\sigma_z+\omega_{c}\hat{a}^{\dagger}\hat{a} +
g(\hat\sigma^{+}\hat a + \hat a^{\dag}\hat\sigma^{-}) \nonumber
\\ &&+\varepsilon(\hat{a}
e^{i\omega_{d}t}+\hat{a}^{\dagger}e^{-i\omega_{d}t}),
  \label{Haq}
\end{eqnarray}
where $\hat\sigma^-$ ($\hat\sigma^+$) is the qubit lowering
(raising) operator; $\sigma_{z}=|e\rangle\langle
e|-|g\rangle\langle g|$ is a Pauli operator; and $|g\rangle$
($|e\rangle$) is the ground (excited) state of the two-level
system.

(2) The prototype Hamiltonian for generating unconventional PB is
given by~\cite{Leonski2004,Adam2006,Liew2010,Bamba2011}:
\begin{eqnarray}
  \hat{H}_{ab} &=& \hat{H}_a + \hat{H}_b
  + J\hat{a}^\dagger\hat{b} +J^*
  \hat{a}\hat{b}^\dagger,
  \label{Hab}
\end{eqnarray}
where
\begin{equation}
  \hat{H}_b = \omega'_c\hat{b}^{\dagger}\hat{b}
  +\varepsilon'\left(\hat{b}e^{i\omega'_d t}
  +\hat{b}^{\dagger}e^{-i\omega'_d t}\right)
  +\chi'\hat{b}^{\dagger}\hat{b}^{\dagger}\hat{b}\hat{b},
  \label{Hb}
\end{equation}
where $\hat b$ ($\hat b^\dagger$) is the annihilation (creation)
operator of the optical mode in the second cavity, $\chi'$ is the
Kerr nonlinearity of the second cavity, and the quantities
$\omega'_c$, $\varepsilon'$, and $\omega'_d $ correspond,
respectively, to $\omega_c$, $\varepsilon$, and $\omega_d $ in
Eq.~(\ref{Hamiltonian1}), but for the second cavity.

In analogy to the derivation of the conventional Kerr-nonlinear
Hamiltonian in Eq.~(\ref{Ha}) from Eq.~(\ref{Haq}), also
Eq.~(\ref{Hab}) can be derived from the two linearly coupled
driven Jaynes-Cummings systems in the dispersive limit. Such a
two-cavity system can be described by:
\begin{eqnarray}
    \hat{H}^q_{ab} &=& \hat{H}^q_a + \hat{H}^q_b
  + J\hat{a}^\dagger\hat{b} +J^*
  \hat{a}\hat{b}^\dagger,
  \label{Habq}
\end{eqnarray}
where $\hat{H}^q_b$ is defined analogously to $\hat{H}^q_a$ in
Eq.~(\ref{Haq}), but for the mode $\hat{b}$ of the second cavity.
This two-atom system can be simplified to include only one atom,
which is the case shown in Fig.~\ref{fig01}(c).

The dissipative evolution of such PB systems has been usually
studied assuming their coupling to a thermal reservoir within the
Lindblad master equation,
\begin{equation}
  \frac{d\hat{\rho}}{dt}= -i[\hat H,\hat\rho]+
  \frac{1}{2}\gamma\big\{ (\Nthermal+1)\Gamma_1[\hat a]\hat{\rho}
  +\Nthermal\Gamma_1[\hat a^\dagger]\hat{\rho} \big\},
 \label{ME5}
\end{equation}
for the reduced density matrix  $\hat\rho$, where the Lindblad
superoperator $\Gamma_1[\hat x]\hat{\rho}$ is defined in
Eq.~(\ref{Lindblad1}) in Appendix~\ref{appB}, $\gamma$ is the
damping rate, and $\Nthermal=\{\exp[\hbar\omega/(k_B T)]-1\}^{-1}$
is the mean thermal photon number.

\section{Master equation for the squeezed-vacuum reservoir}
\label{appB}

Here, we show more explicitly the relation between squeezed states
and a squeezed reservoir by studying the master equation for the
squeezed-vacuum reservoir, given in Eq.~(\ref{ME1}) in its special
case for $|M|=\sqrt{n(n+1)}$. Our presentation is based on
Refs.~\cite{Agarwal1973,Perina1991,ScullyBook} (see also,
e.g.,~\cite{An2004}).

The master equation in Eq.~(\ref{ME1}), with the system
Hamiltonian $\hat H$ in Eq.~(\ref{Hamiltonian}), can be rewritten
more compactly as
\begin{eqnarray}
  \frac{d\hat{\rho}}{dt}&=& -i[\hat H,\hat\rho]+
\frac{1}{2}\gamma\Big\{
(n+1)\Gamma_1[\hat a]\hat{\rho} +n\Gamma_1[\hat a^\dagger]\hat{\rho} \nonumber\\
&&-M\Gamma_2[\hat a]\hat{\rho} -M^*\Gamma_2[\hat
a^\dagger]\hat{\rho} \Big\}, \label{ME2}
\end{eqnarray}
using the superoperators defined as
\begin{eqnarray}
\Gamma_1[\hat x]\hat{\rho} &=&
2\hat{x}\hat{\rho}\hat{x}^{\dagger}-\hat{x}^{\dagger}\hat{x}\hat{\rho}-\hat{\rho}\hat{x}^{\dagger}\hat{x},
\label{Lindblad1}\\
\Gamma_2[\hat x]\hat{\rho} &=&
2\hat{x}\hat{\rho}\hat{x}-\hat{x}\hat{x}\hat{\rho}-\hat{\rho}\hat{x}\hat{x}.
\label{Lindblad2}
\end{eqnarray}
This master equation can be derived by considering a system
described by $\hat H'$ in Eq.~(\ref{Hamiltonian1}) or,
equivalently, $\hat H$ in Eq.~(\ref{Hamiltonian}), with its cavity
mode $\hat a$ being linearly coupled to an infinite set of
reservoir modes $\hat b_k$~\cite{An2004}. We assume that the
reservoir modes $\hat b_k$ are initially in the squeezed vacuum
states,
\begin{eqnarray}
  \ket{\vec \xi\;} &=& \prod_k \ket{\xi_k} = \prod_k \hat S_k(\xi)\ket{0_k},
\label{xi}
\end{eqnarray}
where the $k$th-mode squeezing operator is given by
\begin{equation}
  \hat S_{k}=\exp{(\xi^{\ast}\hat b_{k_0+k}\hat b_{k_0-k}-{\rm
  H.c.})}, \label{S12}
\end{equation}
where $k_0=\omega_c/c$, $\xi=r \exp(i\theta)$ is the usual complex
squeezing parameter, and H.c. denotes the corresponding
Hermitian-conjugate term. Thus, $\hat S_{k}$ in Eq.~(\ref{S12}) is
a two-mode squeezing operator for each $k$. Note that the master
equation in Eq.~(\ref{ME2}) can also be derived for a single-mode
squeezing operator acting on each reservoir mode
$k$~\cite{Perina1991}. The total initial state is assumed to be
$\hat \rho_T=\hat \rho(0)\otimes \ket{\vec \xi\;}\bra{\vec
\xi\;}$; and the total system-reservoir Hamiltonian reads
\begin{eqnarray}
\hat H_T=\hat H' +\sum_k \omega_k \hat b^\dagger_k \hat b_k+\sum_k
g_k(\hat a \hat b_k^\dagger+\hat a^\dagger b_k), \label{HT2}
\end{eqnarray}
where $g_k$ is the coupling strength between the system mode $\hat
a$ and the reservoir mode $\hat b_k$. The standard procedure of
deriving the equation of motion for the reduced density matrix
$\hat \rho$ under the Markov approximation results in the master
equation, given in Eq.~(\ref{ME2}) in the interaction picture,
where
\begin{eqnarray}
  \mean{\hat b^\dagger_k \hat b_{k'}}&=&n \delta_{k {k'}}=\sinh^2(r)\delta_{k
  k'},\nonumber \\
 \mean{\hat b_k\hat
 b_{k'}}&=&-M^*\delta_{{k'}k''}=-\cosh{(r)}\sinh{(r)}
{\rm e}^{i\theta}\delta_{{k'}k''},\quad \label{NM2}
\end{eqnarray}
with $k''=2k_0-k$. By applying the Bogoliubov transformation,
\begin{eqnarray}
  \hat a_s&=&\hat S^\dagger \hat a \hat S=\cosh{(r)}\hat a-\sinh{(r)}e^{i\theta}\hat
  a^\dagger,\nonumber \\
  \hat a^\dagger_s&=&\hat S^\dagger \hat a^\dagger \hat S=\cosh{(r)}\hat a^\dagger-\sinh{(r)}e^{-i\theta}\hat
  a,
\label{Bogoliubov}
\end{eqnarray}
where $\hat S(\xi)$ is the squeezing operator defined in
Eq.~(\ref{SqueezingOp}), the master equation in Eq.~(\ref{ME2})
for $\Delta=0$ reduces, in the squeezed-vacuum frame, to the
standard-form master equation without $\Gamma_2$ terms, i.e.:
\begin{equation}
  \frac{d\hat{\rho}}{dt} = -i[\hat
  H_{s},\hat\rho]+\frac{\gamma}{2}\Gamma_1[\hat{a}_{s}]\hat \rho,
 \label{ME3}
\end{equation}
or, equivalently,
\begin{equation}
  \frac{d\hat{\rho}_s}{dt}
  =-i[\hat H,\hat\rho_{s}]+\frac{\gamma}{2}\Gamma_1[\hat a]\hat{\rho}_s,
 \label{ME4}
\end{equation}
where $\hat{\rho}_s=\hat{S}\rho\hat S^\dagger$ and
\begin{equation}
  \hat{H}_s=\hat S^\dagger \hat H\hat S=\varepsilon(\hat
  a^\dagger_s+\hat a_s).
\end{equation}
As mentioned above, the resonant case $\Delta=0$ is assumed here.
Note that for $\Delta\neq 0$, terms proportional to $\hat a^2$ and
$(\hat a^\dagger)^2$ should be added to the master equations
in~(\ref{ME3}) and~(\ref{ME4}).

\section{Nonclassicality of photon blockade}
\label{appC}

Here we recall that PB is a nonclassical effect. First we show
this for single-PB using the $P$-function approach. And then we
apply another approach for any multi-PB.

We first recall that $(\hat a^\dagger)^2\hat a^2 = \hat n (\hat
n-1)=:\hat n^2:$, where $::$ means the normal ordering of the
creation and annihilation operators. The photon-number variance
$\mean{:(\Delta \hat n)^2:}$ is simply related to $\cor{2}$ as
follows:
\begin{equation}
  \mean{:(\Delta \hat n)^2:} = \mean{:\hat n^2:}- \mean{\hat n}^2
  =[\cor{2}-1]\mean{\hat n}^2,
\end{equation}
where $\Delta \hat n=\hat n-\mean{\hat n}$. So, $\cor{2}<1$ if and
only if the variance is negative:
\begin{equation}
  \mean{:(\Delta \hat n)^2:} = \int {\rm d}^2{\beta}\;
  P(\beta,\beta^*) (|\beta|^2-\mean{\hat n})^2 <0.
\end{equation}
Because  the terms $(|\beta|^2-\mean{\hat n})^2\ge 0$ and
$\mean{:(\Delta \hat n)^2:}<0$, then $P(\beta,\beta^*)$ must also
be negative in some regions of phase space. This means that the
state $\hat\rho$,  which exhibits single-PB, is described by a
non-positive-semidefinite $P(\beta,\beta^*),$ and, thus, has to be
nonclassical.

The nonclassicality of single- and multi-PB can be shown even
faster by recalling the following facts: (1) classical states of
light are either coherent states or their mixtures; (2) coherent
states are characterized by $\cor{k}=1$, for any $k\ge 1$; (3)
$k$-PB requires $\cor{k+1}<1$, or even the sharper condition
$\cor{k+1}<A\equiv \exp(-\mean{\hat n})\le 1$, according to the
refined PB criterion \#1 in Eq.~(\ref{refined_criteria}). So,
single- and multi-PB can occur only for photon-number
distributions which are sharper~\cite{PerinaJr2019a} than that of
a coherent state and, therefore, also sharper than any mixtures of
coherent states. This completes our proofs.

\section{Classical simulation of photon-induced tunneling with thermal states}
\label{appD}

Here we show that usual thermal states can simulate the PIT of an
arbitrary number of photons.

The thermal-state probability $P_n$ of measuring $n$ photons can
be compactly written as $P_n=y x^n$, where $x=\langle \hat
n\rangle y$, $y=1/(1+\langle\hat n\rangle)$, and $\langle\hat
n\rangle\equiv\Bar{n}_{\rm th}=\{\exp[\hbar\omega/(k_B
T)]-1\}^{-1}$. Then the geometric series for the second- and
higher-order correlation functions $\cor{k}$ can be easily
calculated as:
\begin{eqnarray}
 \cor{2}&=&\frac{y}{\mean{\hat n}^{2}}\sum_n x^n n(n-1)=2,\nonumber \\
 \cor{3}&=&\frac{y}{\mean{\hat n}^{3}}\sum_n x^n n(n-1)(n-2)=6,
\end{eqnarray}
These values can also be obtained from Eqs.~(\ref{g2_term})
and~(\ref{g3_term}) in their special cases for $\alpha=r=0$.

By induction, we conclude that for any order $k>1$, the
correlation function $\cor{k}$ for the thermal state with the mean
photon number $\mean{\hat n}$ becomes
\begin{equation}
 \cor{k}=\mean{\hat n}^{-k}\sum_n P_n n^{[k]}=k!,
\end{equation}
where $n^{[k]}=n(n-1)\cdots(n-k+1).$ This implies that for any
$k>1$ and $\mean{\hat n} >0$, the following holds
\begin{equation}
 1 < \cor{k} < \cor{k+1}.
\end{equation}
Thus, thermal states can simulate the PIT of any number of
(thermal) photons. In particular, this includes two- and three-PT,
which are characterized by the conditions: $1< \cor{2}$ and
Eq.~(\ref{3PT}), respectively.

\section{Nonclassical and classical regimes of displaced
squeezed thermal states} \label{appE}

For completeness of our presentation, we show explicitly that the
DSTS, given by $\hat{\rho}(\alpha,\xi,\Nthermal)$, are
nonclassical if the inequality $|\xi| > r_0$, given in
Eq.~(\ref{r_0}) is satisfied.

By defining a phase-dependent quadrature operator
\begin{eqnarray}
  \hat X_{\varphi} &=& \tfrac12
  [\hat a\exp(i\varphi)+\hat a^\dagger\exp(-i\varphi)],
\label{X}
\end{eqnarray}
the minimum value of the normally-ordered variance $\mean{:(\Delta
\hat X_{\varphi})^2:}$ for the DSTS is given by
\begin{eqnarray}
  \min_{\varphi}\mean{:(\Delta \hat X_{\varphi})^2:}
  &\equiv& \mean{:(\Delta \hat  X_{\varphi_0})^2:} \nonumber \\
  &=& \tfrac14 \exp[-2(r-r_0)]-\tfrac14,
\label{varX}
\end{eqnarray}
where $\varphi_0$ denotes the optimal value of the quadrature
phase $\varphi$. In particular, $\varphi_0=0$ for the squeezing
phase $\theta=0$. Moreover, $::$ denotes normal ordering and
$\Delta \hat X_{\varphi}=\hat X_{\varphi}-\mean{\hat
X_{\varphi}}$. It is seen that Eq.~(\ref{varX}) is independent of
the displacement parameter $\alpha$ and, thus, equivalent to the
variance for the squeezed thermal states first derived in
Ref.~\cite{Fearn1988}.

Squeezing occurs if $\mean{:(\Delta X_{\varphi})^2:}<0$. This
normally-ordered variance can be directly calculated from the
corresponding
 $P$-function:\begin{equation}
  \mean{:(\Delta X_{\varphi_0})^2:} = \int {\rm d}^2{\beta}\;
  P(\beta,\beta^*) [X_{\varphi_0}(\beta,\beta^*)-\mean{\hat X_{\varphi_0}}]^2
  <0,
\end{equation}
where
\begin{equation}
   X_{\varphi_{0}} = \tfrac12 [\beta\exp(i\varphi_{0})+ \beta^*\exp(-i\varphi_{0})].
\end{equation}
Because  the term $[...]^2$ is nonnegative and $\mean{:(\Delta
X_{\varphi_0})^2:}$ is negative for any squeezed state, then
$P(\beta,\beta^*)$ has to be negative  in some regions of phase
space. This means that the DSTS for $r>r_0$ are nonclassical. This
result is confirmed by Eq.~(\ref{EP_DSTS}) for the entanglement
potential, and shown in Fig.~\ref{fig12}. Thus, the requirement
$r>r_0$ is the necessary and sufficient condition of the
$P$-function-based nonclassicality for the DSTS. This implies that
any nonclassical DSTS exhibits quadrature squeezing.

In a special case of the SCS, given by $|\alpha,\xi\rangle= \hat
D(\alpha) \hat S(\xi)|0\rangle$, we recover the well-known result
that $r_0=0$, which means that any SCS with  a nonzero squeezing
parameter is nonclassical~\cite{Loudon1987}.

Thus, to show the nonclassicality of the DSTS, we have plotted the
entanglement potential and the squeezing variance in
Figs.~\ref{fig12}(a) and~\ref{fig12}(b), respectively. Moreover,
we plotted the red vertical line at $r=r_0$ in Fig.~\ref{fig05} to
show more explicitly the borderline between the classical and
nonclassical regimes of the DSTS.


%

\end{document}